\documentclass[epj]{svjour}
\usepackage{latexsym}
\usepackage{amsmath,amssymb,amsfonts,wasysym}
\usepackage{graphics}
\usepackage{xcolor}
\usepackage{float}
\usepackage{hyperref}
\usepackage{longtable}
\usepackage{verbatim}
\usepackage{cite}
\numberwithin{equation}{section}

\def\({\left(}
\def\){\right)}

\def\be{\begin{equation}}
\def\ee{\end{equation}}

\input{tcilatex}
\begin{document}

\title{The Effect of Composite Resonances on Higgs decay into two photons.}
\author{A. E. C\'arcamo Hern\'andez, Claudio O. Dib and Alfonso R. Zerwekh%
\thanks{antonio.carcamo@usm.cl} \thanks{claudio.dib@usm.cl} \thanks{alfonso.zerwekh@usm.cl} }
\institute{$^{a,b,c}$Universidad T\'ecnica Federico Santa Mar\'{\i}a and Centro Cient%
\'{\i}fico-Tecnol\'ogico de Valpara\'{\i}so, Casilla 110-V, Valpara\'{\i}so,
Chile}
\date{Received: date / Revised version: date}



%
%

%
\abstract{In scenarios of strongly coupled electroweak symmetry breaking,
heavy composite particles of different spin and parity may arise and cause
observable effects on signals that appear at loop levels. The recently observed process of Higgs to $\gamma \gamma$ at the LHC is one of such signals. We study the new constraints that 
are imposed on composite models from  $H\to \gamma\gamma$, together with the existing constraints from the high precision electroweak tests.  We use an effective chiral Lagrangian to
describe the effective theory that contains the Standard Model spectrum and the extra composites
below the electroweak scale. 
Considering the effective theory  cutoff at $\Lambda = 4\pi v \sim 3 $ TeV,  consistency with
the $T$ and $S$ parameters and the newly observed $H\to \gamma
\gamma$  can be found for a rather restricted range of masses of vector and axial-vector composites from $1.5$ TeV to $1.7$ TeV and $1.8$ TeV to $1.9$ TeV, respectively, and only 
provided a non-standard kinetic mixing between the $W^{3}$ and $B^{0}$ fields is
included.}

%
%
\authorrunning{A. C\'arcamo, C. Dib, A. Zerwekh}
\titlerunning{Composite Resonances in $H\to\gamma\gamma$}
\maketitle
\section{Introduction}

\label{intro} 

One of the possible signals of  composite Higgs boson models is the deviation of the $h\rightarrow \gamma \gamma$ channel from the Standard Model (SM) prediction, as it is a loop process sensitive to heavier virtual states. For instance this signal was predicted in the context of Minimal Walking Technicolor \cite{Hapola:2011sd}. Consequently the recent $h\rightarrow \gamma \gamma$ signal reported by ATLAS and CMS collaborations \cite{atlashiggs,cmshiggs,newtevatron,CMS-PAS-HIG-12-020}, which is very close to the SM prediction, implies an additional constraint on composite models. In this regard, it is important to explore the consequences of this new constraint on composite models, in conjunction with  those previously known from electroweak precision measurements.

Given the recent evidence of the Higgs boson, a strongly interacting sector that is phenomenologically viable nowadays should include this scalar boson in its low energy spectrum, but it is also assumed that vector and axial-vector resonances should appear as well, in a way that the so called Weinberg sum rules \cite{WeinbergSumRules} are satisfied \cite{Appelquist:1998xf,Foadi:2007ue,Foadi:2007se}

Here we formulate this kind of scenario in a general way, without referring to the details of the underlying strong dynamics, by using a low energy effective Lagrangian
which incorporates vector and axial-vector resonances, as well as composite scalars. One of these scalars should be the observed Higgs and the others should be heavier as to avoid detection at the LHC.  Our inclusion of the vector and axial resonances is based on a 4-site Hidden Local Symmetry, which requires three scalar sectors (link fields) responsible for the breaking of the hidden local symmetries. This setup naturally leads to a spectrum that contains three physical scalars.


%

The main reason to still consider strongly interacting mechanisms of electroweak symmetry breaking (EWSB) as alternatives to the Standard Model mechanism is the so called hierarchy problem that arises from the Higgs sector of the SM. This problem is indicative that, in a natural scenario, new physics should appear at scales not much higher than the EWSB scale (say, around a few TeV) in order to stabilize the Higgs mass at scales much lower than the Planck scale ($\sim 10^{19}$ GeV).
An underlying strongly interacting dynamics without fundamental scalars, which becomes
non-perturbative somewhere above the EW scale,  is a possible scenario that gives 
an answer to this problem. The strong dynamics causes the breakdown of the
electroweak symmetry through the formation of condensates in the vacuum \cite{Appelquist:2003,Hirn:2007,Hirn:2008tc,Belyaev:2008yj,Quigg:2009,Hill:2003,Andersen:2011}. 

Many models of strong EWSB have been
proposed which predict the existence of composite particles such as
scalars \cite{Kaplan:1983,Luty:1990,Chivukula:1993,Contino:2009,Contino:2010t,Grojean,ggpr,Burdman:2007,Lodone:2008,Anastasiou:2009,Low:2009,Zerwekh:2010,Barbieri:2007,Panico:2011a,Burdman:2011,Shalom:2011,Panico:2011,Carcamo:2012,Marzocca:2012,Azatov:2012,Pomarol:2012,Barbieri:2012,Bertuzzo:2012,Foadi:2012bb,Alonso:2012px,Alonso:2012pz,Moretti:2013,Pappadopulo:2013vca,Doff:2013dda,Cai:2013ira,Contino:2013kra,Delaunay:2013iia}, 
vectors \cite{Bagger,Casalbuoni:1995,Pelaez:1996,SekharChivukula:2001,Csaki:2003,Barbieri:2008,Cata:2009iy,Barbieri:2010,Carcamo:2010a,S-Orgogozo:11,Pich:2012jv,Falkowski:2011,Bernardini:2011,Torre:2011a},
both scalars and vectors 
\cite{Casalbuoni:1985,Casalbuoni:1986vq,Casalbuoni:1988xm,Dominici:1997zh,Zerwekh:2006,Carcamo:2010,Carcamo:ggtoVV,Carcamo:2011,Torre:2010,Torre:2011b,Vignaroli:2011,SDeCurtis:2012,Redi:2012,Bellazini:2012,Orgogozo:2012,Foadi:2012ga,Contino:2012}
and composite fermions \cite{Kaplan:1991dc,Barbieri:2008b}. These predicted
scalar and vector resonances play a very important role in preserving the
unitarity of longitudinal gauge boson scattering up to the cutoff $\Lambda
\simeq 4\pi v$ \cite{Barbieri:2010,Chivukula:2003,Barbieri:2003pr,Nomura2003,Foadi:2003xa,Georgi:2004iy,SekharChivukula:2008mj,Sanino:2008}%
. One should add that a composite scalar does not have the hierarchy problem
since quantum corrections to its mass are cut off at the compositeness
scale, which is assumed to be much lower than the Planck scale.

In this work we assume a scenario where there is a strongly interacting
sector which possesses a global 
$SU(2)_{L}$ $\times SU(2)_{R}$ symmetry. The
strong dynamics spontaneously breaks this global symmetry down to its
diagonal $SU(2)_{L+R}$ subgroup. As the electroweak gauge group is 
assumed to be contained in the $SU(2)_L\times SU(2)_R$ symmetry,  the
breaking of this symmetry down to the $SU(2)_{L+R}$ subgroup is in fact the
realization of electroweak symmetry breaking. Consequently, the interactions
among the Standard Model particles and all extra composite resonances can be
described by an effective chiral Lagrangian where the $SU(2)_{L}\times
SU(2)_{R}$ is non-linearly realized. The explicit $SU(2)_{L+R}$ that remains
plays the role of a custodial symmetry of the strong sector.

Just as in the SM, the custodial symmetry is explicitly broken by the hypercharge 
coupling $g^{\prime }$ and by  the difference between up- and down-type quark Yukawa
couplings. 
The strong dynamics responsible for EWSB 
in our scenario gives rise to composite 
massive vector and axial vector fields ($V_\mu^a$ and $A_\mu^a$, respectively) 
belonging to the triplet representation of the $SU(2)_{L+R}$
custodial group, 
as well as two composite scalars ($h$ and $H$) and
one pseudoscalar ($\eta$), all singlets under that group. 

We will identify the lightest scalar, $h$, with the state of mass $m_{h}=126$ GeV discovered
at the LHC. All of these composite resonances are assumed to be lighter than
the cutoff $\Lambda \simeq 4\pi v$, so that they explicitly appear as fields
in the effective chiral Lagrangian. Composite states of spin $2$ and higher
are assumed to be heavier than the cutoff, and so are disregarded in this
work.

These composite particles are important signatures of the strongly coupled
scenarios of EWSB and they could manifest themselves either by direct
production or as virtual states in loop corrections. The lack of direct
observation of these particles at the LHC or any previous collider is expected if their masses
are large enough, but  their loop effects may still be detectable. In this
work we study two types of quantities where loop effects are important: the
corrections to the oblique parameters $S$ and $T$ \cite{Peskin:1991sw,Peskin:1991sw2,epsilon-approach,epsilon-approach2,Barbieri:2004,Barbieri-book} and the decay
rate $h\to \gamma\gamma$. Specifically, we use the high precision results on 
$S$ and $T$ and the recent ATLAS and CMS results at the LHC 
on $h\to\gamma\gamma$ to constrain the mass and coupling parameters of the model. 
The rate $h\to\gamma\gamma$ is particularly important in our study as it is a one-loop process
which is sensitive to the existence of extra vector and axial-vector particles.
In this sense, we are studying whether composite models are viable alternatives to electroweak symmetry breaking, given the current experimental success of the Standard Model \cite{PDG}. 

Besides the presence of the heavy vectors, another feature of composite scenarios is that the fermion masses may not be exactly proportional to the scalar-fermion couplings as in the SM. In particular, we found coupling of the Higgs to top quarks to be slightly larger than what is obtained in the SM through a Yukawa term.  

The organization of the paper is as follows. In Sec. II we introduce our
effective Lagrangian that describes the spectrum of the theory. In Sec. III
we describe the calculations of our quantities of interest, i.e. the $T$ and 
$S$ oblique parameters and the rate $h\rightarrow \gamma\gamma$, within our
model. In Sec. IV we study numerically the constraints on the model
parameters, mainly masses and couplings of the extra composite fields, in
order to be consistent with the high precision measurements as well as the
two-photon signal recently observed in the LHC experiments. Finally in Sec.
V we state our conclusions.


\section{The effective chiral Lagrangian with spin-0 and spin-1 fields.}

In this work we formulate our strongly coupled sector by means of an
effective chiral Lagrangian that incorporates the heavy composite states by
means of local hidden symmetries \cite{Bando:1985rf}. 
As shown in Appendix \ref{A1} and described in detail in Ref.\, \cite{Barbieri:2010},
this Lagrangian is based on the symmetry $G=SU\left( 2\right) _{L}\times
SU\left( 2\right)_{C}\times SU\left( 2\right) _{D}\times SU\left( 2\right)
_{R}$. The $SU\left( 2\right)_{C}\times SU\left( 2\right) _{D}$ part is a
hidden local symmetry whose gauge bosons are linear combinations of the
vector and axial-vector composites, and the SM gauge fields [\textsl{cf.} Eq.\ (\ref{VAva})]. The SM gauge group, on the other hand, is contained as a
local form of the $SU\left( 2\right)_{L}\times SU\left( 2\right) _{R}$
global symmetry of the underlying dynamics.

As the symmetry $G$ is spontaneously broken down to the diagonal subgroup $%
SU(2)_{L+C+D+R}$, it is realized in a non-linear way with the inclusion of
three link fields (spin-0 multiplets). These link fields contain two
physical scalars $h$ and $H$, one physical pseudoscalar $\eta$, the three
would-be Goldstone bosons absorbed as longitudinal modes of the SM gauge
fields and the six would-be Goldstone bosons absorbed by the composite
triplets $V_\mu$ and $A_\mu$.

The starting point is the lowest order chiral Lagrangian for the $%
SU(2)_{L}\times SU(2)_{R}/SU(2)_{L+R}$ Goldstone fields, with the addition
of the invariant kinetic terms for the $W$ and $B$ bosons: 
\begin{eqnarray}
\mathcal{L}_{\chi }&=&\frac{v^{2}}{4}\left\langle D_{\mu }UD^{\mu
}U^{\dagger }\right\rangle -\frac{1}{2g^{2}}\left\langle W_{\mu \nu }W^{\mu
\nu }\right\rangle -\frac{1}{2g^{\prime 2}}\left\langle B_{\mu \nu }B^{\mu
\nu }\right\rangle  \notag \\
&& +\frac{c_{WB}}{4}\left\langle U^{\dagger }W_{\mu \nu }UB^{\mu \nu
}\right\rangle .  \label{chiralagrangian}
\end{eqnarray}
Here $\left\langle {\ }\right\rangle$ denotes the trace over the $2\times 2$
matrices, while $U$ is the matrix that contains the SM Goldstone boson fields $\pi
^{a}$ ($a=1,2,3$) after the symmetry is spontaneously broken. $U$  transforms
under $SU(2)_L\times SU(2)_R$ as $U\to g_R U g_L^\dagger$ and can be expressed as
\begin{equation}
U=e^{\frac{i}{v}\pi ^{a}\tau ^{a}},
\end{equation}%
where $\tau ^{a}$ the Pauli matrices. $D_\mu U$ is the covariant derivative
with respect to the SM gauge transformations: 
\begin{equation}
D_{\mu }U=\partial _{\mu }U-iB_{\mu }U+iUW_{\mu },
\end{equation}%
and $W_{\mu\nu}$ and $B_{\mu\nu}$ are the matrix form of the SM tensor
fields, respectively, 
\begin{equation}
W_{\mu \nu }=\partial _{\mu }W_{\nu }-\partial _{\nu }W_{\mu }-i\left[
W_{\mu },W_{\nu }\right] ,\quad B_{\mu \nu }=\partial _{\mu }B_{\nu
}-\partial _{\nu }B_{\mu } ,
\end{equation}%
where $W_\mu=g W_\mu^a \ {\tau^a}/2$ and $B_\mu=g^\prime B_\mu^0\ \tau
^{3}/2 $ are the gauge boson fields in matrix form. Note that we added a
kinetic mixing term $W^{3}-B^{0} $, proportional to a (so far arbitrary)
coupling $c_{WB}$.

The vector and axial-vector composite fields formed due to the underlying
strong dynamics are denoted here as $V_{\mu }=V_{\mu }^{a} \tau ^{a}/\sqrt{2}
$ and $A_{\mu }= A_{\mu }^{a}\tau ^{a}/\sqrt{2}$, respectively. They are
assumed to be triplets under the unbroken $SU(2)_{L+R}$ symmetry. 

Their kinetic and mass terms in the effective Lagrangian can be written as:
\begin{equation}
\mathcal{L}_{V}^{kin}=-\frac{1}{4}\left\langle V_{\mu \nu }V^{\mu \nu
}\right\rangle +\frac{1}{2}M_{V}^{2}\left\langle V_{\mu }V^{\mu
}\right\rangle ,
\end{equation}
\begin{equation}
\mathcal{L}_{A}^{kin}=-\frac{1}{4}\left\langle A_{\mu \nu }A^{\mu \nu
}\right\rangle +\frac{1}{2}M_{A}^{2}\left\langle A_{\mu }A^{\mu
}\right\rangle .
\end{equation}
Here the tensor fields 
$V_{\mu \nu }=\bigtriangledown _{\mu }V_{\nu}-\bigtriangledown _{\nu }V_{\mu }$ 
and $A_{\mu\nu }= \bigtriangledown _{\mu}A_{\nu }-\bigtriangledown _{\nu }A_{\mu }$ 
are written in terms of a covariant derivative in order to include the electroweak gauge symmetry
embedded in $SU(2)_L\times SU(2)_R$ \cite{Barbieri:2010}:
\begin{equation}
\bigtriangledown _{\mu }V_{\nu } =\partial _{\mu }V_{\nu }+\left[
\Gamma_{\mu },V_{\nu }\right] , \qquad \bigtriangledown _{\mu }A_{\nu}
=\partial _{\mu }A_{\nu }+\left[ \Gamma _{\mu },A_{\nu }\right] ,
\end{equation}
where the connection 
$\Gamma _{\mu }$ satisfies $\Gamma _{\mu}^{^{\dagger}}=-\Gamma _{\mu }$ 
and is given by 
\begin{eqnarray}
\Gamma _{\mu }&=&\frac{1}{2}\left[ u^{\dagger }\left( \partial _{\mu}-iB_{\mu}\right) u +u\left( \partial _{\mu }-iW_{\mu }\right) u^{\dagger }\right],\notag\\ 
&&\text{with} \ \ u \equiv \sqrt{U}.
\end{eqnarray}

Assuming that the underlying strong dynamics is invariant under parity, 
the composite fields $V_\mu$ and $A_\mu$ can be included in the effective Lagrangian 
as combinations of gauge vectors of a hidden symmetry, also spontaneously broken. 
In that formulation further interaction terms appear in the effective Lagrangian, as 
derived in Appendix \ref{A1}. The terms that contain one power of $V_\mu$ or $A_\mu$, according 
to  Eq. (\ref{Gaugesector}), 
are given by: 
\begin{eqnarray}
\mathcal{L}_{1V}&=&-\frac{f_{V}}{2\sqrt{2}}\left\langle V^{\mu \nu }\left(
uW_{\mu \nu }u^{\dagger }+u^{\dagger }B_{\mu \nu }u\right) \right\rangle 
\notag \\
&&-\frac{ig_{V}}{2\sqrt{2}}\left\langle V^{\mu \nu }\left[ u_{\mu },u_{\nu }%
\right] \right\rangle \\
&&-\frac{i\kappa f_{A}}{2\sqrt{2}}\left\langle \left( \partial _{\mu }u_{\nu
}-\partial _{\nu }u_{\mu }+\left[ \Gamma _{\mu },u_{\nu }\right] -\left[
\Gamma _{\nu },u_{\mu }\right] \right) \left[ V^{\mu },u^{\nu }\right]
\right\rangle ,  \notag  \label{L1V}
\end{eqnarray}
\begin{eqnarray}
\mathcal{L}_{1A}&=&\frac{f_{A}}{2\sqrt{2}}\left\langle \left( \partial _{\mu
}u_{\nu }-\partial _{\nu }u_{\mu }+\left[ \Gamma _{\mu },u_{\nu }\right] -%
\left[ \Gamma _{\nu },u_{\mu }\right] \right) A^{\mu \nu }\right\rangle 
\notag \\
&& -\frac{if_{A}}{2\sqrt{2}}\left\langle \left( uW_{\mu \nu }u^{\dagger
}+u^{\dagger }B_{\mu \nu }u\right) \left[ A^{\mu },u^{\nu }\right]
\right\rangle ,  \label{L1A}
\end{eqnarray}
where $u_{\mu }=u^{\dagger }_{\mu }=iu^{\dagger }D_{\mu }Uu^{\dagger }$ is a
quantity that transforms covariantly under $SU(2)_{L+R}$. For later
convenience we have also redefined the couplings in terms of the
dimensionless quantities $f_{V}$, $g_{V}$ and $f_{A}$ [see Eqs.\ (\ref%
{Gaugesector}) and (\ref{rel1})], which depend on the
masses of $V_\mu$ and $A_\mu$ according to 
\begin{equation}
f_{V}\equiv \frac{1}{g_C} =\sqrt{\frac{1}{1-\kappa }}\frac{v}{M_{V}}, \quad
g_{V}=\frac{1-\kappa ^{2}}{2}f_{V}, \quad f_{A}=-\kappa f_{V},
\label{couplings}
\end{equation}
where $\kappa = M_V^2/M_A^2$ [see Eq. (\ref{rel1})]. In this way, the
interactions of the vector fields $V_{\mu }^a$ with two longitudinal weak
bosons are characterized by the coupling $g_{V}$, while the interactions of $%
V_{\mu }^a$ with one longitudinal and one transverse gauge boson are
characterized by both $g_{V}$ and $f_{V}$. 
In turn, the interactions of the axial-vector fields $A_{\mu }^a$ with one longitudinal 
and one transverse gauge boson are characterized by the coupling $f_{A}$. 
Finally, the mixing of $V_{\mu }^a$ and of $A_{\mu }^a$ with the SM gauge fields 
are proportional to $gf_{V}$ and $gf_{A}$, respectively.

Now, the terms with two powers of $V_\mu$ and $A_\mu$ as shown in Appendix \ref{A1},
are: 
\begin{eqnarray}
\mathcal{L}_{2V} &=&-\frac{1-\kappa ^{2}}{8}\left\langle \left[ V_{\mu
},V_{\nu }\right] \left[ u^{\mu },u^{\nu }\right] \right\rangle  \notag \\
&&+\frac{\kappa ^{2}}{8}\left\langle \left[ V_{\mu },u_{\nu }\right] \left( %
\left[ V^{\mu },u^{\nu }\right] -\left[ V^{\nu },u^{\mu }\right] \right)
\right\rangle  \notag \\
&&+\frac{i}{4}\left\langle \left[ V^{\mu },V^{\nu }\right] \left( uW_{\mu
\nu }u^{\dagger }+u^{\dagger }B_{\mu \nu }u\right) \right\rangle ,
\end{eqnarray}%
\begin{eqnarray}
\mathcal{L}_{2A} &=&-\frac{1-\kappa ^{2}}{8}\left\langle \left[ A_{\mu
},A_{\nu }\right] \left[ u^{\mu },u^{\nu }\right] \right\rangle  \notag \\
&&+\frac{\kappa ^{2}}{8}\left\langle \left[ A_{\mu },u_{\nu }\right] \left( %
\left[ A^{\mu },u^{\nu }\right] -\left[ A^{\nu },u^{\mu }\right] \right)
\right\rangle  \notag \\
&&+\frac{i}{4}\left\langle \left[ A^{\mu },A^{\nu }\right] \left( uW_{\mu
\nu }u^{\dagger }+u^{\dagger }B_{\mu \nu }u\right) \right\rangle ,
\end{eqnarray}%
\begin{eqnarray}
\mathcal{L}_{1V,1A} &=&-\frac{i\kappa }{2}\left\langle V_{\mu \nu }\left[
A^{\mu },u^{\nu }\right] \right\rangle -\frac{i\kappa }{2}\left\langle
A_{\mu \nu }\left[ V^{\mu },u^{\nu }\right] \right\rangle  \notag \\
&&-\frac{i\kappa }{2}\left\langle \left( \partial _{\mu }u_{\nu }-\partial
_{\nu }u_{\mu }+\left[ \Gamma _{\mu },u_{\nu }\right] -\left[ \Gamma _{\nu
},u_{\mu }\right] \right) \right.  \notag \\
&&\times \left. \left[ V^{\mu },A^{\nu }\right] \right\rangle .
\end{eqnarray}%
The terms with three powers of  $V_\mu$ and $A_\mu$, also
derived in Appendix \ref{A1} and included in Eq.\ (\ref{Gaugesector}), are 
\begin{equation}
\mathcal{L}_{3V}=\frac{ig_{C}}{2\sqrt{2}}\left\langle V^{\mu \nu }\left[
V_{\mu },V_{\nu }\right] \right\rangle .  \label{l3v}
\end{equation}%
\begin{equation}
\mathcal{L}_{3A}=-\frac{\kappa g_{C}}{2\sqrt{2}}\left\langle \left[ A_{\mu
},A_{\nu }\right] \left[ A^{\mu },u^{\nu }\right] \right\rangle ,
\label{l3a}
\end{equation}%
\begin{equation}
\mathcal{L}_{V,2A}=\frac{ig_{C}}{2\sqrt{2}}\left\langle V_{\mu \nu }\left[
A^{\mu },A^{\nu }\right] \right\rangle +\frac{ig_{C}}{\sqrt{2}}\left\langle
A_{\mu \nu }\left[ V^{\mu },A^{\nu }\right] \right\rangle ,  \label{lv2a}
\end{equation}%
\begin{eqnarray}
\mathcal{L}_{A,2V} &=&-\frac{\kappa g_{C}}{2\sqrt{2}}\left\langle \left[
V_{\mu },V_{\nu }\right] \left[ A^{\mu },u^{\nu }\right] \right\rangle \\
&&-\frac{\kappa g_{C}}{2\sqrt{2}}\left\langle \left[ V_{\mu },u_{\nu }\right]
\left( \left[ V^{\mu },A^{\nu }\right] -\left[ V^{\nu },A^{\mu }\right]
\right) \right\rangle .  \notag  \label{la2v}
\end{eqnarray}%
The interactions given in (\ref{l3v})-(\ref{la2v}) are controlled by the
dimensionless parameter $g_{C}$, which is the coupling constant of the
hidden local symmetry $SU\left( 2\right) _{C}$ and $SU\left( 2\right) _{D}$.
In particular, $\mathcal{L}_{3V}$ describes the cubic self-interactions of $%
V_{\mu }$. Notice that, since $g_{C}=1/f_{V}$ [cf. Eq. (\ref{couplings})],
these self-interactions are strong when the mixings between the heavy
vectors and the SM gauge bosons [cf. Eqs.\ (\ref{L1V}, \ref{L1A})] are weak.

Continuing with the expansion given in Eq.\ (\ref{Gaugesector}), 
the quartic self-interactions of $V_{\mu }$ and of $A_{\mu }$ are
proportional to $g_{C}^{2}$ and described by the terms: 
\begin{equation}
\mathcal{L}_{4V}=\frac{g_{C}^{2}}{8}\left\langle \left[ V_{\mu },V_{\nu }%
\right] \left[ V^{\mu },V^{\nu }\right] \right\rangle ,
\end{equation}%
\begin{equation}
\mathcal{L}_{4A}=\frac{g_{C}^{2}}{8}\left\langle \left[ A_{\mu },A_{\nu }%
\right] \left[ A^{\mu },A^{\nu }\right] \right\rangle .
\end{equation}%
\begin{eqnarray}
\mathcal{L}_{2V2A} &=&\frac{g_{C}^{2}}{4}\left\langle \left[ V_{\mu },A_{\nu
}\right] \left( \left[ V^{\mu },A^{\nu }\right] -\left[ V^{\nu },A^{\mu }%
\right] \right) \right\rangle  \notag \\
&&+\frac{g_{C}^{2}}{4}\left\langle \left[ V_{\mu },V_{\nu }\right] \left[
A^{\mu },A^{\nu }\right] \right\rangle .
\end{eqnarray}%
Since $V_{\mu }^{a}$ and $A_{\mu }^{a}$ are linear combinations of the gauge
bosons of the hidden local symmetry $SU\left( 2\right) _{C}\times SU\left(
2\right) _{D}$ and of the SM gauge fields [see Eq.\ (\ref{VAva})], the field strength tensors corresponding to the gauge bosons of
this hidden local symmetry will include the field strength tensors of $%
V_{\mu }^{a}$ and $A_{\mu }^{a}$ as well as those of the SM gauge bosons
[cf. Eqs.\ (\ref{Vmunu}, \ref{Amunu})]. Because of this reason, additional
contact interactions involving the SM gauge fields and Goldstone bosons
having couplings depending on $f_{V}$, $f_{A}$ and $g_{V}$ (see Eq.\ \ref%
{couplings}) will automatically emerge from the invariant kinetic terms for
the gauge bosons of the $SU\left( 2\right) _{C}\times SU\left( 2\right) _{D}$
sector. These \emph{contact interactions} are given by: 
\begin{eqnarray}
\mathcal{L}_{contact} &=&-\frac{f_{A}^{2}}{8}\left\langle \left( \left(
\partial _{\mu }u_{\nu }-\partial _{\nu }u_{\mu }+\left[ \Gamma _{\mu
},u_{\nu }\right] -\left[ \Gamma _{\nu },u_{\mu }\right] \right) \right)
\right.  \notag \\
&&\times \left. \left( \left( \partial ^{\mu }u^{\nu }-\partial ^{\nu
}u^{\mu }+\left[ \Gamma ^{\mu },u^{\nu }\right] -\left[ \Gamma ^{\nu
},u^{\mu }\right] \right) \right) \right\rangle  \notag \\
&&+\frac{g_{V}^{2}}{8}\left\langle \left[ u_{\mu },u_{\nu }\right] \left[
u^{\mu },u^{\nu }\right] \right\rangle  \notag \\
&&-\frac{f_{V}^{2}}{8}\left\langle \left( uW_{\mu \nu }u^{\dagger
}+u^{\dagger }B_{\mu \nu }u\right) \left( uW^{\mu \nu }u^{\dagger
}+u^{\dagger }B^{\mu \nu }u\right) \right\rangle  \notag \\
&&-\frac{if_{V}g_{V}}{4}\left\langle \left[ u^{\mu },u^{\nu }\right] \left(
uW_{\mu \nu }u^{\dagger }+u^{\dagger }B_{\mu \nu }u\right) \right\rangle ,
\label{lcontact}
\end{eqnarray}
and they  ensure that the scattering amplitudes involving SM
particles have good behavior at high energies. For example, as shown in
Ref.\ \cite{Carcamo:2011}, the second term in Eq.\ (\ref{lcontact}) which
contains four derivative terms involving only the SM Goldstone bosons, is
crucial for having a consistent description of high energy WW scattering.

In addition to $V_{\mu }$ and $A_{\mu }$, there are
two composite scalar singlets, $h$ and $H$, and one pseudoscalar singlet, $\eta $. 
We will identify the lightest of these fields, $h$, with the $m=126$ GeV boson 
recently discovered at the LHC. 
The kinetic and mass terms for these spin-0 fields, as well as their interaction terms with 
one power in $h$, $H$ or $\eta$, are derived in Eqs.\ (\ref{scalarpotential}), (\ref{SBsector}), (\ref{VEVs}) and (\ref{rel1}) of
Appendix \ref{A1}, and given by 
\begin{eqnarray}
\mathcal{L}_{h} &=&\frac{1}{2}\partial _{\mu }h\partial ^{\mu }h+\frac{%
m_{h}^{2}}{2}h^{2}-\frac{g_{C}^{2}v}{2\sqrt{2\left( 1-\kappa \right) }}%
h\left\langle V_{\mu }V^{\mu }\right\rangle  \notag \\
&&+\frac{g_{C}^{2}\left( \frac{1}{\sqrt{\kappa }}-\frac{1}{2\sqrt{1-\kappa }}%
\right) v}{\sqrt{2}}h\left\langle A_{\mu }A^{\mu }\right\rangle  \notag \\
&&+\frac{\left[ 2\kappa ^{\frac{3}{2}}-\left( 1-\kappa \right) ^{\frac{3}{2}}%
\right] v}{4\sqrt{2}}h\left\langle u_{\mu }u^{\mu }\right\rangle  \notag \\
&&-\frac{g_{C}\left( \sqrt{1-\kappa }+2\sqrt{\kappa }\right) v}{2}%
h\left\langle A_{\mu }u^{\mu }\right\rangle ,
\end{eqnarray}%
\begin{eqnarray}
\mathcal{L}_{H} &=&\frac{1}{2}\partial _{\mu }H\partial ^{\mu }H+\frac{%
m_{H}^{2}}{2}H^{2}+\frac{g_{C}^{2}v}{2\sqrt{2\left( 1-\kappa \right) }}%
H\left\langle V_{\mu }V^{\mu }\right\rangle  \notag \\
&&+\frac{g_{C}^{2}\left( \frac{1}{\sqrt{\kappa }}+\frac{1}{2\sqrt{1-\kappa }}%
\right) v}{\sqrt{2}}H\left\langle A_{\mu }A^{\mu }\right\rangle  \notag \\
&&+\frac{\left[ 2\kappa ^{\frac{3}{2}}+\left( 1-\kappa \right) ^{\frac{3}{2}}%
\right] v}{4\sqrt{2}}H\left\langle u_{\mu }u^{\mu }\right\rangle  \notag \\
&&+\frac{g_{C}\left( \sqrt{1-\kappa }-2\sqrt{\kappa }\right) v}{2}%
H\left\langle A_{\mu }u^{\mu }\right\rangle ,
\end{eqnarray}%
\begin{eqnarray}
\mathcal{L}_{\eta } &=&\frac{1}{2}\partial _{\mu }\eta \partial ^{\mu }\eta +%
\frac{m_{\eta }^{2}}{2}\eta ^{2}+\frac{g_{C}^{2}v}{\sqrt{1-\kappa }}%
\left\langle V_{\mu }A^{\mu }\right\rangle \eta  \notag \\
&&+\frac{g_{C}\sqrt{1-\kappa }v}{\sqrt{2}}\left\langle V_{\mu }u^{\mu
}\right\rangle \eta .
\end{eqnarray}

In turn, the interaction terms with two powers of these fields, according to Eqs. (\ref{SBsector}), (\ref{VEVs}) and (\ref{rel1}), are given by
\begin{eqnarray}
\mathcal{L}_{2h} &=&\frac{g_{C}^{2}}{16}h^{2}\left\langle V_{\mu }V^{\mu
}\right\rangle +\frac{5g_{C}^{2}}{8}h^{2}\left\langle A_{\mu }A^{\mu
}\right\rangle\notag \\
&&+\frac{1}{32}\left[ \left( 1-\kappa \right) ^{2}+4\kappa ^{2}\right]
h^{2}\left\langle u_{\mu }u^{\mu }\right\rangle\notag \\
&&+\frac{g_{C}\left( 1-5\kappa \right) }{8\sqrt{2}}h^{2}\left\langle A_{\mu
}u^{\mu }\right\rangle ,
\label{L2h}
\end{eqnarray}
\begin{eqnarray}
\mathcal{L}_{2H} &=&\frac{g_{C}^{2}}{16}H^{2}\left\langle V_{\mu }V^{\mu
}\right\rangle +\frac{5g_{C}^{2}}{8}H^{2}\left\langle A_{\mu }A^{\mu
}\right\rangle  \notag \\
&&+\frac{1}{32}\left[ \left( 1-\kappa \right) ^{2}+4\kappa ^{2}\right]
H^{2}\left\langle u_{\mu }u^{\mu }\right\rangle  \notag \\
&&+\frac{g_{C}\left( 1-5\kappa \right) }{8\sqrt{2}}H^{2}\left\langle A_{\mu
}u^{\mu }\right\rangle ,
\label{L2H}
\end{eqnarray}
\begin{eqnarray}
\mathcal{L}_{2\eta } &=&\frac{g_{C}^{2}}{8}\eta ^{2}\left\langle V_{\mu
}V^{\mu }\right\rangle +\frac{g_{C}^{2}}{8}\eta ^{2}\left\langle A_{\mu
}A^{\mu }\right\rangle  \notag \\
&&+\frac{\left( 1-\kappa \right) ^{2}}{16}\eta ^{2}\left\langle u_{\mu
}u^{\mu }\right\rangle  \notag \\
&&+\frac{g_{C}\left( 1-\kappa \right) }{4\sqrt{2}}\eta ^{2}\left\langle
A_{\mu }u^{\mu }\right\rangle ,
\end{eqnarray}
\begin{eqnarray}
\mathcal{L}_{hH} &=&-\frac{g_{C}^{2}}{8}hH\left\langle V_{\mu }V^{\mu
}\right\rangle +\frac{3g_{C}^{2}}{8}hH\left\langle A_{\mu }A^{\mu
}\right\rangle  \notag \\
&&+\frac{4\kappa ^{2}-\left( 1-\kappa \right) ^{2}}{16}hH\left\langle u_{\mu
}u^{\mu }\right\rangle  \notag \\
&&-\frac{g_{C}\left( 1+3\kappa \right) }{4\sqrt{2}}hH\left\langle A_{\mu
}u^{\mu }\right\rangle ,
\end{eqnarray}
\begin{equation}
\mathcal{L}_{h\eta }=-\frac{g_{C}^{2}}{2\sqrt{2}}\left\langle V_{\mu }A^{\mu
}\right\rangle h\eta -\frac{\left( 1-\kappa \right) g_{C}}{4}\left\langle
V_{\mu }u^{\mu }\right\rangle h\eta ,
\end{equation}
\begin{equation}
\mathcal{L}_{H\eta }=\frac{g_{C}^{2}}{2\sqrt{2}}\left\langle V_{\mu }A^{\mu
}\right\rangle H\eta +\frac{\left( 1-\kappa \right) g_{C}}{4}\left\langle
V_{\mu }u^{\mu }\right\rangle H\eta .
\end{equation}

Finally, we also consider the fermion mass and Yukawa terms: 
\begin{eqnarray}
\mathcal{L}_{Y} &=&-\frac{v}{\sqrt{2}}\sum_{i,j}\left( \bar{u}%
_{L}^{(i)}d_{L}^{(i)}\right) U\left( 1+a_{hff}\,\frac{h}{v}+a_{Hff}\,\frac{H%
}{v}\right.  \notag \\
&&+\left. ia_{\eta ff}\,\frac{\eta }{v}\right) 
\begin{pmatrix}
\lambda _{ij}^{u}\,u_{R}^{(j)} \\[0.1cm] 
\lambda _{ij}^{d}\,d_{R}^{(j)}%
\end{pmatrix}%
+h.c.,
\end{eqnarray}%
where $\lambda _{ij}^{u}$ and $\lambda _{ij}^{d}$ are the \emph{up}- and 
\emph{down}-type quarks Yukawa couplings, respectively. Here $a_{hff}$
parametrizes in our model a deviation factor from the SM Higgs-fermion
coupling (in the SM this factor is unity).

Since $V_{\mu}$, $h$ and $H$ contribute to
the elastic WW scattering amplitude, a good asymptotic behavior of the
latter at high energies will depend on the $a_{hWW}$, $a_{HWW}$ and $g_V$ parameters. 
Because of the extra contributions of $H$ and $V_{\mu}$, $a_{hWW}$ will turn out to be different from unity, in contrast to the SM.

Summarizing, in the framework of strongly interacting dynamics for EWSB, the
interactions below the EWSB scale among the SM particles and the extra composites  
can be described by the effective Lagrangian: 
\begin{eqnarray}
\mathcal{L}_{eff} &=&\mathcal{L}_{\chi }+\mathcal{L}_{V}^{kin}+\mathcal{L}%
_{A}^{kin}+\mathcal{L}_{1V}+\mathcal{L}_{1A}+\mathcal{L}_{2V}+\mathcal{L}%
_{2A}  \notag \\
&&+\mathcal{L}_{1V,1A}+\mathcal{L}_{3V}+\mathcal{L}_{3A}+\mathcal{L}_{V,2A}+%
\mathcal{L}_{A,2V}+\mathcal{L}_{4V}  \notag \\
&&+\mathcal{L}_{4A}+\mathcal{L}_{2V2A}+\mathcal{L}_{contact}+\mathcal{L}_{h}+%
\mathcal{L}_{H}+\mathcal{L}_{\eta }  \notag \\
&&+\mathcal{L}_{2h}+\mathcal{L}_{2H}+\mathcal{L}_{2\eta }+\mathcal{L}_{hH}+%
\mathcal{L}_{h\eta }+\mathcal{L}_{H\eta }+\mathcal{L}_{Y}.  \notag \\
&&  \label{Leff}
\end{eqnarray}%

Our effective theory is based on the following assumptions:

\begin{enumerate}
\item The Lagrangian responsible for EWSB has an underlying strong dynamics
with a global $SU(2)_{L}\times SU(2) _{R}$ symmetry which is spontaneously
broken by the strong dynamics down to the $SU(2)_{L+R}$ custodial group. The
SM electroweak gauge symmetry $SU(2)_L\times U(1)_Y$ is assumed to be
embedded as a local part of the $SU(2)_{L}\times SU(2) _{R}$ symmetry. Thus
the spontaneous breaking of $SU(2)_{L}\times SU(2) _{R}$ also leads to the
breaking of the electroweak gauge symmetry down to $U(1)_{em}$.

\item The strong dynamics produces composite heavy vector fields $V_{\mu
}^{a}$ and axial-vector fields $A_{\mu }^{a}$, triplets under the custodial $%
SU(2)_{L+R} $, as well as a composite scalar singlet $h$ with mass $%
m_{h}=126 $ GeV, a heavier scalar singlet $H$, and a heavier pseudoscalar
singlet $\eta$. These fields are assumed to be the only composites lighter
than the symmetry breaking cutoff $\Lambda \simeq 4\pi v$.

\item The heavy fields $V_{\mu }^{a}$ and $A_{\mu }^{a}$ couple to SM
fermions only through their kinetic mixings with the SM gauge bosons.

\item The spin-0 fields $h$, $H$ and $\eta$ interacts with the fermions
only via (proto)-Yukawa couplings.
\end{enumerate}

Our Lagrangian has in total eight extra free parameters: the modified kinetic 
$W^{3}-B^{0}$ mixing coupling $c_{WB}$, the scalar top quark couplings $a_{htt}$, $a_{Htt}$, 
the pseudoscalar top quark coupling $a_{\eta tt}$, the heavy vector and heavy axial-vector 
masses $M_V$ and $M_A$, and the heavy scalar and heavy pseudoscalar
masses $m_H$ and $m_{\eta}$. However, from the expressions in Appendix B we
can see that the oblique $T$ and $S$ parameters have little sensitivity to
the masses of $H$ and $\eta$. 
Therefore, taking into account the experimental bound $600$ GeV $\lesssim$ 
$m_{H},m_{\eta}$ $\lesssim$ $1$ TeV for heavy spin-0 particles, we can constrain 
the couplings of the heavy $H$ and $\eta$ to the top quark, $a_{Htt}$ and $a_{\eta tt}$, 
that enter in the radiative corrections to the masses of $H$ and $\eta$. We are then left with 
six free parameters: $c_{WB}$, $a_{htt}$, $a_{Htt}$, $a_{\eta tt}$, $M_V$ and $M_A$. 
In what follows, we will constrain these parameters by setting the mass $m_h$ at $125.5$ GeV (the recently discovered 
Higgs at the LHC), imposing the aforementioned experimental bound on $m_H$ and $m_\eta$, and imposing consistency 
with the high precision results on the $T$ and $S$ parameters and the current ATLAS and CMS results on the $h\to \gamma\gamma$ rate.


\section{Calculations of the rate $h\rightarrow \protect\gamma \protect\gamma$, the parameters $T$ and $S$ and the masses of $h$, $H$ and $\eta$.}

\begin{figure*}[tbh]\resizebox{1\textwidth}{!}{
\includegraphics{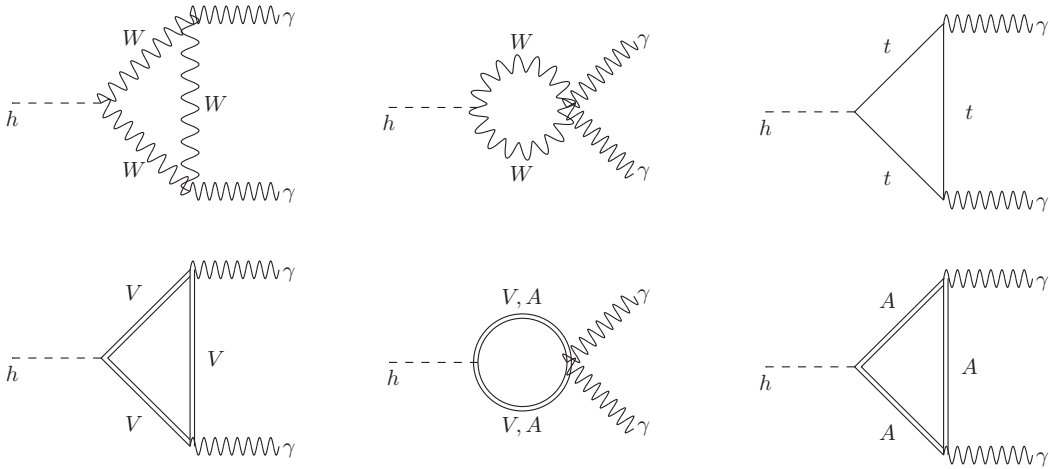}}\vspace{-15cm}
\caption{One loop Feynman diagrams in the Unitary Gauge contributing to the $%
h\rightarrow \protect\gamma \protect\gamma $ decay.}
\label{hto2photons}
\end{figure*}

In the Standard Model, the $h\rightarrow \gamma \gamma $ decay is dominated
by $W$ loop diagrams which can interfere destructively with the subdominant
top quark loop. In our strongly coupled model, the $h\rightarrow \gamma
\gamma $ decay receives additional contributions from loops with charged $V_{\mu
} $ and $A_{\mu }$, as shown in Fig.\ref{hto2photons}. The explicit form
for the $h\rightarrow \gamma \gamma $ decay rate is: 
\begin{eqnarray*}
\Gamma \left( h\rightarrow \gamma \gamma \right) =\frac{\alpha
_{em}^{2}m_{h}^{3}}{256\pi ^{3}v^{2}} &&\left\vert
\sum_{f}a_{hff}N_{c}Q_{f}^{2}F_{1/2}\left( \beta _{f}\right) \right. \\
&&+\left. a_{hWW}F_{1}\left( \beta _{W}\right) +a_{hVV}F_{1}\left( \beta
_{V}\right) \right. \\
&&+\left. a_{hAA}F_{1}\left( \beta _{A}\right) \right\vert ^{2},
\end{eqnarray*}%
where: 
\begin{equation}
a_{hWW}=\frac{2\kappa ^{\frac{3}{2}}-\left( 1-\kappa \right) ^{\frac{3}{2}}}{%
2\sqrt{2}},  \label{ahWW}
\end{equation}%
\begin{equation}
a_{hVV}=-\frac{\left( 1-\kappa \right) ^{\frac{1}{2}}}{2\sqrt{2}},\quad
a_{hAA}=\kappa \left( 1-2\sqrt{\frac{1-\kappa }{\kappa }}\right) a_{hVV}.
\end{equation}%
Here $\beta _{i}$ are the mass ratios $\beta _{i}=m_{h}^{2}/4M_{i}^{2}$,
with $M_{i}=m_{f},M_{W},M_{V}$ and $M_{A}$, respectively, $\alpha _{em}$ is
the fine structure constant, $N_{C}$ is the color factor ($N_{C}=1$ for
leptons, $N_{C}=3$ for quarks), and $Q_{f}$ is the electric charge of the
fermion in the loop. We should recall that $\kappa =M_{V}^{2}/M_{A}^{2}$ and 
$M_{V}=g_{C}v/\sqrt{1-\kappa }$, as shown in Eq.~(\ref{rel1}). From the fermion-loop 
contributions we will keep only the dominant term,
which is the one involving the top quark. 

The dimensionless loop factors $F_{1/2}\left( \beta \right) $ and $%
F_{1}\left( \beta \right) $ (for particles of spin-$1/2$ and spin-$1$ in the loop,
respectively) are \cite{Ellis:1975ap,HppBorn,HppBorn0,HppAnnecy,Gunion:1989we,Spira:1997dg,Djouadi2008,Marciano:2011gm}%
: 
\begin{equation}
F_{1/2}\left( \beta \right) =2\left[ \beta +\left( \beta -1\right) f\left(
\beta \right) \right] \beta ^{-2},
\end{equation}
\begin{equation}
F_{1}\left( \beta \right) =-\left[ 2\beta ^{2}+3\beta +3\left( 2\beta
-1\right) f\left( \beta \right) \right] \beta ^{-2},  \label{F}
\end{equation}
with 
\begin{equation}
f\left( \beta \right) =%
\begin{cases}
\arcsin ^{2}\sqrt{\beta },\hspace{0.5cm}\mathit{for}\hspace{0.2cm}\beta \leq
1 \\ 
-\frac{1}{4}\left[ \ln \left( \frac{1+\sqrt{1-\beta ^{-1}}}{1-\sqrt{1-\beta
^{-1}}}\right) -i\pi \right] ^{2},\hspace{0.5cm}\mathit{for}\hspace{0.2cm}%
\beta >1.%
\end{cases}%
\end{equation}
From the previous expressions it follows that the contribution of heavy vectors  
to $h\rightarrow \gamma \gamma $ strongly dominates over that of axial
vectors when $M_{V}\ll M_{A}$, since in this case we have $%
a_{hVV} \gg a_{hAA}$.

Notice that we have not considered the contribution from contact interactions of gluons, such as
\begin{equation}
\tciLaplace _{ggVV}
=\frac{a_{ggVV}}{\Lambda ^{2}}G_{\mu \nu }G^{\mu
\nu }V_{\alpha }V^{\alpha }.
\end{equation}
to the Higgs production mechanism at the LHC, $gg\to h$, which could have a sizable effect that might contradict the current experiments. Nevertheless, we have checked that this contribution is negligible provided the effective coupling $a_{ggVV}<0.5$. We recall that the heavy vector and heavy axial-vector resonances are colorless, and therefore they do not have renormalizable interactions with gluons.%

Here we want to determine the range of values for $M_V$ and $M_A$ which
is consistent with the $h\rightarrow \gamma \gamma $ results
at the LHC. To this end, we will introduce the ratio $%
R_{\gamma \gamma }$, which normalises the $\gamma\gamma$ signal predicted
by our model relative to that of the SM: 
\begin{eqnarray}
R_{\gamma \gamma }&=&\frac{\sigma\left(pp\rightarrow h \right)\Gamma \left(
h\rightarrow \gamma \gamma \right) }{\sigma\left(pp\rightarrow h
\right)_{SM}\Gamma \left( h\rightarrow \gamma \gamma \right) _{SM}}\notag\\
&\simeq&a^2_{htt}\frac{\Gamma \left( h\rightarrow \gamma \gamma \right) }{\Gamma
\left( h\rightarrow \gamma \gamma \right) _{SM}}.  \label{R_gamma}
\end{eqnarray}
This normalization for  $h\rightarrow \gamma \gamma$ was also done 
in Ref.~\cite{Wang}. Here we have used 
the fact that in our model, single Higgs production is also dominated
by gluon fusion as in the Standard Model.

The inclusion of the extra composite particles also modifies the oblique
corrections of the SM, the values of which have been extracted from high precision
experiments. Consequently, the validity of our model depends on the
condition that the extra particles do not contradict those experimental
results. These oblique corrections are parametrized in terms of the two well
known quantities $T$ and $S$. The $T$ parameter is defined as \cite{Peskin:1991sw,epsilon-approach,Barbieri:2004,Barbieri-book}: 
\begin{equation}
T=\frac{\Pi _{33}\left( 0\right) -\Pi _{11}\left( 0\right) } {M_{W}^{2} \
\alpha _{em}\left( m_{Z}\right) }.
\end{equation}
where $\Pi _{11}\left( 0\right) $ and $\Pi _{33}\left( 0\right) $ are the
vacuum polarization amplitudes at $q^2 =0$ for loop diagrams having gauge bosons $W_{\mu
}^{1}$, $W_{\mu }^{1}$ and $W_{\mu }^{3}$, $W_{\mu }^{3}$ in the external
lines, respectively. 

The one-loop diagrams that contribute 
to the $T$ parameter should include the hypercharge gauge
boson $B_{\mu }^{0}$, since the $g^{\prime }$ coupling is one of the sources
of custodial symmetry  breaking. The other source comes from the difference between 
up- and down-type quark Yukawa couplings.

In turn, the $S$ parameter  is defined by \cite{Peskin:1991sw,epsilon-approach,Barbieri:2004,Barbieri-book}: 
\begin{equation}
S=\frac{4\sin ^{2}\theta _{W}}{\alpha_{em}\left( m_{Z}\right) } \frac{g}{%
g^{\prime }}\frac{d\Pi _{30}\left( q^{2}\right) }{dq^{2}}\biggl|_{q^{2}=0},
\end{equation}
where $\Pi _{30}\left( q^{2}\right) $ is the vacuum polarization amplitude
for a loop diagram having $W_{\mu }^{3}$ and $B_{\mu }$ in the external
lines.

The corresponding Feynman diagrams and details of the lengthy calculation of 
$T$ and $S$ that includes the extra particles in the loops are included in
Appendix \ref{A2}.

\begin{figure*}[htb]
\includegraphics{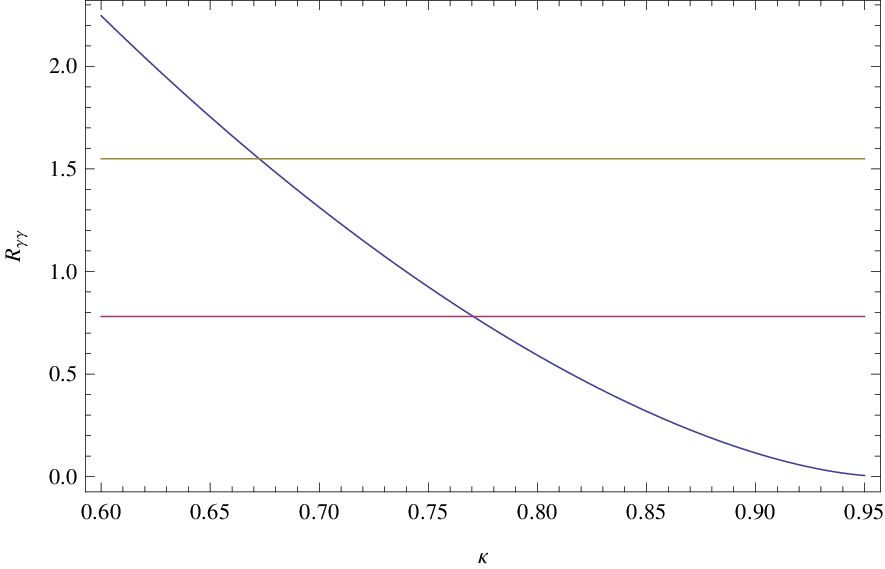}
\caption{The ratio $R_{\protect\gamma \protect\gamma }$ as a function of $%
\protect\kappa$ for $g_Cv=0.8$ TeV and $a_{htt}=2.6$. The horizontal lines
are the $R_{\protect\gamma \protect\gamma }$ experimental values given by
CMS and ATLAS, which are equal to $0.78^{+0.28}_{-0.26}$ and $1.55\pm 0.23$,
respectively \protect\cite{CMS-2013,ATLAS-2013,CMStwiki}.}
\label{fig2}
\end{figure*}

Let us now address the masses of the composite scalars $h$, $H$ and $\eta$.  
In order to fit the particle spectrum observed so far, the model should contain one
scalar with mass at  $125.5$ GeV, which we call $h$, while the heavier $H$ and $\eta $ should have masses satisfying the experimental bound $600$ GeV $\lesssim$ $m_{H},m_{\eta}$ $\lesssim$ $1$~TeV. 
These masses have tree-level
contributions directly from the scalar potential, but also important one-loop contributions from the
Feynman diagrams shown in Appendix \ref{Apendix3}. All these one-loop diagrams have quadratic and
some have also quartic sensitivity to the ultraviolet cutoff $\Lambda$ of the effective theory. 
The calculation details are included in Appendix \ref{Apendix3}. 
As shown there, the contact interaction diagrams
involving $V_\mu$ and $A_\mu$ in the internal lines interfere destructively with
those involving trilinear couplings between the heavy spin-0 and spin-1
bosons. As shown in Eqs.~(\ref{L2h}) and (\ref{L2H}), the quartic couplings of a pair of spin-1 fields with two $h$'s are equal to those with two $H$'s. This implies that contact interactions contribute at one-loop level equally to the $h$ and $H$ masses. On the other hand, since the couplings of two spin-1 fields with one $h$ or one $H$ are different, i.e., $a_{hWW}\neq a_{HWW}$, $a_{hAA}\neq a_{HAA}$, $a_{hWA}\neq a_{HWA}$, $a_{hZA}\neq a_{HZA}$, these loop contributions cause the masses $m_h$ and $m_H$ to be significantly different,  the former being much smaller than the latter 
 (notice that in the Standard Model, $a_{hWW}=b_{hhWW}=1$, implying an exact cancelation of the quartic divergences in the one-loop contributions to the Higgs mass). 
As it turns out, one can easily find conditions where the terms that are quartic in the cutoff cause partial cancelations in $m_h$, but not so in $m_H$ and $m_\eta$, making $m_h$ much lighter that the cutoff $\Lambda$ (e.g. $m_h\sim$ 126 GeV) while $m_H$ and $m_\eta$ remain heavy. 

\begin{figure*}[tbh]
\resizebox{1\textwidth}{!}{
\includegraphics{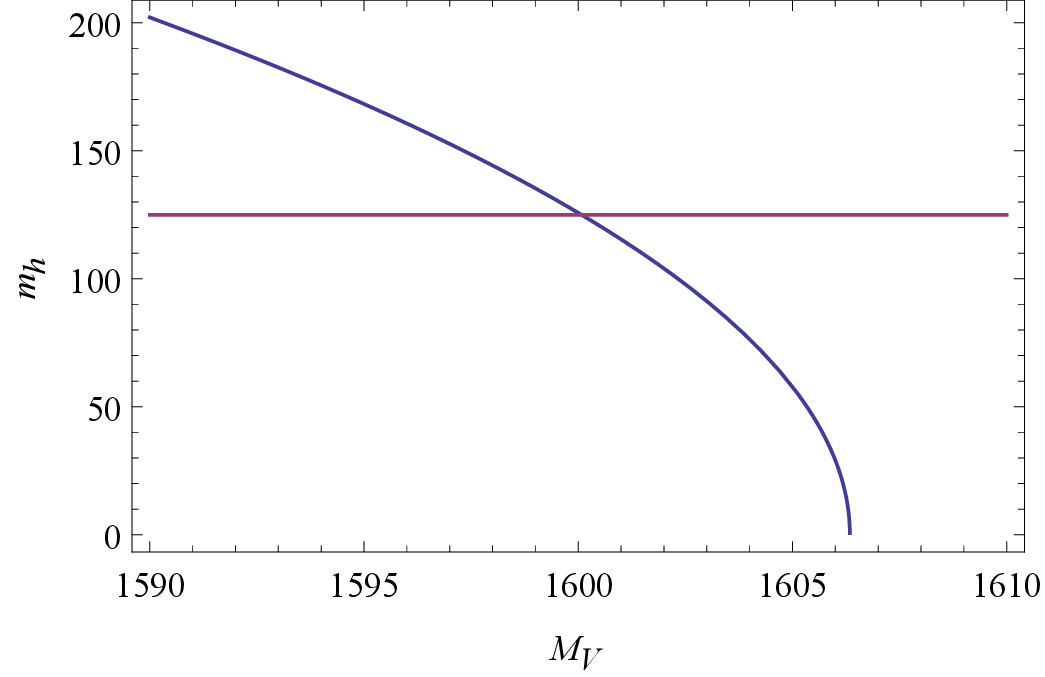}
\includegraphics{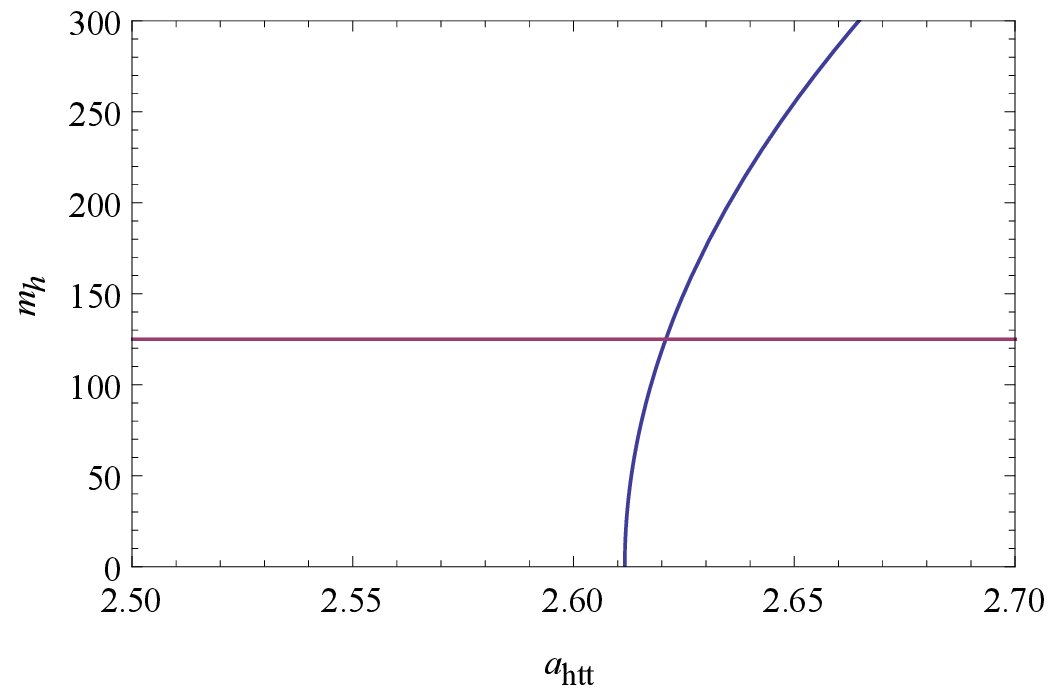}
}
\begin{tabular}{cccc}
\hspace{3cm} &
$\kappa=0.76$, $a_{htt}=2.62$ &
\hspace{4cm} &
$\kappa=0.76$, $M_{V}=1.6$ TeV\\
\hspace{3cm} &
(\ref{fig0}.a) &  
\hspace{4cm} &
(\ref{fig0}.b)
\end{tabular}
\caption{Light scalar mass $m_h$ as function of $M_V$ for $\kappa=0.76$, $a_{htt}=2.62$ TeV  (Fig.\ \protect\ref{fig0}.a), $a_{htt}$ for $\kappa=0.76$, $M_{V}=1.6$ TeV (Fig.\ \protect\ref{fig0}.b). The horizontal line corresponds to the value $126$ GeV for the light Higgs boson mass.}
\label{fig0}
\end{figure*}

%

In Figs.\ \ref{fig0}.a and \ref{fig0}.b we show the sensitivity of the light scalar mass $m_h$ to variations of $M_V$ and $a_{htt}$, respectively. These Figures show that the values of $M_V$ and $a_{htt}$ have an important effect on $m_h$. We can see that these models with composite vectors and axial vectors have the potential to generate scalar masses well below the supposed value around the cutoff, but only in a rather restricted range of parameters. The high sensitivity to the parameters, however, does not exhibit a fine tuning in the usual sense: that deviations from the adjusted point would always bring the mass back to a ``naturally high'' value near the cutoff. Here, the adjustment of parameters could bring the light scalar mass either back up or further below the actual value of 126 GeV.


\section{Numerical study of the effects on $T$, $S$ and $h\rightarrow \protect\gamma \protect\gamma $.}
Let us first study the masses of $h$, $H$ and $\eta$ up to one loop. The one-loop 
diagrams are shown in Appendix \ref{Apendix3}. In order to reduce the parameter space, we assume approximate universality in the quartic couplings of the scalar potential, i.e.  
$\kappa _{1}=\kappa _{2}=\kappa _{3}=\lambda _{1}=\lambda _{3}=1$,  with the sole exception of $\lambda _{2}$ which is given by Eq. (\ref{Relquarticcouplings}) in order that $h$, $H$ and $\eta $ become mass eigenstates (see Appendix \ref{A1} for details). 
As stated in the previous section, we define $h$ to be the recently discovered Higgs boson of mass $125.5$ GeV, while $H$ and 
$\eta$ should be heavier, their masses satisfying the experimental bound $600$ GeV $\lesssim$ $m_{H},m_{\eta}$ $\lesssim$ $1$~TeV. The masses $m_h$, $m_H$ and 
$m_{\eta}$ depend on five free parameters: $a_{htt}$, $a_{Htt}$, $a_{\eta tt}$, $M_V$ and $\kappa=M^2_V/M^2_A$. We will constrain $a_{htt}$, $M_V$ and $\kappa$  by the following observables: the Higgs boson mass $m_h=125.5$ GeV, the two-photon signal $0.78\lesssim R_{\gamma \gamma }\lesssim 1.55$ (where we use $0.78$ and $1.55$, the central values of CMS and ATLAS recent results, respectively) and the oblique parameter $T$. On the other hand, $a_{Htt}$ and $a_{\eta tt}$ will be more loosely constrained by the masses of $H$ and $\eta$. Finally, regarding the modified $W^3-B^0$ mixing coupling $c_{WB}$ [see Eq.~(\ref{chiralagrangian})], it will be constrained by the $S$ parameter.

Let us now analyze in more detail the constraints imposed on our parameters by the values
of $T$ and $S$ obtained from experimental high precision tests. 
The definitions of $T$ and $S$ and their calculation within
our model are given in Appendix \ref{A2} [see Eqs.\ (\ref{T}) and (\ref{Sparameter})]. 
As shown there, our expressions for 
$T$ and $S$ exhibit quartic, quadratic and logarithmic dependence on the
cutoff $\Lambda \sim 3$ TeV. However, the contributions from loops
containing $h$, $H$ and $\eta $ are not very sensitive to the cutoff,
as they do not contain quartic terms in $\Lambda$.
As a consequence, $T$ and $S$ happen to have a rather mild dependence on $m_H$ and $m_\eta$. In contrast, most of the other diagrams, i.e. those containing the spin-1 fields (SM gauge bosons and composite $V_\mu$ or $A_\mu$) have quartic dependence on the cutoff, and as a consequence they are very sensitive to
the masses $M_V$ and $M_A$.

We can separate the contributions to $T$ and $S$ as 
$T = T_{SM} +\Delta T$ and $S = S_{SM} +\Delta S$, where 
\begin{equation}
T_{SM} = -\frac{3}{16\pi \cos ^{2}\theta _{W}}\ln \left( \frac{m_{h}^{2}}{%
m_{W}^{2}}\right), \hspace{0.2cm}S _{SM}= \frac{1}{12\pi}\ln \left( \frac{m_{h}^{2}%
}{m_{W}^{2}}\right)
\end{equation}
are the contributions within the SM, while $\Delta T$ and $\Delta S$ contain
all the contributions involving the extra particles.

\begin{figure*}[tbh]
\resizebox{1\textwidth}{!}{
\includegraphics{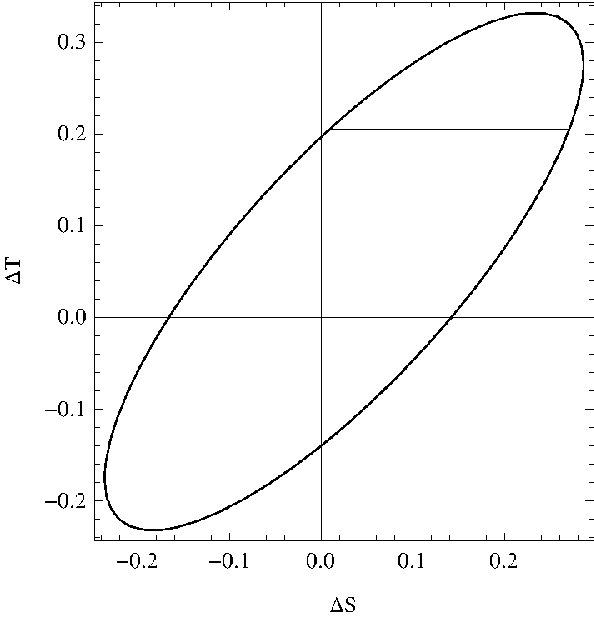}
\includegraphics{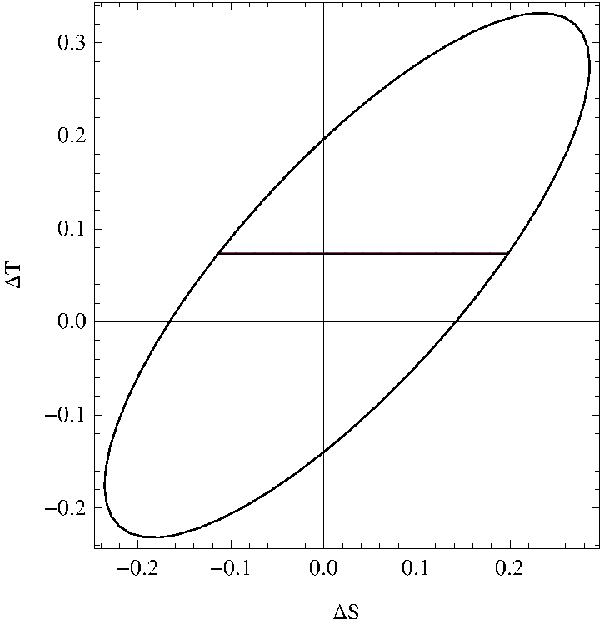}
\includegraphics{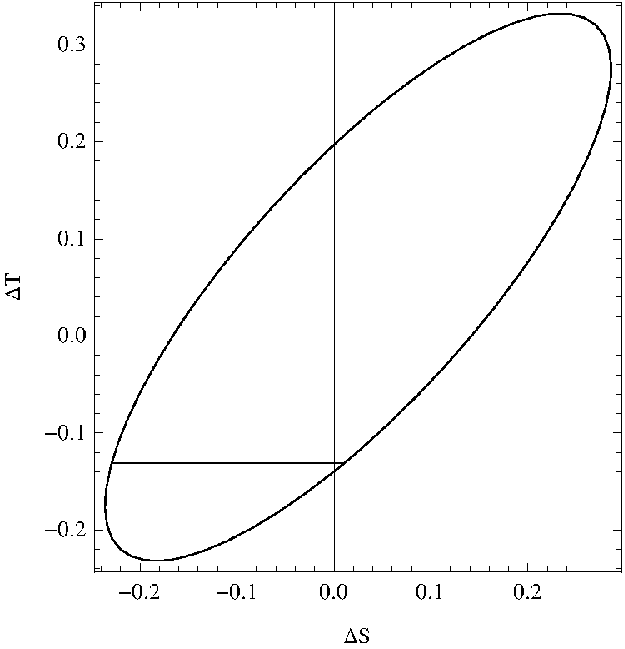}}
{\footnotesize {$M_{A}=1.78$ TeV, $M_{V}=1.55$ TeV}}\hspace{1.7cm}%
{\footnotesize {$M_{A}=1.8$ TeV, $M_{V}=1.6$ TeV}}\hspace{1.7cm}%
{\footnotesize {$M_{A}=1.9$ TeV, $M_{V}=1.7$ TeV}}\newline
{\footnotesize{\ \ \ (\ref{fig1}.a)}}\hspace{7.7cm}{\footnotesize {(\ref{fig1}.b)}%
}\hspace{5.5cm}{\footnotesize {(\ref{fig1}.c)}}\newline
\caption{The $\Delta S-\Delta T$ plane in our model with composite scalars
and vector fields. The ellipses denote the experimentally allowed region at $%
95\%$CL taken from \protect\cite{GFitter}. The origin $\Delta S=\Delta T=0$
corresponds to the Standard Model value, with $m_{h}=125.5$~GeV and $%
m_{t}=176$~GeV. Figures a, b and c correspond to three different sets of
values for the masses $M_V$ and $M_A$, as indicated. The horizontal line
shows the values of $\Delta S$ and $\Delta T$ in the model, as the mixing
parameter $c_{WB}$ varies over the ranges $0.228\leq c_{WB}\leq 0.231$
(Fig.\ \protect\ref{fig1}.a), $0.208\leq c_{WB}\leq 0.212$ (Fig.\ \protect
\ref{fig1}.b), and $0.180\leq c_{WB}\leq 0.182$ (Fig.\ \protect\ref{fig1}%
.c). }
\label{fig1}
\end{figure*}

The experimental results on $T$ and $S$ restrict 
$\Delta T$ and $\Delta S$ to lie inside a region in the $\Delta S-\Delta T$
plane. At the $95\%$C.L. (confident level), these regions are the elliptic contours shown in Figs.\ \ref{fig1}. The origin $\Delta S=\Delta T=0$
corresponds to the Standard Model value, with $m_{h}=125.5$~GeV and $m_{t}=176$~GeV.

We can now study the restrictions on $a_{htt}$, $M_{V}$ and  $\kappa$  
imposed by the value of the Higgs mass $m_h = 125.5$ GeV, by the 
$h\to \gamma\gamma$ signal within the range $0.78\lesssim R_{\gamma \gamma }\lesssim 1.55$, and 
the previously described bounds imposed by the $T$ and $S$ parameters 
at $95\%$ CL.

After scanning the parameter space we find that the heavy
vector mass has to be in the range $1.51$ TeV$\lesssim M_{V}\lesssim 1.75$
TeV in order for the $T$ parameter to be within its bounds. 
Regarding the mass ratio $\kappa
=M_{V}^{2}/M_{A}^{2}$ and the Higgs-top coupling $a_{htt}$, we
find that they have to be in the ranges $0.75\lesssim \kappa \lesssim 0.78$
and $2.53\lesssim a_{htt}\lesssim 2.72$, respectively. Therefore, the Higgs boson, $h$, in this model 
couples strongly with the top quark, yet without spoiling
the perturbative regime in the sense that the condition $\frac{a_{htt}^{2}}{4\pi }\lesssim 1$ is still
fulfilled. 

Concerning the coupling of the top quark to the heavy pseudoscalar $\eta$, 
by imposing the experimental bound on heavy spin-0 particles $600$ GeV $\lesssim m_{\eta}\lesssim 1$ TeV, 
we find that the coupling has the bound $a_{\eta tt}\lesssim 1.39$ for 
$M_{V}\simeq 1.51$ TeV, $\kappa \simeq 0.75$ (lower bounds),  and $ a_{\eta tt}\lesssim 1.46$ for 
$M_{V}\simeq 1.75$ TeV, $\kappa \simeq 0.78$ (upper bounds). 

Regarding the coupling of the top quark to the heavy scalar $H$, we find that it grows with $m_H$ 
and, at the lower bound $m_H \sim 600$ GeV, it is restricted to be $a_{Htt}\simeq
3.53$, which implies that $H$ also couples strongly to the top
quark. 
Lower values of  the coupling $a_{Htt}$ will result if $H$ were lighter than $600$ GeV, 
the experimental bound for heavy spin-0 particles. 
Nevertheless, as before, this large coupling $a_{Htt}$ is still consistent
with the perturbative regime as it satisfies $\frac{a_{Htt}^{2}}{4\pi }\lesssim 1$.

Besides jeopardizing the perturbative regime, these lar-ge couplings may cause violation of unitarity in longitudinal gauge boson scattering. Accordingly, we also checked that the aforementioned values of top quark couplings $a_{htt}$, $a_{Htt}$ and $a_{\eta tt}$ do not cause violation of the unitarity constraint for the scattering of gauge fields into fermion pairs for any energy up to $\sqrt{s}=\Lambda\simeq 3$ TeV.

Let us now study the restrictions imposed by the $h\to \gamma\gamma$ signal, expressed in 
Eq.~(\ref{R_gamma}). We explored the parameter space of $M_V$ and $\kappa$ 
($\kappa=M_V^2/M_A^2$) trying to find values for $R_{\gamma \gamma }$ within a range more or less 
consistent with the ATLAS and CMS results. 
In Fig.~\ref{fig2} we show $R_{\gamma \gamma }$ as a function of 
$\kappa$, for the fixed values $g_Cv=0.8$ TeV and $a_{htt}=2.6$. We chose $a_{htt}=2.6$, which is near the center of the range $2.53\lesssim a_{htt}\lesssim 2.72$ 
imposed by a light Higgs boson mass of $m_h =125.5$ GeV, as previously described.  
In turn, the value $g_Cv$ was chosen in order to fulfill the condition $\frac{g^2_C}{4\pi}\lesssim 1$, which
implies $g_Cv\lesssim 0.9$ TeV. In any case, we checked that our prediction on
$R_{\gamma \gamma }$ stays almost at the same value when the scale $g_Cv$ is
varied from $0.8$ TeV to $1$ TeV. This occurs because the loop function $%
F_{1}\left( \beta \right)$ [see Eq.~(\ref{F})] is rather insensitive to 
$\beta$ in the corresponding range.

Considering the bounds for $\kappa$ shown in Fig.~\ref{fig2}, together with the 
restriction imposed by $T$ to be within its $95\%$ CL, we found that $M_A$ should
have a value in a rather narrow range $1.78$~TeV$- 1.9$~TeV, while  $M_V\lesssim 0.9M_A$. To arrive at this conclusion, we selected three representative values of the axial  vector mass $M_{A}$, namely  at $1.78 $ TeV, $1.8$ TeV and $1.9$ TeV, and then we computed the resulting $T$ and 
$S$ parameters.  We recall that SM point, which corresponds to $\Delta T=\Delta S=0$ is included in the allowed parameter space identified in our analysis. 

For each of these values of $M_A$, we found that the corresponding values of $M_V$ have to be in the ranges $1.54$
TeV $\lesssim M_V\lesssim$ $1.57$ TeV, $1.56$ TeV $\lesssim M_V\lesssim$ $1.59$ TeV  and $1.65$ TeV $\lesssim M_V\lesssim$ $1.68$ TeV in order to have $R_{\gamma\gamma}$ within the range $0.78\lesssim R_{\gamma \gamma }\lesssim 1.55$ and the light Higgs to have a mass $m_h =125.5$ GeV, without spoiling the condition $\frac{a_{htt}^{2}}{4\pi }\lesssim 1$.

Now, continuing with the analysis of the constraints in the $\Delta T -
\Delta S$ plane, we also find that, in order to fulfill the constraint on $%
\Delta S$ as well, an additional condition must be met: for the
aforementioned range of values of $M_V$ and $M_A$, the $S$ parameter turns
out to be unacceptably large, unless a modified $W^3 - B^0$ mixing is added.
Here we introduce this mixing in terms of a coupling $c_{WB}$ [see Eq.~(\ref{chiralagrangian})]. While $\Delta T$ does not depend much on this coupling, 
$\Delta S$ does depend on it, because this coupling enters in the
quadratically divergent loop diagrams involving the $\pi ^{1}\pi
^{1}W^{3}B^{0}$ and $\pi ^{2}\pi ^{2}W^{3}B^{0}$ contact interactions (where 
$\pi^i$ are the SM Goldstone bosons), as well as in the $W^{3}-B^{0}$ tree-level mixing diagram.

In Figs.\ \ref{fig1}.a, \ref{fig1}.b and \ref{fig1}.c we show the allowed
regions for the $\Delta T$ and $\Delta S$ parameters, for the three
sets of values of $M_V$ and $M_A$ previously indicated. 
The ellipses denote the experimentally
allowed region at 95\% C.L., while the horizontal line shows the values of $%
\Delta T$ and $\Delta S$ in the model, as the mixing parameter $c_{WB}$ is
varied over the specified range in each case. The lines are horizontal because $\Delta T$ does not
depend on $c_{WB}$. As seen in the figures, $c_{WB}$
must be in the ranges $0.228\leq c_{WB}\leq 0.231$, $0.208\leq c_{WB}\leq
0.212$ and $0.180\leq c_{WB}\leq 0.182$ for the cases $M_A =$$1.78$ TeV, $%
1.8$ TeV and $1.9$ TeV, respectively.
Notice that  the case $c_{WB}=0$ is clearly excluded, as 
$\Delta S$ would be smaller than its lower bound (the point would be further
to the left of the corresponding ellipse).

As a final remark, we should notice that the model of Ref.~\cite{Pich:2012} is different from ours in the sense that they use a tensor
formulation instead of a vector formulation to describe the heavy spin-1 fields, their spectrum does not include a pseudoscalar and, more important,
the interactions involving more than one heavy spin-1 field are not
considered, so that vertices like $hVV$ and $hAA$ are absent. This implies
that the heavy spin-1 particles do not play a role in the $%
h\to\gamma\gamma $ decay. However, that model does consider an interaction
between the scalar, the SM gauge bosons and the axial vector involving a
covariant derivative of the scalar field, which we do not consider in the present work.


\section{Conclusions.}

We studied a framework of electroweak symmetry breaking without
fundamental scalars, based on an underlying dynamics that becomes strong at
a scale which we assume $\Lambda = 4\pi v \sim 3$ TeV. 
In general, below this scale there could be composite states bound by the strong dynamics. 
The spectrum of composite fields with masses below $\Lambda$ was assumed to consist of
spin-0 and spin-1 fields only, and the interactions among these particles
and those of the Standard Model was described by means of a $SU(2)_{L}\times
SU(2)_{R}/SU(2)_{L+R}$ effective chiral Lagrangian. Specifically, the
composite fields included here were two scalars, $h$ and $H$, one
pseudoscalar $\eta$, a vector triplet $V_\mu^a$ and an axial-vector triplet 
$A_\mu^a$. The lightest scalar, $h$, was taken to be the newly discovered
state at the LHC, with mass $\sim 125.5$ GeV. In this
scenario, in general one must include a deviation of the Higgs-fermion
couplings with respect to the SM, which we denote here as $a_{hff}$. 
In particular, the  coupling of the light Higgs to the top quark, $a_{htt}$, is constrained from the
requirement of having $m_h \simeq 125.5$ GeV and a $h\to\gamma\gamma$ signal in the range $0.78\lesssim R_{\gamma \gamma }\lesssim 1.55$ (where we use $0.78$ and $1.55$, the central values of CMS and ATLAS recent results, respectively).

Our main goal within this framework was to study the consistency of having this
spectrum of composite particles, regarding the loop processes that these extra particles may affect, specifically 
the rate $h\rightarrow \gamma \gamma $, which is a crucial signal for the Higgs, and the high precision 
electroweak parameters $T$ and $S$. 

Besides requiring that  the scalar spectrum 
in our model includes a $125.5$ GeV Higgs boson, the other two spin-0 states, namely $H$ and $\eta$, must be heavier and
within the experimental bounds $600$ GeV $\lesssim$ $m_{H},m_{\eta}$ $\lesssim$ $1$ TeV.
 
We found that the known value of the $T$ parameter at the $95\%$C.L., together with the observed 
$h\rightarrow \gamma \gamma $ rate, restrict the mass of the axial vector to be in the range 
$1.8$ TeV $\lesssim M_{A}\lesssim $ $1.9$ TeV and imply that the mass ratio $\kappa=M_{V}^{2}/M_{A}^{2}$ should satisfy the bound $0.75\lesssim \kappa \lesssim 0.78$.  

In addition, consistency with the experimental value on the $S$ parameter
required the presence of a modified $W^{3}-B^{0}$ mixing, which we
parametrized in terms of a coupling $c_{WB}$. We found that a non-zero value
for this coupling was necessary. The precise value depends on the masses $M_V$
and $M_A$, but within the ranges quoted above, $c_{WB}$ is about 0.2. 

We also found that the $T$ and $S$ parameters have low sensitivity to the
masses of the scalar and pseudoscalar composites, because the dominant
contributions to $T$ and $S$ arise from quartic divergent terms, which only
depend on the heavy vector and axial-vector masses, not on the scalars.
Consequently, from the point of view of the $T$ and $S$ values, the masses
of the heavy scalars and pseudoscalars are not restricted.

Furthermore, we have found that one-loop effects are crucial to account for the mass 
hierarchy between the $125.5$ GeV Higgs boson, $h$, and the heavier states $H$ and  $\eta$. 

The requirement of having a light $125.5$ GeV Higgs boson without spoiling the $T$ parameter and the $h\to\gamma\gamma$ constraints implies that this Higgs boson must couple strongly to the top quark by a factor of about $2$ larger than the Standard Model case. More precisely, the  bound $0.78\lesssim R_{\gamma \gamma }\lesssim 1.55$ constrains the $h$ to top quark coupling to be in the range $2.53\lesssim a_{htt}\lesssim 2.72$. Regarding the heavy scalar $H$, we find that it should have a mass close to its lower bound of $600$ GeV for a $H$ to top quark coupling as low as $a_{Htt}\sim 3.5$. This value implies that  $H$ also couples strongly to the top quark. Lower values of  $a_{Htt}$ will result in an $H$ lighter than the $600$ GeV experimental lower bound. 
On the other hand, we found that the value of the  $\eta$ to top quark coupling $a_{\eta tt}$ can vary from $0$ to about $1.5$.

In summary, we find that composite vectors and axial vectors do have an
important effect on the rate $h\rightarrow\gamma\gamma$, and on the $T$ and $S$ parameters, 
and that one can find values for their masses that are consistent with the experimental values on the previous parameters.
However, one does require an extra $W^3 - B^0$ mixing, which in any
case can be included in the Lagrangian still respecting all the symmetries.
We also find that modified top quark to scalar and to pseudoscalar couplings may appear, in order to have a  spectrum with a light $125.5$ GeV Higgs boson, and with heavier scalar and pseudoscalar states consistent with the experimental allowed range $600$ GeV $\lesssim m_{H},m_{\eta }\lesssim 1$ TeV.

Note that we find quartic and quadratic divergences in both $T$ and $S$,
while deconstructed models only yield logarithmic divergences for both
parameters. This is due to the kinetic mixings between the SM gauge bosons
and the heavy spin-1 fields, which modify their propagators, introducing
different loop contributions to the oblique parameters. Also worth
mentioning is that we did not include composite fermions below the cutoff
scale $\Lambda \sim 3$ TeV, which may affect the oblique $T$ and $S$
parameters as well. An extension of the model could include composite
quarks, a fourth quark generation and/or vector-like quarks. Their effects
on the oblique parameters and on the $h\to\gamma\gamma$ decay rate may be
worth studying. Since the inclusion of extra quarks gives a positive
contribution to the $T$ parameter as shown in Refs.\cite{Barbieri:2007,Lodone:2008,Barbieri:2008b,Barbieri:2012}, we expect that an
extension of the quark sector will increase the upper bound on the axial-vector mass obtained from oblique parameter constraints, because the $T$
parameter takes negative values when the heavy axial-vector mass is
increased. Addressing all these issues requires an additional and careful
analysis that we have left outside the scope of this work.

\subsection*{Acknowledgements}
This work was supported in part by Conicyt (Chile) grant ACT-119 ``Institute
for advanced studies in Science and Technology''. C.D. also received support
from Fondecyt (Chile) grant No.~1130617, and A.Z. from Fondecyt grant
No.~1120346 and Conicyt grant ACT-91 ``Southern Theoretical Physics
Laboratory''. A.E.C.H was partially supported by Fondecyt (Chile), Grant No. 11130115.

\section*{Appendices}

\appendix


\section{: Spontaneously broken gauge theory based on $SU\left( 2\right)_{L}\times
SU\left( 2\right) _{C} \times SU\left( 2\right) _{D} \times U\left(1\right)
_{Y}$ }

\label{A1}

Let us consider a theory with a gauge group of 4 sites, $SU\left( 2\right)
_{L}\times SU\left( 2\right) _{C}\times SU\left( 2\right) _{D}\times U\left(
1\right) _{Y}$. We will assume that the interactions at some energy scale
above a few TeV will cause the condensation of fermion bilinears, in a way
somewhat analogous to what happens in QCD at the chiral symmetry breaking
scale. The gauge symmetry is thus spontaneously broken to $U(1)_{em}$. The
dynamical fields that are left below the symmetry breaking scale will obey
an effective non-linear sigma model Lagrangian of the form 
\begin{equation}
\mathcal{L}=\mathcal{L}_{gauge}+\mathcal{L}_{\chi }^{gauge}-V\left( \Sigma
_{LC},\Sigma _{CD},\Sigma _{DY}\right) ,
\end{equation}%
where $\mathcal{L}_{gauge}$ is the Lagrangian of the gauge fields, $\mathcal{%
L}_{\chi }^{gauge}$ contains the kinetic terms for the Higgs fields that
will break the gauge symmetry when the Higgses acquire vacuum expectation
values, and $V\left( \Sigma_{LC},\Sigma _{CD},\Sigma _{DY}\right) $ is the
Higgs interaction potential. They are given by 
\begin{equation}
\mathcal{L}_{gauge}=-\sum_{I}\frac{1}{2g_{I}^{2}}\left\langle \omega _{\mu
\nu }^{I}\omega ^{\mu \nu I}\right\rangle , \quad \text{with\ } I= L,\ C,\
D,\ Y ,  \label{Lgauge}
\end{equation}

\begin{eqnarray}
\mathcal{L}_{\chi }^{gauge} &=&2v_{LC}^{2}\left\langle D_{\mu }\Sigma
_{LC}D^{\mu }\Sigma _{LC}^{\dagger }\right\rangle  \notag \\
&&+2v_{CD}^{2}\left\langle D_{\mu }\Sigma _{CD}D^{\mu }\Sigma _{CD}^{\dagger
}\right\rangle  \notag \\
&&+2v_{DY}^{2}\left\langle D_{\mu }\Sigma _{DY}D^{\mu }\Sigma _{DY}^{\dagger
}\right\rangle ,  \label{l1}
\end{eqnarray}%
and 
\begin{eqnarray}
V\left( \Sigma _{LC},\Sigma _{CD},\Sigma _{DY}\right) &=&-\frac{\mu
_{1}^{2}v_{LC}^{2}}{2}\left\langle \Sigma _{LC}\Sigma _{LC}^{\dag
}\right\rangle \label{scalarpotential}  \\
&&-\frac{\mu _{2}^{2}v_{CD}^{2}}{2}\left\langle \Sigma _{CD}\Sigma
_{CD}^{\dag }\right\rangle  \notag \\
&&-\frac{\mu _{3}^{2}v_{DY}^{2}}{2}\left\langle \Sigma _{DY}\Sigma
_{DY}^{\dag }\right\rangle  \notag \\
&&+\frac{\lambda _{1}v_{LC}^{4}}{4}\left( \left\langle \Sigma _{LC}\Sigma
_{LC}^{\dag }\right\rangle \right) ^{2}  \notag \\
&&+\frac{\lambda _{2}v_{CD}^{4}}{4}\left( \left\langle \Sigma _{CD}\Sigma
_{CD}^{\dag }\right\rangle \right) ^{2}  \notag \\
&&+\frac{\lambda _{3}v_{DY}^{4}}{4}\left( \left\langle \Sigma _{DY}\Sigma
_{DY}^{\dag }\right\rangle \right) ^{2}  \notag \\
&&+\kappa _{1}v_{CD}^{2}v_{LC}^{2}  \notag \\
&&\times \left\langle \Sigma _{CD}\Sigma _{LC}^{\dag }\Sigma _{LC}\Sigma
_{CD}^{\dag }\right\rangle  \notag \\
&&+\kappa _{2}v_{LC}^{2}v_{DY}^{2}  \notag \\
&&\times \left\langle \Sigma _{DY}\Sigma _{LC}^{\dag }\Sigma _{LC}\Sigma
_{DY}^{\dag }\right\rangle  \notag \\
&&+\kappa _{3}v_{CD}^{2}v_{DY}^{2}  \notag \\
&&\times \left\langle \Sigma _{DY}\Sigma _{CD}^{\dag }\Sigma _{CD}\Sigma
_{DY}^{\dag }\right\rangle .\nonumber  
\end{eqnarray}%
The covariant derivates are defined as 
\begin{equation}
D_{\mu }\Sigma _{IJ}=\partial _{\mu }\Sigma _{IJ}-i\omega _{\mu }^{I}\Sigma
_{IJ}+i\Sigma _{IJ}\omega _{\mu }^{J},  \label{l2}
\end{equation}%
where $\omega _{\mu }^{I}=\left( W_{\mu },\widetilde{v}_{\mu },\widetilde{a}%
_{\mu },B_{\mu }\right) $ with 
\begin{eqnarray}
B_{\mu } &=&\frac{g^{\prime }}{2}B_{\mu }^{0}\tau ^{3},\quad W_{\mu }=\frac{g%
}{2}W_{\mu }^{a}\tau ^{a},\quad \\
v_{\mu } &=&\frac{g_{C}}{2}v_{\mu }^{a}\tau ^{a}\quad a_{\mu }=\frac{g_{C}}{2%
}a_{\mu }^{a}\tau ^{a},
\end{eqnarray}%
where it has been assumed that $g_{C}=g_{D}$ and the indices $I,J$ stand for 
$I,J=L,C,D,Y$. In turn, the field strength tensors are generically given by 
\begin{equation}
\omega _{\mu \nu }^{I}=\mathcal{\partial }_{\mu }\omega _{\nu }^{I}-\mathcal{%
\partial }_{\nu }\omega _{\mu }^{I}-i\left[ \omega _{\mu }^{I},\omega _{\nu
}^{I}\right] .
\end{equation}%
To ensure the correct normalization for the Goldstone bosons kinetic terms, $%
\Sigma _{LC}$, $\Sigma _{DY}$ and $\Sigma _{CD}$ are defined as 
\begin{eqnarray}
\Sigma _{LC} &=&\left( 1+\frac{\eta -\frac{1}{\sqrt{2}}h+\frac{1}{\sqrt{2}}H%
}{4v_{LC}}\right) U_{LC},  \\
\quad \text{with }\ U_{LC} &=&\exp \left[ \frac{i}{4v_{LC}}\left( \pi -\frac{%
1}{\sqrt{2}}\sigma +\frac{1}{\sqrt{2}}\rho \right) \right] ,\nonumber
\end{eqnarray}

\begin{eqnarray}
\Sigma _{DY} &=&\left( 1+\frac{-\eta -\frac{1}{\sqrt{2}}h+\frac{1}{\sqrt{2}}H%
}{4v_{DY}}\right) U_{DY},\quad   \\
\text{with }\ U_{DY} &=&\exp \left[ \frac{i}{4v_{DY}}\left( -\pi -\frac{1}{%
\sqrt{2}}\sigma +\frac{1}{\sqrt{2}}\rho \right) \right] ,\nonumber
\end{eqnarray}

\begin{eqnarray}
\Sigma _{CD} &=&\left( 1+\frac{h+H}{4v_{CD}}\right) U_{CD},\quad  \notag \\
\text{with }\ U_{CD} &=&\exp \left[ \frac{i}{4v_{CD}}\left( \sigma +\rho
\right) \right] ,\quad  \notag \\
v_{LC} &=&v_{DY},
\end{eqnarray}%
where $\pi =\pi ^{a}\tau ^{a}$, $\sigma =\sigma ^{a}\tau ^{a}$ and $\rho
=\rho ^{a}\tau ^{a}$, being $\pi ^{a}$, $\sigma ^{a}$ and $\rho ^{a}$ the
Goldstone bosons associated with the SM gauge bosons, the heavy vectors and
heavy axial vectors, respecttively, and $\tau ^{a}$ the usual Pauli
matrices. In turn, $h$ and $H$ are the massive scalars and $\eta $ is the
massive pseudoscalar.

It is worth mentioning that $h$, $H$ and $\eta $ are physical scalar fields
when the following relations are fulfilled: 
\begin{equation}
\lambda _{1}=\lambda _{3},\quad\kappa _{1}=\kappa _{3},\quad%
\lambda _{2}=\sqrt{2\kappa _{2}+\lambda _{3}^{2}}\frac{v_{LC}}{v_{CD}}.
\label{Relquarticcouplings}
\end{equation}%

The three Higgs doublets acquire vacuum expectation values, thus causing the
spontaneous breaking of the $SU(2)_{L}$\newline
$\times SU(2)_{C}\times SU(2)_{D}\times
U(1)_{Y}$ local symmetry down to $U(1)_{\text{em}}$, while the global group $%
G=SU(2)_{L}\times SU(2)_{C}\times SU(2)_{D}\times SU(2)_{R}$ is broken to
the diagonal subgroup $H=$ $SU(2)_{L+C+D+R}$. The Goldstone boson fields $%
U_{IJ} $ can be put in the form 
\begin{eqnarray}
U_{IJ} &=&\xi _{I}\xi _{J}^{\dagger },\quad \text{where }\ U_{IJ}\in \frac{%
SU(2)_{I}\times SU(2)_{J}}{H},\   \notag \\
I,J &=&L,C,D,Y.
\end{eqnarray}%
These $\xi _{I}$ transform under the full $SU(2)_{L}\times SU(2)_{C}\times
SU(2)_{D}\times U(1)_{Y}$\ as $\xi _{I}\rightarrow g_{I}\xi _{I}h^{^{\dagger
}}$. Choosing a gauge transformation $g_{I}=\xi _{I}^{\dagger }$ we can
transfer the would-be Goldstone bosons to degrees of freedom of the gauge
fields: 
\begin{equation}
U_{IJ}\rightarrow \xi _{I}^{\dagger }U_{IJ}\xi _{J}=1,\quad \omega _{\mu
}^{I}\rightarrow \xi _{I}^{\dagger }\omega _{\mu }^{I}\xi _{I}+i\xi
_{I}^{\dagger }\partial _{\mu }\xi _{I}=\Omega _{I}^{\mu },
\end{equation}%
and the Lagrangian of Eq.\ (\ref{l1}) reduces to: 
\begin{eqnarray}
\mathcal{L}_{\chi }^{gauge} &=&2v_{LC}^{2}\left( 1+\frac{\eta -\frac{1}{%
\sqrt{2}}h+\frac{1}{\sqrt{2}}H}{4v_{LC}}\right) ^{2}\left\langle \left(
\Omega _{\mu }^{L}-\Omega _{\mu }^{C}\right) ^{2}\right\rangle  \notag \\
&&+2v_{LC}^{2}\left( 1+\frac{-\eta -\frac{1}{\sqrt{2}}h+\frac{1}{\sqrt{2}}H}{%
4v_{LC}}\right) ^{2}  \notag \\
&&\times \left\langle \left( \Omega _{\mu }^{D}-\Omega _{\mu }^{Y}\right)
^{2}\right\rangle  \notag \\
&&+2v_{CD}^{2}\left( 1+\frac{h+H}{4v_{CD}}\right) ^{2}\left\langle \left(
\Omega _{\mu }^{C}-\Omega _{\mu }^{D}\right) ^{2}\right\rangle  \notag \\
&&+\frac{1}{2}\partial _{\mu }h\partial ^{\mu }h+\frac{1}{2}\partial _{\mu
}H\partial ^{\mu }H+\frac{1}{2}\partial _{\mu }\eta \partial ^{\mu }\eta .
\end{eqnarray}%
Specifically, we will do a partial gauge fixing resulting in $\xi _{Y}=\xi
_{L}^{^{\dagger }}=e^{{i\pi }/{4v_{LC}}}$ and $\xi _{C}=\xi _{D}=1$, which
implies that $\sigma =\rho =0$ and $U_{YD}=U_{CL}$. This gauge fixing
corresponds to the unitary gauge where the Goldstone boson triplets $\sigma $
and $\rho $ are absorbed as longitudinal modes of $\Omega _{\mu }^{C}$ and $%
\Omega _{\mu }^{D}$. These fields now transform under $SU\left( 2\right)
_{L}\times SU\left( 2\right) _{R}$ according to 
\begin{equation}
\Omega _{\mu }^{C,D}\rightarrow h\Omega _{\mu }^{C,D}h^{\dagger }+ih\partial
_{\mu }h^{\dagger }.
\end{equation}%
The $\Omega _{\mu }^{C}$ and $\Omega _{\mu }^{D}$ can be decomposed with
respect to parity as 
\begin{equation}
\Omega _{\mu }^{C}=v_{\mu }+a_{\mu },\qquad \Omega _{\mu }^{D}=v_{\mu
}-a_{\mu },
\end{equation}%
so that under $SU\left( 2\right) _{L}\times SU\left( 2\right) _{R}$ one has
the following transformations: 
\begin{equation}
v_{\mu }\rightarrow hv_{\mu }h^{\dagger }+ih\partial _{\mu }h^{\dagger
},\qquad a_{\mu }\rightarrow ha_{\mu }h^{\dagger }.
\end{equation}%
Defining 
\begin{eqnarray}
v_{\mu \nu } &=&\partial _{\mu }v_{\nu }-\partial _{\nu }v_{\mu }-i\left[
v_{\mu },v_{\nu }\right] \quad \text{and}  \notag \\
\quad D_{\mu }^{V}a_{\nu } &=&\partial _{\mu }a_{\nu }-i\left[ v_{\mu
},a_{\nu }\right] ,
\end{eqnarray}%
we can write the interactions of the gauge sector of Eq.\ (\ref{Lgauge}) in
the form \cite{Barbieri:2008,Barbieri:2010}: 
\begin{eqnarray}
\mathcal{L}_{gauge} &=&\mathcal{L}_{gauge,SM}-\frac{1}{2g_{C}^{2}}%
\left\langle \left( v_{\mu \nu }-i\left[ a_{\mu },a_{\nu }\right] \right)
^{2}\right\rangle  \notag \\
&&-\frac{1}{2g_{C}^{2}}\left\langle \left( D_{\mu }^{V}a_{\nu }-D_{\nu
}^{V}a_{\mu }\right) ^{2}\right\rangle .
\end{eqnarray}

Now, due to mixing with the SM fields, $v_{\mu }$ and $a_{\mu }$ are not
mass eigenstates. The vector and axial-vector mass eigenstates as $V_{\mu }$
and $A_{\mu }$, respectively, are actually given by the following relations 
\cite{Barbieri:2008,Barbieri:2010}: 
\begin{equation}
V_{\mu }=v_{\mu }-i\Gamma _{\mu },\qquad A_{\mu }=a_{\mu }+\frac{\kappa }{2}%
\,u_{\mu },  \label{VAva}
\end{equation}%
where $\kappa $ will be determined below, and $\Gamma _{\mu }$ is defined
as 
\begin{eqnarray}
\Gamma _{\mu } &\equiv &\frac{1}{2i}\left( \Omega _{\mu }^{Y}+\Omega _{\mu
}^{L}\right)  \notag \\
&=&\frac{1}{2}\left[ u^{\dagger }\left( \partial _{\mu }-iB_{\mu }\right)
u+u\left( \partial _{\mu }-iW_{\mu }\right) u^{\dagger }\right] .
\end{eqnarray}%
Considering these definitions, the strength tensors satisfy the following
identities: 
\begin{equation}
v_{\mu \nu }=V_{\mu \nu }-i\left[ V_{\mu },V_{\nu }\right] +\frac{i}{4}\left[
u_{\mu },u_{\nu }\right] +\frac{1}{2}\left( uW_{\mu \nu }u^{\dagger
}+u^{\dagger }B_{\mu \nu }u\right) ,  \label{Vmunu}
\end{equation}%
\begin{equation}
a_{\mu \nu }=A_{\mu \nu }-\frac{\kappa }{2}u_{\mu \nu }-i\left[ V_{\mu
},A_{\nu }-\frac{\kappa }{2}u_{\nu }\right] +i\left[ V_{\nu },A_{\mu }-\frac{%
\kappa }{2}u_{\mu }\right] ,  \label{Amunu}
\end{equation}%
where 
\begin{equation}
W_{\mu \nu }=\partial _{\mu }W_{\nu }-\partial _{\nu }W_{\mu }-i\left[
W_{\mu },W_{\nu }\right] ,\quad B_{\mu \nu }=\partial _{\mu }B_{\nu
}-\partial _{\nu }B_{\mu },
\end{equation}%
\begin{equation}
V_{\mu \nu }=\bigtriangledown _{\mu }V_{\nu }-\bigtriangledown _{\nu }V_{\mu
}=\partial _{\mu }V_{\nu }-\partial _{\nu }V_{\mu }+\left[ \Gamma _{\mu
},V_{\nu }\right] -\left[ \Gamma _{\nu },V_{\mu }\right] ,
\end{equation}%
\begin{equation}
A_{\mu \nu }=\bigtriangledown _{\mu }A_{\nu }-\bigtriangledown _{\nu }A_{\mu
}=\partial _{\mu }A_{\nu }-\partial _{\nu }A_{\mu }+\left[ \Gamma _{\mu
},A_{\nu }\right] -\left[ \Gamma _{\nu },A_{\mu }\right] ,
\end{equation}%
\begin{equation}
u_{\mu \nu }=\partial _{\mu }u_{\nu }-\partial _{\nu }u_{\mu }+\left[ \Gamma
_{\mu },u_{\nu }\right] -\left[ \Gamma _{\nu },u_{\mu }\right] .
\end{equation}

With these definitions and the aforementioned gauge fixing, the symmetry
breaking sector of the Lagrangian becomes 
\begin{eqnarray}
\mathcal{L}_{\chi }^{gauge} &=&4v_{LC}^{2}\left( 1+\frac{H-h}{2\sqrt{2}v_{LC}%
}+\frac{\left( H-h\right) ^{2}}{32v_{LC}^{2}}+\frac{\eta ^{2}}{16v_{LC}^{2}}%
\right)  \notag \\
&\times &\left[ \left\langle V_{\mu }V^{\mu }\right\rangle +\left\langle
A_{\mu }A^{\mu }\right\rangle \right.  \notag \\
&&+\left. \left( \frac{1-\kappa }{2}\right) ^{2}\left\langle u_{\mu }u^{\mu
}\right\rangle +\left( 1-\kappa \right) \left\langle A_{\mu }u^{\mu
}\right\rangle \right]  \notag \\
&&+4v_{LC}\eta \left( 1+\frac{H-h}{4\sqrt{2}v_{LC}}\right)  \notag \\
&&\times \left[ \left\langle V_{\mu }A^{\mu }\right\rangle +\left( \frac{%
1-\kappa }{2}\right) \left\langle V_{\mu }u^{\mu }\right\rangle \right] 
\notag \\
&&+8v_{CD}^{2}\left( 1+\frac{h+H}{2v_{CD}}+\frac{h^{2}+2hH+H^{2}}{%
16v_{CD}^{2}}\right)  \notag \\
&&\times \left[ \left\langle A_{\mu }A^{\mu }\right\rangle +\frac{\kappa ^{2}%
}{4}\left\langle u_{\mu }u^{\mu }\right\rangle -\kappa \left\langle A_{\mu
}u^{\mu }\right\rangle \right]  \notag \\
&&+\frac{1}{2}\partial _{\mu }\eta \partial ^{\mu }\eta +\frac{1}{2}\partial
_{\mu }h\partial ^{\mu }h+\frac{1}{2}\partial _{\mu }H\partial ^{\mu }H,
\end{eqnarray}%
where one defines: 
\begin{equation}
u_{\mu }\equiv \Omega _{\mu }^{Y}-\Omega _{\mu }^{L}=iu^{\dagger }D_{\mu
}Uu^{\dagger },\quad \text{with}\quad U=u^{2}=e^{\frac{i}{v}\pi ^{a}\tau
^{a}},
\end{equation}

\begin{equation}
\text{and}\quad D_{\mu }U=\partial _{\mu }U-iB_{\mu }U+iUW_{\mu },
\end{equation}

and where $D_{\mu }$ is a covariant derivative containing the SM gauge
fields only.

With the further replacement $V_{\mu }\rightarrow \frac{g_{C}}{\sqrt{2}}%
V_{\mu }$, $A_{\mu }\rightarrow \frac{g_{C}}{\sqrt{2}}A_{\mu }$, the gauge
sector of the Lagrangian becomes: \vspace{-0.2cm}
\begin{eqnarray}
\mathcal{L}_{gauge} &=&\mathcal{L}_{gauge,SM}-\frac{1}{4}\left\langle V_{\mu
\nu }V^{\mu \nu }\right\rangle -\frac{1}{4}\left\langle A_{\mu \nu }A^{\mu
\nu }\right\rangle   \notag \\
&&-\frac{i\left( 1-\kappa ^{2}\right) }{8g_{C}^{2}}\left\langle \left[
u^{\mu },u^{\nu }\right] \left( uW_{\mu \nu }u^{\dagger }+u^{\dagger }B_{\mu
\nu }u\right) \right\rangle   \notag \\
&&-\frac{\kappa ^{2}}{8g_{C}^{2}}\left\langle u_{\mu \nu }u^{\mu \nu
}\right\rangle +\frac{\left( 1-\kappa ^{2}\right) ^{2}}{32g_{C}^{2}}%
\left\langle \left[ u_{\mu },u_{\nu }\right] \left[ u^{\mu },u^{\nu }\right]
\right\rangle   \notag \\
&&-\frac{1}{2\sqrt{2}g_{C}}\left\langle V^{\mu \nu }\left( uW_{\mu \nu
}u^{\dagger }+u^{\dagger }B_{\mu \nu }u\right) \right\rangle   \notag \\
&&+\frac{\kappa }{2\sqrt{2}g_{C}}\left\langle u_{\mu \nu }A^{\mu \nu
}\right\rangle -\frac{i\left( 1-\kappa ^{2}\right) }{4\sqrt{2}g_{C}}%
\left\langle V^{\mu \nu }\left[ u_{\mu },u_{\nu }\right] \right\rangle  
\notag \\
&&+\frac{i\kappa ^{2}}{2\sqrt{2}g_{C}}\left\langle u_{\mu \nu }\left[ V^{\mu
},u^{\nu }\right] \right\rangle\notag\\
&& +\frac{\kappa \left( 1-\kappa ^{2}\right) }{4%
\sqrt{2}g_{C}}\left\langle \left[ u_{\mu },u_{\nu }\right] \left[ A^{\mu
},u^{\nu }\right] \right\rangle   \notag \\
&&-\frac{i\kappa }{2\sqrt{2}g_{C}}\left\langle \left( uW_{\mu \nu
}u^{\dagger }+u^{\dagger }B_{\mu \nu }u\right) \left[ A^{\mu },u^{\nu }%
\right] \right\rangle   \notag \\
&&-\frac{1}{8g_{C}^{2}}\Big\langle \left( uW_{\mu \nu }u^{\dagger
}+u^{\dagger }B_{\mu \nu }u\right)\notag\\
&&\times\left( uW^{\mu \nu }u^{\dagger}+u^{\dagger }B^{\mu \nu }u\right) \Big\rangle   \notag \\
&&-\frac{i\kappa }{2}\left\langle u_{\mu \nu }\left[ V^{\mu },A^{\nu }\right]
\right\rangle -\frac{i\kappa }{2}\left\langle V_{\mu \nu }\left[ A^{\mu
},u^{\nu }\right] \right\rangle   \notag \\
&&-\frac{1-\kappa ^{2}}{8}\left\langle \left[ V_{\mu },V_{\nu }\right] \left[
u^{\mu },u^{\nu }\right] \right\rangle -\frac{i\kappa }{2}\left\langle
A_{\mu \nu }\left[ V^{\mu },u^{\nu }\right] \right\rangle   \notag \\
&&+\frac{\kappa ^{2}}{8}\left\langle \left[ V_{\mu },u_{\nu }\right] \left( %
\left[ V^{\mu },u^{\nu }\right] -\left[ V^{\nu },u^{\mu }\right] \right)
\right\rangle   \notag \\
&&-\frac{1-\kappa ^{2}}{8}\left\langle \left[ A_{\mu },A_{\nu }\right] \left[
u^{\mu },u^{\nu }\right] \right\rangle   \label{Gaugesector} \\
&&+\frac{\kappa ^{2}}{8}\left\langle \left[ A_{\mu },u_{\nu }\right] \left( %
\left[ A^{\mu },u^{\nu }\right] -\left[ A^{\nu },u^{\mu }\right] \right)
\right\rangle   \notag \\
&&+\frac{i}{4}\left\langle \left[ V^{\mu },V^{\nu }\right] \left( uW_{\mu
\nu }u^{\dagger }+u^{\dagger }B_{\mu \nu }u\right) \right\rangle   \notag \\
&&+\frac{i}{4}\left\langle \left[ A^{\mu },A^{\nu }\right] \left( uW_{\mu
\nu }u^{\dagger }+u^{\dagger }B_{\mu \nu }u\right) \right\rangle   \notag \\
&&+\frac{ig_{C}}{2\sqrt{2}}\left\langle V^{\mu \nu }\left[ V_{\mu },V_{\nu }%
\right] \right\rangle +\frac{ig_{C}}{2\sqrt{2}}\left\langle V_{\mu \nu }%
\left[ A^{\mu },A^{\nu }\right] \right\rangle   \notag \\
&&+\frac{ig_{C}}{\sqrt{2}}\left\langle A_{\mu \nu }\left[ V^{\mu },A^{\nu }%
\right] \right\rangle -\frac{\kappa g_{C}}{2\sqrt{2}}\left\langle \left[
A_{\mu },A_{\nu }\right] \left[ A^{\mu },u^{\nu }\right] \right\rangle  
\notag \\
&&-\frac{\kappa g_{C}}{2\sqrt{2}}\left\langle \left[ V_{\mu },u_{\nu }\right]
\left( \left[ V^{\mu },A^{\nu }\right] -\left[ V^{\nu },A^{\mu }\right]
\right) \right\rangle   \notag \\
&&+\frac{g_{C}^{2}}{8}\left\langle \left[ V_{\mu },V_{\nu }\right] \left[
V^{\mu },V^{\nu }\right] \right\rangle +\frac{g_{C}^{2}}{4}\left\langle %
\left[ V_{\mu },V_{\nu }\right] \left[ A^{\mu },A^{\nu }\right]
\right\rangle   \notag \\
&&+\frac{g_{C}^{2}}{8}\left\langle \left[ A_{\mu },A_{\nu }\right] \left[
A^{\mu },A^{\nu }\right] \right\rangle -\frac{\kappa g_{C}}{2\sqrt{2}}%
\left\langle \left[ V_{\mu },V_{\nu }\right] \left[ A^{\mu },u^{\nu }\right]
\right\rangle   \notag \\
&&+\frac{g_{C}^{2}}{4}\left\langle \left[ V_{\mu },A_{\nu }\right] \left( %
\left[ V^{\mu },A^{\nu }\right] -\left[ V^{\nu },A^{\mu }\right] \right)
\right\rangle .  \notag
\end{eqnarray}%
where the correct normalization of the kinetic terms of the heavy spin-1
resonances implies \cite{Barbieri:2008,Barbieri:2010}: 
\begin{equation}
V_{\mu }=\frac{1}{\sqrt{2}}\tau ^{a}V_{\mu }^{a},\hspace{2cm}A_{\mu }=\frac{1%
}{\sqrt{2}}\tau ^{a}A_{\mu }^{a},
\end{equation}%
while the symmetry breaking sector of the Lagrangian takes the following
form: 
\begin{eqnarray}
\mathcal{L}_{\chi }^{gauge} &=&4v_{LC}^{2}\left( 1+\frac{H-h}{2\sqrt{2}v_{LC}%
}+\frac{\left( H-h\right) ^{2}}{32v_{LC}^{2}}+\frac{\eta ^{2}}{16v_{LC}^{2}}%
\right)   \notag \\
&&\times \left[ \frac{g_{C}^{2}}{2}\left\langle V_{\mu }V^{\mu
}\right\rangle +\frac{g_{C}^{2}}{2}\left\langle A_{\mu }A^{\mu
}\right\rangle \right.   \notag \\
&&+\left. \left( \frac{1-\kappa }{2}\right) ^{2}\left\langle u_{\mu }u^{\mu
}\right\rangle +\frac{2g_{C}}{\sqrt{2}}\left( \frac{1-\kappa }{2}\right)
\left\langle A_{\mu }u^{\mu }\right\rangle \right]   \notag \\
&&+4v_{LC}\eta \left( 1+\frac{H-h}{4\sqrt{2}v_{LC}}\right)   \notag \\
&&\times \left[ \frac{g_{C}^{2}}{2}\left\langle V_{\mu }A^{\mu
}\right\rangle +\frac{g_{C}}{\sqrt{2}}\left( \frac{1-\kappa }{2}\right)
\left\langle V_{\mu }u^{\mu }\right\rangle \right]   \notag \\
&&+8v_{CD}^{2}\left( 1+\frac{h+H}{2v_{CD}}+\frac{h^{2}+2hH+H^{2}}{%
16v_{CD}^{2}}\right)   \notag \\
&&\times \left[ \frac{g_{C}^{2}}{2}\left\langle A_{\mu }A^{\mu
}\right\rangle +\frac{\kappa ^{2}}{4}\left\langle u_{\mu }u^{\mu
}\right\rangle -\frac{\kappa g_{C}}{\sqrt{2}}\left\langle A_{\mu }u^{\mu
}\right\rangle \right]   \notag \\
&&+\frac{1}{2}\partial _{\mu }h\partial ^{\mu }h+\frac{1}{2}\partial _{\mu
}H\partial ^{\mu }H+\frac{1}{2}\partial _{\mu }\eta \partial ^{\mu }\eta .
\label{SBsector}
\end{eqnarray}%
Since $V_{\mu }$ and $A_{\mu }$ define the mass eigenstates, the term $%
A_{\mu }u^{\mu }$ should be absent in the previous expression, yielding the
following relation: 
\begin{equation}
\left( \frac{1-\kappa }{2}\right) v_{LC}^{2}-\kappa v_{CD}^{2}=0.
\end{equation}%
In addition, the requirement of having the correct $W$ gauge boson mass
implies 
\begin{equation}
\left( \frac{1-\kappa }{2}\right) ^{2}v_{LC}^{2}+\frac{\kappa ^{2}}{2}%
v_{CD}^{2}=\frac{v^{2}}{16}.
\end{equation}%
The previous equations have the following solutions: 
\begin{equation}
v_{LC}=\frac{v}{2\sqrt{1-\kappa }},\quad v_{CD}=\frac{v}{2\sqrt{2\kappa }}%
,\quad \text{with}\ \ 0<\kappa <1.  \label{VEVs}
\end{equation}%
Then from the expressions (\ref{SBsector}) and (\ref{VEVs}) it follows that
the masses of $V_{\mu }^{a}$ and $A_{\mu }^{a}$ are determined by the
parameters $g_{C}$ and $\kappa$ as 
\begin{equation}
M_{V}=\frac{g_{C}v}{\sqrt{1-\kappa }},\qquad M_{A}=\frac{M_{V}}{\sqrt{\kappa 
}}.  \label{rel1}
\end{equation}%
We now see that the diagonalization procedure determines $\kappa $ in Eq.\ (%
\ref{VAva}) as the mass ratio $\kappa =M_{V}^{2}/M_{A}^{2}$. On the other
hand, the strength of the gauge coupling $g_{C}$ determines the absolute
value of these masses. The coupling $g_{C}$ also controls the kinetic mixing
between $V_{\mu }^{a}$ and the SM gauge bosons, while the kinetic mixing
between $A_{\mu }^{a}$ and the SM gauge bosons is controlled by both $\kappa 
$ and $g_{C}$, as seen in Eq.\ (\ref{Gaugesector}).

Consequently the Lagrangian that describes the interactions among the
composite spin zero fields, the composite spin one fields and the SM gauge
bosons and SM Goldstone bosons is given by 
\begin{eqnarray}
\mathcal{L}&=&\mathcal{L}_{gauge}+\mathcal{L}_{\chi }^{gauge} -V\left(
\Sigma _{LC},\Sigma _{CD},\Sigma _{DY}\right) .
\end{eqnarray}
This same Lagrangian is described in Eq.\ (\ref{Leff}), where the scalar
potential has been expanded to quadratic factors of the scalar fields. We
did not include the cubic and quartic scalar interactions in Eq.\ (\ref{Leff}%
) as they are irrelevant to our calculations of the $h\to\gamma\gamma$ decay
rate and the oblique $T$ and $S$ parameters.


\newpage

\section{: Calculation of the $T$ and $S$ parameters.}

\label{A2} The $T$ parameter is defined as \cite{Peskin:1991sw,epsilon-approach,Barbieri:2004,Barbieri-book}: 
\begin{equation}
T=\frac{\widehat{T}}{\alpha _{em}\left( m_{Z}\right) },\qquad \widehat{T}=%
\frac{\Pi _{33}\left( 0\right) -\Pi _{11}\left( 0\right) }{M_{W}^{2}},
\label{T}
\end{equation}%
where $\Pi _{11}\left( 0\right) $ and $\Pi _{33}\left( 0\right) $ are the
vacuum polarization amplitudes for loop diagrams having gauge bosons $W_{\mu
}^{1}$, $W_{\mu }^{1}$ and $W_{\mu }^{3}$, $W_{\mu }^{3}$ in the external
lines, respectively. These vacuum polarization amplitudes are evaluated at $%
q^{2}=0$, where $q$ the external momentum. 
\begin{figure*}[tbh]\resizebox{0.8\textwidth}{!}{
\includegraphics{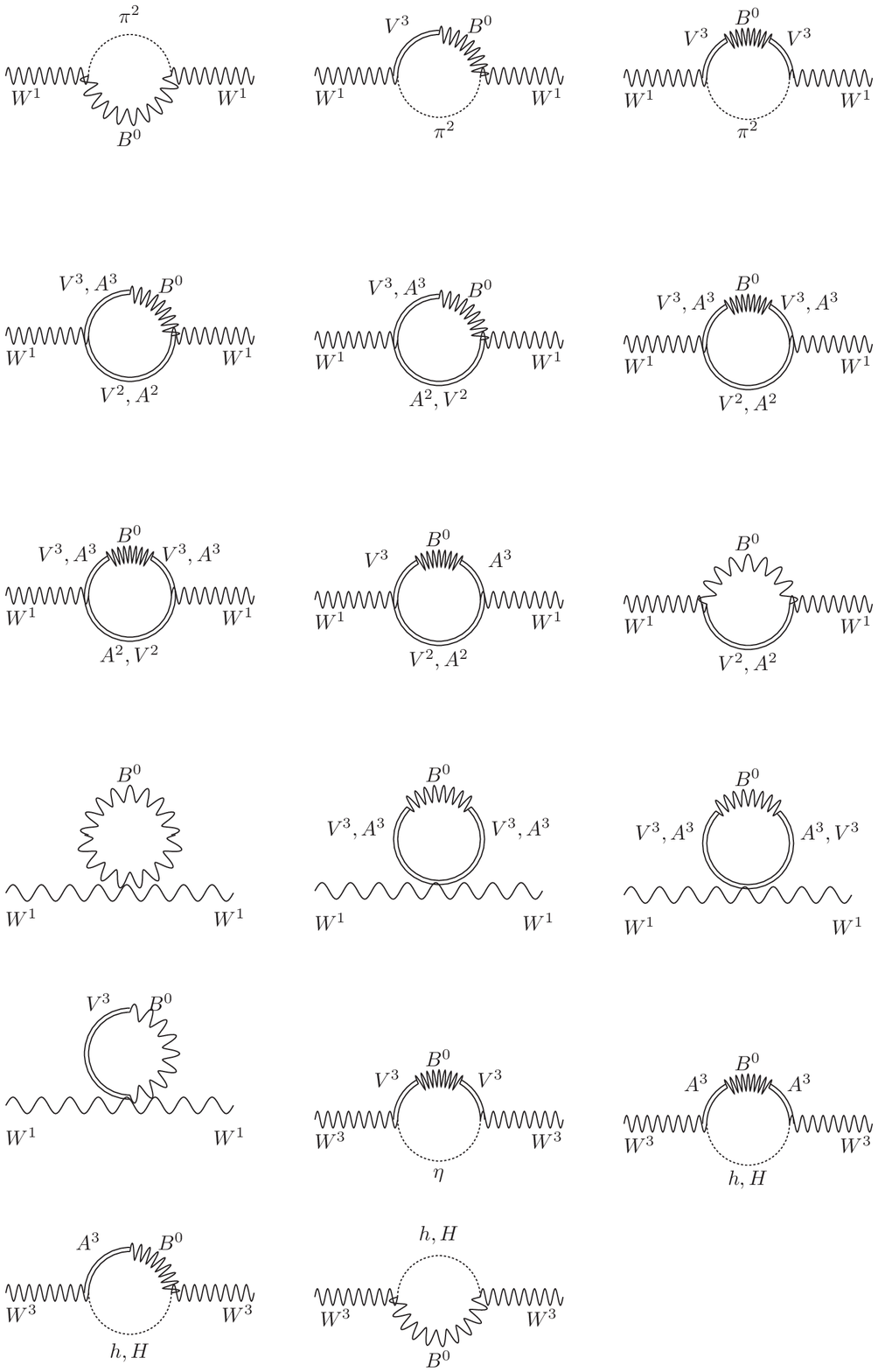}}\vspace{-1.5cm}
\caption{One loop Feynman diagrams contributing to the $T$ parameter.}
\label{figT}
\end{figure*}
The one-loop diagrams that give contributions to the $T$ parameter should
include the hypercharge gauge boson $B_{\mu }^{0}$ since the $g^{\prime }$
coupling is one of the sources of the breaking of the custodial symmetry.
The other source of custodial symmetry breaking comes from the difference
between up- and down-type quark Yukawa couplings. The corresponding Feynman
diagrams are shown in Figure \ref{figT} and we computed them in the Landau
gauge for the SM gauge bosons and Goldstone bosons, where the global $%
SU(2)_{L}\times U(1)_{Y}$ symmetry is preserved. Regarding the heavy
composite spin-1 resonances, we use the unitary gauge for their propagators
since the Lagrangian given in Eq. (\ref{Leff}) does not include the
Goldstone bosons associated to the longitudinal components of these heavy
parity even and parity odd spin-1 resonances. From the Feynman diagrams
shown in Figure \ref{figT}, it follows that the $\widehat{T}$ parameter is
given by 
\begin{eqnarray}
\widehat{T} &=&\widehat{T}_{\left( \pi ^{2}B^{0}\right) }+\widehat{T}%
_{\left( V^{3}B^{0},\pi ^{2}\right) }+\widehat{T}_{\left(
V^{3}B^{0}V^{3},\pi ^{2}\right) }+\widehat{T}_{\left( V^{2}B^{0}\right) } 
\notag \\
&&+\widehat{T}_{\left( A^{2}B^{0}\right) }+\widehat{T}_{\left(
V^{3}B^{0},V^{2}\right) }+\widehat{T}_{\left( A^{3}B^{0},A^{2}\right) } 
\notag \\
&&+\widehat{T}_{\left( V^{3}B^{0}V^{3},V^{2}\right) }+\widehat{T}_{\left(
A^{3}B^{0}A^{3},A^{2}\right) }+\widehat{T}_{\left( V^{3}B^{0},A^{2}\right) }
\notag \\
&&+\widehat{T}_{\left( A^{3}B^{0},V^{2}\right) }+\widehat{T}_{\left(
V^{3}B^{0}V^{3},A^{2}\right) }+\widehat{T}_{\left(
A^{3}B^{0}A^{3},V^{2}\right) }  \notag \\
&&+\widehat{T}_{\left( V^{3}B^{0}A^{3},V^{2}\right) }+\widehat{T}_{\left(
V^{3}B^{0}A^{3},A^{2}\right) }+\widehat{T}_{\left( B^{0}\right) }+\widehat{T}%
_{\left( V^{3}B^{0}V^{3}\right) }  \notag \\
&&+\widehat{T}_{\left( A^{3}B^{0}A^{3}\right) }+\widehat{T}_{\left(
V^{3}B^{0}A^{3}\right) }+\widehat{T}_{\left( V^{3}-B^{0}\right) }+\widehat{T}%
_{\left( hB^{0}\right) }  \notag \\
&&+\widehat{T}_{\left( HB^{0}\right) }+\widehat{T}_{\left( \eta
,V^{3}B^{0}V^{3}\right) }+\widehat{T}_{\left( h,A^{3}B^{0}A^{3}\right) }+%
\widehat{T}_{\left( H,A^{3}B^{0}A^{3}\right) }  \notag \\
&&+\widehat{T}_{\left( h,A^{3}B^{0}\right) }+\widehat{T}_{\left(
H,A^{3}B^{0}\right) },  \label{hatt}
\end{eqnarray}%
where the different one-loop contributions to the $\widehat{T}$ parameter
are
\begin{eqnarray}
\widehat{T}_{\left( \pi ^{2}B^{0}\right) } &\simeq &-\frac{3\alpha _{em}}{%
16\pi \cos ^{2}\theta _{W}}\ln \left( \frac{\Lambda ^{2}}{M_{W}^{2}}\right) 
\notag \\
&&-\frac{3\alpha _{em}\left( f_{V}g_{V}-\frac{1}{2}f_{A}^{2}\right) ^{2}}{%
32\pi v^{4}\cos ^{2}\theta _{W}}\Lambda ^{4}  \notag \\
&&+\frac{3\alpha _{em}\left( f_{V}g_{V}-\frac{1}{2}f_{A}^{2}\right) }{8\pi
v^{2}\cos ^{2}\theta _{W}}\Lambda ^{2},
\end{eqnarray}

\begin{eqnarray}
\widehat{T}_{\left( V^{3}B^{0},\pi ^{2}\right) } &\simeq &-\frac{3\alpha
_{em}f_{V}g_{V}\left( f_{V}g_{V}-\frac{1}{2}f_{A}^{2}\right) }{16\pi
v^{4}\cos ^{2}\theta _{W}}\Lambda ^{4}  \notag \\
&&-\frac{3\alpha _{em}f_{V}g_{V}}{8\pi v^{2}\cos ^{2}\theta _{W}}\left[
1-\left( f_{V}g_{V}-\frac{1}{2}f_{A}^{2}\right) \frac{M_{V}^{2}}{v^{2}}%
\right]  \notag \\
&&\times \left[ \allowbreak \Lambda ^{2}-M_{V}^{2}\ln \left( \frac{\Lambda
^{2}+M_{V}^{2}}{M_{V}^{2}}\right) \right] ,
\end{eqnarray}

\begin{eqnarray}
\widehat{T}_{\left( V^{3}B^{0}V^{3},\pi ^{2}\right) } &\simeq &-\frac{%
3\alpha _{em}f_{V}^{2}g_{V}^{2}}{32\pi v^{4}\cos ^{2}\theta _{W}}\left[
\Lambda ^{4}-4M_{V}^{2}\Lambda ^{2}\right. \\
&&+\left. 6M_{V}^{4}\ln \left( \frac{\Lambda ^{2}+M_{V}^{2}}{M_{V}^{2}}%
\right) -\frac{2M_{V}^{4}\Lambda ^{2}}{\Lambda ^{2}+M_{V}^{2}}\right] , 
\notag
\end{eqnarray}

\begin{eqnarray}
\widehat{T}_{\left( V^{2}B^{0}\right) } &\simeq &\frac{3\alpha _{em}\left(
1-\kappa ^{2}\right) ^{2}f_{V}^{2}}{128\pi v^{2}M_{V}^{2}\cos ^{2}\theta _{W}%
}\Lambda ^{4}  \notag \\
&&+\frac{3\alpha _{em}\left[ \left( 1-\kappa ^{2}\right) f_{V}-2g_{V}\right]
^{2}}{64\pi v^{2}\cos ^{2}\theta _{W}}\Bigg\{\Lambda ^{2}  \notag \\
&&-M_{V}^{2}\ln \left( \frac{\Lambda ^{2}+M_{V}^{2}}{M_{V}^{2}}\right) %
\Bigg\},
\end{eqnarray}

\begin{eqnarray}
\widehat{T}_{\left( A^{2}B^{0}\right) } &\simeq &\frac{3\alpha _{em}f_{A}^{2}%
}{128\pi v^{2}M_{A}^{2}\cos ^{2}\theta _{W}}\Lambda ^{4} \\
&&+\frac{3\alpha _{em}f_{A}^{2}}{64\pi v^{2}\cos ^{2}\theta _{W}}\left[
\Lambda ^{2}-M_{A}^{2}\ln \left( \frac{\Lambda ^{2}+M_{A}^{2}}{M_{A}^{2}}%
\right) \right] ,  \notag
\end{eqnarray}

\begin{eqnarray}
\widehat{T}_{\left( V^{3}B^{0},V^{2}\right) } &\simeq &\frac{3\alpha
_{em}f_{V}}{64\pi v^{2}M_{V}^{2}\cos ^{2}\theta _{W}}\Bigg\{\left( 1-\kappa
^{2}\right) f_{V}\Lambda ^{4}  \notag \\
&&-2\left[ 4g_{V}-\left( 1-\kappa ^{2}\right) f_{V}\right] M_{V}^{2}\Lambda
^{2}  \notag \\
&&+2\left[ 8g_{V}-3\left( 1-\kappa ^{2}\right) f_{V}\right] M_{V}^{4}\ln
\left( \frac{\Lambda ^{2}+M_{V}^{2}}{M_{V}^{2}}\right)  \notag \\
&&-\frac{4\left[ 2g_{V}-\left( 1-\kappa ^{2}\right) f_{V}\right]
M_{V}^{4}\Lambda ^{2}}{\Lambda ^{2}+M_{V}^{2}}\Bigg\},
\end{eqnarray}

\begin{eqnarray}
\widehat{T}_{\left( A^{3}B^{0},A^{2}\right) } &\simeq &-\frac{3\alpha
_{em}f_{A}^{2}}{64\pi v^{2}M_{A}^{2}\cos ^{2}\theta _{W}}\Bigg\{\Lambda
^{4}+2M_{A}^{2}\Lambda ^{2} \\
&&-\left. 6M_{A}^{4}\ln \left( \frac{\Lambda ^{2}+M_{A}^{2}}{M_{A}^{2}}%
\right) +\frac{4M_{A}^{4}\Lambda ^{2}}{\Lambda ^{2}+M_{A}^{2}}\right] \Bigg\}%
,  \notag
\end{eqnarray}

\begin{eqnarray}
\widehat{T}_{\left( V^{3}B^{0}V^{3},V^{2}\right) } &\simeq &\frac{3\alpha
_{em}f_{V}^{2}}{128\pi v^{2}M_{V}^{2}\cos ^{2}\theta _{W}}\Bigg\{\Lambda
^{4}+4M_{V}^{2}\Lambda ^{2}  \notag \\
&&-2M_{V}^{6}\frac{11\Lambda ^{2}+9M_{V}^{2}}{\left( \Lambda
^{2}+M_{V}^{2}\right) ^{2}}\Bigg\},
\end{eqnarray}

\begin{eqnarray}
\widehat{T}_{\left( A^{3}B^{0}A^{3},A^{2}\right) } &\simeq &\frac{3\alpha
_{em}f_{A}^{2}}{128\pi v^{2}M_{A}^{2}\cos ^{2}\theta _{W}}\Bigg\{\Lambda
^{4}+4M_{A}^{2}\Lambda ^{2}  \notag \\
&&+18M_{A}^{4}\left[ 1-\ln \left( \frac{\Lambda ^{2}+M_{A}^{2}}{M_{A}^{2}}%
\right) \right]  \notag \\
&&-2M_{A}^{6}\frac{11\Lambda ^{2}+9M_{A}^{2}}{\left( \Lambda
^{2}+M_{A}^{2}\right) ^{2}}\Bigg\},
\end{eqnarray}

\begin{eqnarray}
\widehat{T}_{\left( V^{3}B^{0},A^{2}\right) } &\simeq &\frac{3\alpha
_{em}\kappa f_{V}f_{A}}{64\pi v^{2}M_{A}^{2}\cos ^{2}\theta _{W}}\Bigg\{%
\Lambda ^{4}+2\left( 2M_{A}^{2}-M_{V}^{2}\right) \Lambda ^{2}  \notag \\
&&+\frac{2M_{V}^{4}\left( M_{V}^{2}-3M_{A}^{2}\right) }{M_{V}^{2}-M_{A}^{2}}%
\ln \left( \frac{\Lambda ^{2}+M_{V}^{2}}{M_{V}^{2}}\right)  \notag \\
&&-\frac{4M_{A}^{6}}{M_{A}^{2}-M_{V}^{2}}\ln \left( \frac{\Lambda
^{2}+M_{A}^{2}}{M_{A}^{2}}\right) \Bigg\},
\end{eqnarray}

\begin{eqnarray}
\widehat{T}_{\left( A^{3}B^{0},V^{2}\right) } &\simeq &-\frac{3\alpha
_{em}\kappa f_{A}}{64\pi v^{2}M_{V}^{2}\cos ^{2}\theta _{W}}\Bigg\{\left(
1-\kappa ^{2}\right) f_{V}\Lambda ^{4}  \notag \\
&&+\left[ 4\left( \left( 1-\kappa ^{2}\right) f_{V}-2g_{V}\right)
M_{V}^{2}\right.  \notag \\
&&-\left. 2\left( 1-\kappa ^{2}\right) f_{V}M_{A}^{2}\right] \Lambda ^{2} \\
&&-\frac{4\left( \left( 1-\kappa ^{2}\right) f_{V}-2g_{V}\right) M_{V}^{6}}{%
M_{V}^{2}-M_{A}^{2}}\ln \left( \frac{\Lambda ^{2}+M_{V}^{2}}{M_{V}^{2}}%
\right)  \notag \\
&&+\frac{M_{A}^{4}}{M_{A}^{2}-M_{V}^{2}}\left[ 2\left( 1-\kappa ^{2}\right)
f_{V}M_{A}^{2}\right.  \notag \\
&&-\left. 2\left( 3\left( 1-\kappa ^{2}\right) f_{V}-4g_{V}\right) M_{V}^{2} 
\right] \ln \left( \frac{\Lambda ^{2}+M_{A}^{2}}{M_{A}^{2}}\right) \Bigg\}, 
\notag
\end{eqnarray}

\begin{eqnarray}
\widehat{T}_{\left( V^{3}B^{0}V^{3},A^{2}\right) } &\simeq &\frac{3\alpha
_{em}\kappa ^{2}f_{V}^{2}}{128\pi v^{2}M_{A}^{2}\cos ^{2}\theta _{W}}\Bigg\{%
\Lambda ^{4}  \notag \\
&&+4\left( 2M_{A}^{2}-M_{V}^{2}\right) \Lambda ^{2}  \notag \\
&&+\frac{2M_{V}^{4}}{\left( M_{A}^{2}-M_{V}^{2}\right) {}^{2}}\left[
3M_{V}^{4}+15M_{A}^{4}\right.  \notag \\
&&-\left. 14M_{A}^{2}M_{V}^{2}\right] \ln \left( \frac{\Lambda ^{2}+M_{V}^{2}%
}{M_{V}^{2}}\right)  \notag \\
&&-\frac{8M_{A}^{8}}{\left( M_{A}^{2}-M_{V}^{2}\right) {}^{2}}\ln \left( 
\frac{\Lambda ^{2}+M_{A}^{2}}{M_{A}^{2}}\right)  \notag \\
&&-\frac{2M_{V}^{4}\left( 5M_{A}^{2}-M_{V}^{2}\right) \Lambda ^{2}}{\left(
M_{A}^{2}-M_{V}^{2}\right) \left( \Lambda ^{2}+M_{V}^{2}\right) }\Bigg\},
\end{eqnarray}

\begin{eqnarray}
\widehat{T}_{\left( A^{3}B^{0}A^{3},V^{2}\right) } &\simeq &\frac{3\alpha
_{em}\kappa ^{2}f_{A}^{2}}{128\pi v^{2}M_{V}^{2}\cos ^{2}\theta _{W}}\Bigg\{%
\Lambda ^{4}  \notag \\
&&+4\left( 2M_{V}^{2}-M_{A}^{2}\right) \Lambda ^{2}  \notag \\
&&+\frac{2M_{A}^{4}}{\left( M_{A}^{2}-M_{V}^{2}\right) {}^{2}}\left[
3M_{A}^{4}+15M_{V}^{4}\right.  \notag \\
&&-\left. 14M_{A}^{2}M_{V}^{2}\right] \ln \left( \frac{\Lambda ^{2}+M_{A}^{2}%
}{M_{A}^{2}}\right)  \notag \\
&&-\frac{8M_{V}^{8}}{\left( M_{A}^{2}-M_{V}^{2}\right) {}^{2}}\ln \left( 
\frac{\Lambda ^{2}+M_{V}^{2}}{M_{V}^{2}}\right)  \notag \\
&&-\frac{2M_{A}^{4}\left( 5M_{V}^{2}-M_{A}^{2}\right) \Lambda ^{2}}{\left(
M_{V}^{2}-M_{A}^{2}\right) \left( \Lambda ^{2}+M_{A}^{2}\right) }\Bigg\},
\end{eqnarray}

\begin{eqnarray}
\widehat{T}_{\left( V^{3}B^{0}A^{3},V^{2}\right) } &\simeq &-\frac{3\alpha
_{em}\kappa f_{V}f_{A}}{64\pi v^{2}M_{V}^{2}\cos ^{2}\theta _{W}}\Bigg\{%
\Lambda ^{4}  \notag \\
&&-2\left( M_{A}^{2}-3M_{V}^{2}\right) \Lambda ^{2}  \notag \\
&&+\frac{2M_{A}^{6}\left( M_{A}^{2}-5M_{V}^{2}\right) }{\left(
M_{A}^{2}-M_{V}^{2}\right) ^{2}}\ln \left( \frac{\Lambda ^{2}+M_{A}^{2}}{%
M_{A}^{2}}\right)  \notag \\
&&+\frac{2M_{V}^{4}}{\left( M_{V}^{2}-M_{A}^{2}\right) ^{2}}\left[ 3\left(
M_{V}^{2}-M_{A}^{2}\right) ^{2}\right.  \notag \\
&&-\left. \left( M_{A}^{2}-5M_{V}^{2}\right) \left(
3M_{A}^{2}-2M_{V}^{2}\right) \right]  \notag \\
&&\times \ln \left( \frac{\Lambda ^{2}+M_{V}^{2}}{M_{V}^{2}}\right)  \notag
\\
&&+\frac{8M_{V}^{6}\Lambda ^{2}}{\left( M_{V}^{2}-M_{A}^{2}\right) \left(
\Lambda ^{2}+M_{V}^{2}\right) }\Bigg\},
\end{eqnarray}

\begin{eqnarray}
\widehat{T}_{\left( V^{3}B^{0}A^{3},A^{2}\right) } &\simeq &-\frac{3\alpha
_{em}\kappa f_{V}f_{A}}{64\pi v^{2}M_{A}^{2}\cos ^{2}\theta _{W}}\Bigg\{%
\Lambda ^{4}  \notag \\
&&-2\left( M_{V}^{2}-3M_{A}^{2}\right) \Lambda ^{2}  \notag \\
&&+\frac{2M_{A}^{4}}{\left( M_{V}^{2}-M_{A}^{2}\right) ^{2}}\left[ 3\left(
M_{V}^{2}-M_{A}^{2}\right) ^{2}\right.  \notag \\
&&-\left. \left( M_{V}^{2}-5M_{A}^{2}\right) \left(
3M_{V}^{2}-2M_{A}^{2}\right) \right]  \notag \\
&&\times \ln \left( \frac{\Lambda ^{2}+M_{A}^{2}}{M_{A}^{2}}\right)  \notag
\\
&&+\frac{2M_{V}^{6}\left( M_{V}^{2}-5M_{A}^{2}\right) }{\left(
M_{A}^{2}-M_{V}^{2}\right) ^{2}}\ln \left( \frac{\Lambda ^{2}+M_{V}^{2}}{%
M_{V}^{2}}\right)  \notag \\
&&+\frac{8M_{A}^{6}\Lambda ^{2}}{\left( M_{A}^{2}-M_{V}^{2}\right) \left(
\Lambda ^{2}+M_{A}^{2}\right) }\Bigg\},
\end{eqnarray}

\begin{equation}
\widehat{T}_{\left( B^{0}\right) }\simeq -\frac{9\alpha _{em}g_{V}^{2}}{%
16\pi v^{2}\cos ^{2}\theta _{W}}\allowbreak \Lambda ^{2},
\end{equation}

\begin{eqnarray}
\widehat{T}_{\left( V^{3}B^{0}V^{3}\right) } &\simeq &-\frac{9\alpha
_{em}\left( 1+\kappa ^{2}\right) f_{V}^{2}}{64\pi v^{2}\cos ^{2}\theta _{W}}%
\Bigg\{\Lambda ^{2} \\
&&-2M_{V}^{2}\ln \left( \frac{\Lambda ^{2}+M_{V}^{2}}{M_{V}^{2}}\right) +%
\frac{\Lambda ^{2}M_{V}^{2}}{\Lambda ^{2}+M_{V}^{2}}\Bigg\},  \notag
\end{eqnarray}

\begin{eqnarray}
\widehat{T}_{\left( A^{3}B^{0}A^{3}\right) } &\simeq &-\frac{9\alpha
_{em}\left( 1+\kappa ^{2}\right) f_{A}^{2}}{64\pi v^{2}\cos ^{2}\theta _{W}}%
\Bigg\{\Lambda ^{2} \\
&&-2M_{A}^{2}\ln \left( \frac{\Lambda ^{2}+M_{A}^{2}}{M_{A}^{2}}\right) +%
\frac{\Lambda ^{2}M_{A}^{2}}{\Lambda ^{2}+M_{A}^{2}}\Bigg\},  \notag
\end{eqnarray}

\begin{eqnarray}
\widehat{T}_{\left( V^{3}B^{0}A^{3}\right) } &\simeq &\frac{9\alpha
_{em}\kappa f_{V}f_{A}}{16\pi v^{2}\cos ^{2}\theta _{W}}\Bigg\{\Lambda ^{2}
\\
&&-\frac{M_{V}^{4}}{M_{V}^{2}-M_{A}^{2}}\ln \left( \frac{\Lambda
^{2}+M_{V}^{2}}{M_{V}^{2}}\right)  \notag \\
&&-\frac{M_{A}^{4}}{M_{A}^{2}-M_{V}^{2}}\ln \left( \frac{\Lambda
^{2}+M_{A}^{2}}{M_{A}^{2}}\right) \Bigg\}  \notag
\end{eqnarray}

\begin{equation}
\widehat{T}_{\left( V^{3}-B^{0}\right) }\simeq \frac{9\alpha _{em}f_{V}g_{V}%
}{16\pi v^{2}\cos ^{2}\theta _{W}}\left[ \Lambda ^{2}-M_{V}^{2}\ln \left( 
\frac{\Lambda ^{2}+M_{V}^{2}}{M_{V}^{2}}\right) \right] ,
\end{equation}%
\begin{equation}
\widehat{T}_{\left( hB^{0}\right) }\simeq \frac{3\alpha _{em}\left[ \left(
1-\kappa \right) ^{\frac{3}{2}}-2\kappa ^{\frac{3}{2}}\right] ^{2}}{128\pi
\cos ^{2}\theta _{W}}\ln \left( \frac{\Lambda ^{2}}{m_{h}^{2}}\right) ,
\end{equation}%
\begin{equation}
\widehat{T}_{\left( HB^{0}\right) }\simeq \frac{3\alpha _{em}\left[ \left(
1-\kappa \right) ^{\frac{3}{2}}+2\kappa ^{\frac{3}{2}}\right] ^{2}}{128\pi
\cos ^{2}\theta _{W}}\ln \left( \frac{\Lambda ^{2}}{m_{H}^{2}}\right) ,
\label{TH1}
\end{equation}%
\begin{eqnarray}
\widehat{T}_{\left( \eta ,V^{3}B^{0}V^{3}\right) } &\simeq &\frac{3\alpha
_{em}f_{V}^{2}\left( 1-\kappa \right) ^{2}M_{V}^{2}}{64\pi v^{2}\left(
M_{V}^{2}-m_{\eta }^{2}\right) \cos ^{2}\theta _{W}}  \notag \\
&&\times \Bigg\{\frac{M_{V}^{4}-2M_{V}^{2}m_{\eta }^{2}}{M_{V}^{2}-m_{\eta
}^{2}}\ln \left( \frac{\Lambda ^{2}+M_{V}^{2}}{M_{V}^{2}}\right)  \notag \\
&&-\frac{m_{\eta }^{4}}{m_{\eta }^{2}-M_{V}^{2}}\ln \left( \frac{\Lambda
^{2}+m_{\eta }^{2}}{m_{\eta }^{2}}\right)  \notag \\
&&-\frac{M_{V}^{2}\Lambda ^{2}}{\Lambda ^{2}+M_{V}^{2}}\Bigg\},
\label{Teta1}
\end{eqnarray}%
\begin{eqnarray}
\widehat{T}_{\left( h,A^{3}B^{0}A^{3}\right) } &\simeq &\frac{3\alpha
_{em}\kappa \left( 1-\kappa \right) \left( \sqrt{1-\kappa }+2\sqrt{\kappa }%
\right) ^{2}f_{A}^{2}M_{A}^{2}}{128\pi v^{2}\left(
M_{A}^{2}-m_{h}^{2}\right) \cos ^{2}\theta _{W}}  \notag \\
&&\times \Bigg\{\frac{M_{A}^{4}-2M_{A}^{2}m_{h}^{2}}{M_{A}^{2}-m_{h}^{2}}\ln
\left( \frac{\Lambda ^{2}+M_{A}^{2}}{M_{A}^{2}}\right)  \notag \\
&&-\frac{m_{h}^{4}}{m_{h}^{2}-M_{A}^{2}}\ln \left( \frac{\Lambda
^{2}+m_{h}^{2}}{m_{h}^{2}}\right)  \notag \\
&&-\frac{M_{A}^{2}\Lambda ^{2}}{\Lambda ^{2}+M_{A}^{2}}\Bigg\},
\end{eqnarray}%
\begin{eqnarray}
\widehat{T}_{\left( H,A^{3}B^{0}A^{3}\right) } &\simeq &\frac{3\alpha
_{em}\kappa \left( 1-\kappa \right) \left( \sqrt{1-\kappa }-2\sqrt{\kappa }%
\right) ^{2}f_{A}^{2}M_{A}^{2}}{128\pi v^{2}\left(
M_{A}^{2}-m_{H}^{2}\right) \cos ^{2}\theta _{W}}  \notag \\
&&\times \Bigg\{\frac{M_{A}^{4}-2M_{A}^{2}m_{H}^{2}}{M_{A}^{2}-m_{H}^{2}}\ln
\left( \frac{\Lambda ^{2}+M_{A}^{2}}{M_{A}^{2}}\right)  \notag \\
&&-\frac{m_{H}^{4}}{m_{H}^{2}-M_{A}^{2}}\ln \left( \frac{\Lambda
^{2}+m_{H}^{2}}{m_{H}^{2}}\right)  \notag \\
&&-\frac{M_{A}^{2}\Lambda ^{2}}{\Lambda ^{2}+M_{A}^{2}}\Bigg\},  \label{TH2}
\end{eqnarray}%
\begin{eqnarray}
\widehat{T}_{\left( h,A^{3}B^{0}\right) } &\simeq &\frac{3\alpha _{em}\sqrt{%
\kappa }\sqrt{1-\kappa }\left( \sqrt{1-\kappa }+2\sqrt{\kappa }\right)
a_{h}f_{A}M_{A}}{16\sqrt{2}\pi v\cos ^{2}\theta _{W}}  \notag \\
&&\times \Bigg\{\frac{M_{A}^{2}}{M_{A}^{2}-m_{h}^{2}}\ln \left( \frac{%
\Lambda ^{2}+M_{A}^{2}}{M_{A}^{2}}\right)  \notag \\
&&+\frac{m_{h}^{2}}{m_{h}^{2}-M_{A}^{2}}\ln \left( \frac{\Lambda
^{2}+m_{h}^{2}}{m_{h}^{2}}\right) \Bigg\},
\end{eqnarray}

\begin{eqnarray}
\widehat{T}_{\left( H,A^{3}B^{0}\right) } &\simeq &\frac{3\alpha _{em}\sqrt{%
\kappa }\sqrt{1-\kappa }\left( \sqrt{1-\kappa }+2\sqrt{\kappa }\right)
a_{h}f_{A}M_{A}}{16\sqrt{2}\pi v\cos ^{2}\theta _{W}}  \notag \\
&&\times \Bigg\{\frac{M_{A}^{2}}{M_{A}^{2}-m_{H}^{2}}\ln \left( \frac{%
\Lambda ^{2}+M_{A}^{2}}{M_{A}^{2}}\right)  \notag \\
&&+\frac{m_{H}^{2}}{m_{H}^{2}-M_{A}^{2}}\ln \left( \frac{\Lambda
^{2}+m_{H}^{2}}{m_{H}^{2}}\right) \Bigg\}.
\end{eqnarray}
\newpage
\begin{figure*}[tbh]\resizebox{0.9\textwidth}{!}{
\includegraphics{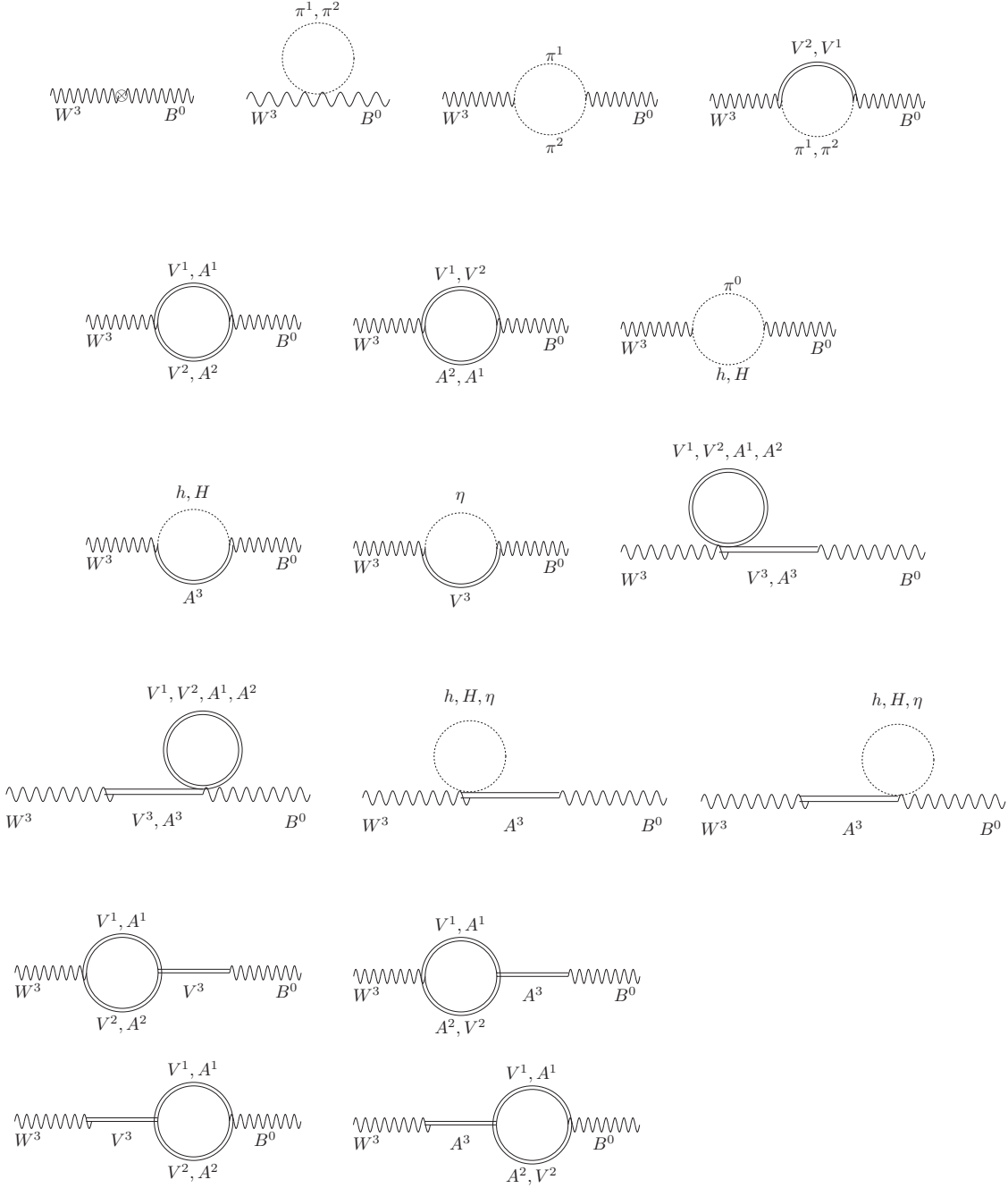}}\vspace{-1.7cm}
\caption{One loop Feynman diagrams contributing to the $S$ parameter.}
\label{figS}
\end{figure*}\newpage
Now, considering the $S$ parameter, it is defined as \cite{Peskin:1991sw,epsilon-approach,Barbieri:2004,Barbieri-book}: 
\begin{equation}
S=\frac{4\sin ^{2}\theta _{W}}{\alpha_{em}\left( m_{Z}\right) }\widehat{S},
\qquad \widehat{S}=\frac{g}{g^{\prime }}\frac{d\Pi _{30}\left( q^{2}\right) 
}{dq^{2}}\biggl|_{q^{2}=0},  \label{Sparameter}
\end{equation}
where $\Pi _{30}\left( q^{2}\right) $ is the vacuum polarization amplitude
for a loop diagram having $W_{\mu }^{3}$ and $B_{\mu }$ in the external
lines. As before, here $q$ is the external momentum.

Corresponding to the Feynman diagrams shown in Figure \ref{figS}, we
decompose the $\widehat{S}$ parameter as: 
\begin{eqnarray}
\widehat{S} &=&\widehat{S}_{\left( W^{3}-B^{0}\right) }^{\left( tree\right)
}+\widehat{S}_{\left( \pi ^{1}\pi ^{2}\right) }+\widehat{S}_{\left( \pi
^{1}V^{2}\right) }+\widehat{S}_{\left( \pi ^{2}V^{1}\right) }+\widehat{S}%
_{\left( V^{1}V^{2}\right) }  \notag \\
&&+\widehat{S}_{\left( A^{1}A^{2}\right) }+\widehat{S}_{\left(
V^{1}A^{2}\right) }+\widehat{S}_{\left( V^{2}A^{1}\right) }+\widehat{S}%
_{\left( h\pi ^{0}\right) }+\widehat{S}_{\left( H\pi ^{0}\right) }  \notag \\
&&+\widehat{S}_{\left( hA^{3}\right) }+\widehat{S}_{\left( HA^{3}\right) }+%
\widehat{S}_{\left( \eta V^{3}\right) }+\widehat{S}_{\left( A^{2}\right)
\left( W^{3}-A^{3}\right) }  \notag \\
&&+\widehat{S}_{\left( A^{1}\right) \left( W^{3}-A^{3}\right) }+\widehat{S}%
_{\left( A^{2}\right) \left( A^{3}-B^{0}\right) }+\widehat{S}_{\left(
A^{1}\right) \left( A^{3}-B^{0}\right) }  \notag \\
&&+\widehat{S}_{\left( V^{2}\right) \left( A^{3}-B^{0}\right) }+\widehat{S}%
_{\left( V^{1}\right) \left( A^{3}-B^{0}\right) }+\widehat{S}_{\left(
V^{2}\right) \left( W^{3}-A^{3}\right) }  \notag \\
&&+\widehat{S}_{\left( V^{1}\right) \left( W^{3}-A^{3}\right) }+\widehat{S}%
_{\left( A^{2}\right) \left( W^{3}-V^{3}\right) }+\widehat{S}_{\left(
A^{1}\right) \left( W^{3}-V^{3}\right) }  \notag \\
&&+\widehat{S}_{\left( A^{2}\right) \left( V^{3}-B^{0}\right) }+\widehat{S}%
_{\left( A^{1}\right) \left( V^{3}-B^{0}\right) }+\widehat{S}_{\left(
V^{2}\right) \left( V^{3}-B^{0}\right) }  \notag \\
&&+\widehat{S}_{\left( V^{1}\right) \left( V^{3}-B^{0}\right) }+\widehat{S}%
_{\left( V^{2}\right) \left( W^{3}-V^{3}\right) }+\widehat{S}_{\left(
V^{1}\right) \left( W^{3}-V^{3}\right) }  \notag \\
&&+\widehat{S}_{\left( V^{1}V^{2}\right) }^{\left( W^{3}-V^{3}\right) }+%
\widehat{S}_{\left( V^{1}V^{2}\right) }^{\left( V^{3}-B^{0}\right) }+%
\widehat{S}_{\left( A^{1}A^{2}\right) }^{\left( W^{3}-V^{3}\right) }+%
\widehat{S}_{\left( A^{1}A^{2}\right) }^{\left( V^{3}-B^{0}\right) }  \notag
\\
&&+\widehat{S}_{\left( A^{1}V^{2}\right) }^{\left( W^{3}-A^{3}\right) }+%
\widehat{S}_{\left( V^{1}A^{2}\right) }^{\left( W^{3}-A^{3}\right) }+%
\widehat{S}_{\left( A^{1}V^{2}\right) }^{\left( A^{3}-B^{0}\right) }+%
\widehat{S}_{\left( V^{1}A^{2}\right) }^{\left( A^{3}-B^{0}\right) }  \notag
\\
&&+\widehat{S}_{\left( h\right) \left( W^{3}-A^{3}\right) }+\widehat{S}%
_{\left( h\right) \left( A^{3}-B^{0}\right) }+\widehat{S}_{\left( H\right)
\left( W^{3}-A^{3}\right) }  \notag \\
&&+\widehat{S}_{\left( H\right) \left( A^{3}-B^{0}\right) }+\widehat{S}%
_{\left( \eta \right) \left( W^{3}-A^{3}\right) }+\widehat{S}_{\left( \eta
\right) \left( A^{3}-B^{0}\right) }  \notag \\
&&+\widehat{S}_{\left( \pi ^{1}\right) }+\widehat{S}_{\left( \pi ^{2}\right)
},  \label{Shat}
\end{eqnarray}

where the different one-loop contributions are 
\begin{equation}
\widehat{S}_{\left( W^{3}-B^{0}\right) }^{\left( tree\right) }=\frac{\pi
\alpha_{em} \left( f_{V}^{2}-f_{A}^{2}-c_{WB}\right) }{\sin ^{2}\theta _{W}},
\end{equation}

\begin{equation}
\widehat{S}_{\left( \pi ^{1}\pi ^{2}\right) }\simeq \frac{\alpha_{em} }{%
48\pi \sin ^{2}\theta _{W}}\ln \left( \frac{\Lambda ^{2}}{M_{W}^{2}}\right) +%
\frac{\alpha_{em} f_{V}g_{V}}{8\pi v^{2}\sin ^{2}\theta _{W}}\Lambda ^{2},
\end{equation}

\begin{eqnarray}
\widehat{S}_{\left( V^{1}V^{2}\right) } &=&-\frac{\alpha _{em}\Lambda ^{2}}{%
24\pi M_{V}^{2}\sin ^{2}\theta _{W}}+\frac{29\alpha _{em}}{96\pi \sin
^{2}\theta _{W}}\ln \left( \frac{\Lambda ^{2}+M_{V}^{2}}{M_{V}^{2}}\right) 
\notag \\
&&+\frac{9\alpha _{em}}{64\pi \sin ^{2}\theta _{W}}-\frac{35\alpha
_{em}\Lambda ^{2}}{96\pi \left( \Lambda ^{2}+M_{V}^{2}\right) \sin
^{2}\theta _{W}}  \notag \\
&&-\frac{\alpha _{em}\left( 34M_{V}^{2}\Lambda ^{2}+27M_{V}^{4}\right) }{%
192\pi \left( \Lambda ^{2}+M_{V}^{2}\right) ^{2}\sin ^{2}\theta _{W}},
\end{eqnarray}

\begin{equation}
\widehat{S}_{\left( \pi ^{2}V^{1}\right) }=\widehat{S}_{\left( \pi
^{1}V^{2}\right) },\qquad \widehat{S}_{\left( V^{2}A^{1}\right) }=\widehat{S}%
_{\left( V^{1}A^{2}\right) },
\end{equation}

\begin{eqnarray}
\widehat{S}_{\left( \pi ^{1}V^{2}\right) } &=&-\frac{\alpha _{em}}{256\pi
v^{2}M_{V}^{2}\sin ^{2}\theta _{W}}\Bigg[6g_{V}\left( 2g_{V}+\kappa
^{2}f_{V}\right)  \notag \\
&&-\frac{16}{3}g_{V}^{2}-8\kappa ^{2}f_{V}g_{V}\Bigg]\Lambda ^{4}  \notag \\
&&+\frac{\alpha _{em}}{64\pi v^{2}\sin ^{2}\theta _{W}}\Bigg[2\left(
f_{V}-2g_{V}-\kappa ^{2}f_{V}\right)  \notag \\
&&\times \left( f_{V}+2g_{V}+\kappa ^{2}f_{V}\right)  \notag \\
&&+14g_{V}\left( 2g_{V}+\kappa ^{2}f_{V}\right)  \notag \\
&&-16\kappa ^{2}f_{V}g_{V}-\frac{32}{3}g_{V}^{2}\Bigg]\Lambda ^{2}  \notag \\
&&-\frac{\alpha _{em}}{64\pi v^{2}\sin ^{2}\theta _{W}}\int_{0}^{1}dx\Bigg[%
2\left( f_{V}-2g_{V}-\kappa ^{2}f_{V}\right)  \notag \\
&&\times \left( f_{V}+2g_{V}+\kappa ^{2}f_{V}\right)  \notag \\
&&+24g_{V}\left( 2g_{V}+\kappa ^{2}f_{V}\right) x-12\kappa
^{2}f_{V}g_{V}-44g_{V}^{2}x^{2}\Bigg]  \notag \\
&&\times \Bigg[2M_{V}^{2}\left( 1-x\right) \ln \left( \frac{\Lambda
^{2}+M_{V}^{2}\left( 1-x\right) }{M_{V}^{2}\left( 1-x\right) }\right)  \notag
\\
&&-\frac{M_{V}^{2}\Lambda ^{2}\left( 1-x\right) }{\Lambda
^{2}+M_{V}^{2}\left( 1-x\right) }\Bigg]  \notag \\
&&-\frac{\alpha _{em}}{256\pi v^{2}M_{V}^{2}\sin ^{2}\theta _{W}}%
\int_{0}^{1}dx\Bigg[6g_{V}\left( 2g_{V}+\kappa ^{2}f_{V}\right) x  \notag \\
&&-24g_{V}^{2}x^{2}-8\kappa ^{2}f_{V}g_{V}\Bigg]  \notag \\
&&\times \Bigg[6M_{V}^{4}\left( 1-x\right) ^{2}\ln \left( \frac{\Lambda
^{2}+M_{V}^{2}\left( 1-x\right) }{M_{V}^{2}\left( 1-x\right) }\right)  \notag
\\
&&-\frac{2M_{V}^{4}\Lambda ^{2}\left( 1-x\right) ^{2}}{\Lambda
^{2}+M_{V}^{2}\left( 1-x\right) }\Bigg]  \notag \\
&&-\frac{26\alpha _{em}g_{V}^{2}M_{V}^{2}}{320\pi v^{2}\sin ^{2}\theta _{W}}-%
\frac{3\alpha _{em}g_{V}^{2}}{8\pi v^{2}\sin ^{2}\theta _{W}}%
\int_{0}^{1}dx\,x\left( 1-x\right)  \notag \\
&&\times \Bigg[3M_{V}^{2}\left( 1-x\right) \ln \left( \frac{\Lambda
^{2}+M_{V}^{2}\left( 1-x\right) }{M_{V}^{2}\left( 1-x\right) }\right)  \notag
\\
&&+\frac{M_{V}^{4}\left( 1-x\right) ^{2}\left[ 5M_{V}^{2}\left( 1-x\right)
+6\Lambda ^{2}\right] }{2\left[ \Lambda ^{2}+M_{V}^{2}\left( 1-x\right) %
\right] ^{2}}\Bigg]  \notag \\
&&-\frac{\alpha _{em}g_{V}^{2}}{16\pi v^{2}M_{V}^{2}\sin ^{2}\theta _{W}}%
\int_{0}^{1}dx\,x\left( 1-x\right)  \notag \\
&&\Bigg\{12M_{V}^{4}\left( 1-x\right) ^{2}\ln \left( \frac{\Lambda
^{2}+M_{V}^{2}\left( 1-x\right) }{M_{V}^{2}\left( 1-x\right) }\right)  \notag
\\
&&-\frac{6M_{V}^{4}\Lambda ^{2}\left( 1-x\right) ^{2}}{\Lambda
^{2}+M_{V}^{2}\left( 1-x\right) }  \notag \\
&&+\frac{M_{V}^{6}\left( 1-x\right) ^{3}\left[ M_{V}^{2}\left( 1-x\right)
+2\Lambda ^{2}\right] }{\left[ M_{V}^{2}\left( 1-x\right) +\Lambda ^{2}%
\right] ^{2}}\Bigg\},
\end{eqnarray}%

\begin{equation}
\widehat{S}_{\left( h\pi ^{0}\right) }\simeq -\frac{\alpha_{em} \left[
\left( 1-\kappa \right) ^{\frac{3}{2}}-2\kappa ^{\frac{3}{2}}\right]^{2}}{%
384\pi \sin ^{2}\theta _{W}}\left[ \ln \left( \frac{\Lambda ^{2}}{m_{h}^{2}}%
\right) -\frac{1}{6}\right],
\end{equation}

\begin{eqnarray}
\widehat{S}_{\left( A^{1}A^{2}\right) } &=&-\frac{\alpha _{em}\Lambda ^{2}}{%
24\pi M_{A}^{2}\sin ^{2}\theta _{W}}+\frac{29\alpha _{em}}{96\pi \sin
^{2}\theta _{W}}\ln \left( \frac{\Lambda ^{2}+M_{A}^{2}}{M_{A}^{2}}\right) 
\notag \\
&&+\frac{9\alpha _{em}}{64\pi \sin ^{2}\theta _{W}}-\frac{35\alpha
_{em}\Lambda ^{2}}{96\pi \left( \Lambda ^{2}+M_{A}^{2}\right) \sin
^{2}\theta _{W}}  \notag \\
&&-\frac{\alpha _{em}\left( 34M_{A}^{2}\Lambda ^{2}+27M_{A}^{4}\right) }{%
192\pi \left( \Lambda ^{2}+M_{A}^{2}\right) ^{2}\sin ^{2}\theta _{W}},
\end{eqnarray}

\begin{equation}
\widehat{S}_{\left( H\pi ^{0}\right) }\simeq -\frac{\alpha_{em}\left[ \left(
1-\kappa \right) ^{\frac{3}{2}}+2\kappa ^{\frac{3}{2}}\right]^{2}}{384\pi
\sin ^{2}\theta _{W}}\left[ \ln \left( \frac{\Lambda ^{2}}{m_{H}^{2}}\right)
-\frac{1}{6}\right],  \label{SH1}
\end{equation}

\begin{eqnarray}
\widehat{S}_{\left( V^{1}A^{2}\right) } &=&\frac{20\,\alpha _{em}\kappa
^{2}\left( M_{A}^{2}+M_{V}^{2}\right) }{192\pi M_{V}^{2}M_{A}^{2}\sin
^{2}\theta _{W}}\Lambda ^{2}-\frac{9\,\alpha _{em}\kappa ^{2}}{64\pi \sin
^{2}\theta _{W}}  \notag \\
&&+\frac{5\,\alpha _{em}\kappa ^{2}\left( M_{A}^{2}+M_{V}^{2}\right) ^{2}}{%
256\pi M_{V}^{2}M_{A}^{2}\sin ^{2}\theta _{W}} \\
&&-\frac{\alpha _{em}\kappa ^{2}}{8\pi \sin ^{2}\theta _{W}}%
\int_{0}^{1}dx\left( x^{2}-x+\frac{5}{2}\right)  \notag \\
&&\times \Bigg[\ln \left( \frac{\Lambda ^{2}+M_{V}^{2}-\left(
M_{V}^{2}-M_{A}^{2}\right) x}{M_{V}^{2}-\left( M_{V}^{2}-M_{A}^{2}\right) x}%
\right)  \notag \\
&&-\frac{\Lambda ^{2}}{\Lambda ^{2}+M_{V}^{2}-\left(
M_{V}^{2}-M_{A}^{2}\right) x}\Bigg]  \notag \\
&&-\frac{\alpha _{em}\kappa ^{2}}{64\pi M_{V}^{2}M_{A}^{2}\sin ^{2}\theta
_{W}}\int_{0}^{1}dx\bigg[10M_{V}^{2}  \notag \\
&&+\left( 26x-3\right) M_{A}^{2}-13\left( M_{A}^{2}+M_{V}^{2}\right) x^{2}%
\bigg]  \notag \\
&&\times \Bigg[\allowbreak 2\left[ M_{V}^{2}-\left(
M_{V}^{2}-M_{A}^{2}\right) x\right]  \notag \\
&&\times \ln \left( \frac{\Lambda ^{2}+\left[ M_{V}^{2}-\left(
M_{V}^{2}-M_{A}^{2}\right) x\right] }{M_{V}^{2}-\left(
M_{V}^{2}-M_{A}^{2}\right) x}\right)  \notag \\
&&-\frac{\left[ M_{V}^{2}-\left( M_{V}^{2}-M_{A}^{2}\right) x\right] \Lambda
^{2}}{\Lambda ^{2}+\left[ M_{V}^{2}-\left( M_{V}^{2}-M_{A}^{2}\right) x%
\right] }\Bigg]  \notag \\
&&+\frac{9\alpha _{em}\kappa ^{2}}{16\pi \sin ^{2}\theta _{W}}%
\int_{0}^{1}dx\,x\left( 1-x\right)  \notag \\
&&\times \Bigg\{\ln \left( \frac{\Lambda ^{2}+\left[ M_{V}^{2}-\left(
M_{V}^{2}-M_{A}^{2}\right) x\right] }{\left[ M_{V}^{2}-\left(
M_{V}^{2}-M_{A}^{2}\right) x\right] }\right)  \notag \\
&&+\frac{4\left[ M_{V}^{2}-\left( M_{V}^{2}-M_{A}^{2}\right) x\right]
\Lambda ^{2}}{2\left( \Lambda ^{2}+M_{V}^{2}-\left(
M_{V}^{2}-M_{A}^{2}\right) x\right) ^{2}}  \notag \\
&&+\frac{3\left[ M_{V}^{2}-\left( M_{V}^{2}-M_{A}^{2}\right) x\right] ^{2}}{%
2\left( \Lambda ^{2}+M_{V}^{2}-\left( M_{V}^{2}-M_{A}^{2}\right) x\right)
^{2}}\Bigg\}  \notag \\
&&-\frac{3\alpha _{em}\kappa ^{2}\left( M_{A}^{2}+M_{V}^{2}\right) }{32\pi
M_{V}^{2}M_{A}^{2}\sin ^{2}\theta _{W}}\int_{0}^{1}dx\,x\left( 1-x\right) 
\notag \\
&&\times \Bigg\{3\left[ M_{V}^{2}-\left( M_{V}^{2}-M_{A}^{2}\right) x\right]
\notag \\
&&\times \ln \left( \frac{\Lambda ^{2}+\left[ M_{V}^{2}-\left(
M_{V}^{2}-M_{A}^{2}\right) x\right] }{\left[ M_{V}^{2}-\left(
M_{V}^{2}-M_{A}^{2}\right) x\right] }\right)  \notag \\
&&+\frac{\left[ M_{V}^{2}-\left( M_{V}^{2}-M_{A}^{2}\right) x\right] ^{2}}{%
2\left( \Lambda ^{2}+\left[ M_{V}^{2}-\left( M_{V}^{2}-M_{A}^{2}\right) x%
\right] \right) ^{2}}  \notag \\
&&\times \left( 5\left[ M_{V}^{2}-\left( M_{V}^{2}-M_{A}^{2}\right) x\right]
+6\Lambda ^{2}\right) \Bigg\},  \notag
\end{eqnarray}

\begin{eqnarray}
\widehat{S}_{\left( hA^{3}\right) } &=&-\frac{\alpha _{em}\kappa \left(
1-\kappa \right) \left( \sqrt{1-\kappa }+2\sqrt{\kappa }\right) ^{2}M_{A}^{2}%
}{64\pi \left( M_{A}^{2}-m_{h}^{2}\right) ^{3}\sin ^{2}\theta _{W}}  \notag
\\
&&\times \left[ M_{A}^{4}-m_{h}^{4}-2m_{h}^{2}M_{A}^{2}\ln \left( \frac{%
M_{A}^{2}}{m_{h}^{2}}\right) \right] ,
\end{eqnarray}

\begin{eqnarray}
\widehat{S}_{\left( HA^{3}\right) } &=&-\frac{\alpha _{em}\kappa \left(
1-\kappa \right) \left( \sqrt{1-\kappa }-2\sqrt{\kappa }\right) ^{2}M_{A}^{2}%
}{64\pi \left( M_{A}^{2}-m_{H}^{2}\right) ^{3}\sin ^{2}\theta _{W}}  \notag
\\
&&\times \left[ M_{A}^{4}-m_{H}^{4}-2m_{H}^{2}M_{A}^{2}\ln \left( \frac{%
M_{A}^{2}}{m_{H}^{2}}\right) \right] ,  \label{SH2}
\end{eqnarray}

\begin{eqnarray}
\widehat{S}_{\left( \eta V^{3}\right) } &=&-\frac{\alpha _{em}\left(
1-\kappa \right) ^{2}M_{V}^{2}}{32\pi \left( M_{V}^{2}-m_{\eta }^{2}\right)
^{3}\sin ^{2}\theta _{W}}  \notag \\
&&\times \left[ M_{V}^{4}-m_{\eta }^{4}-2m_{\eta }^{2}M_{A}^{2}\ln \left( 
\frac{M_{V}^{2}}{m_{\eta }^{2}}\right) \right] ,  \label{Seta1}
\end{eqnarray}

\begin{eqnarray}
\widehat{S}_{\left( A^{2}\right) \left( W^{3}-A^{3}\right) } &=&\widehat{S}%
_{\left( A^{1}\right) \left( W^{3}-A^{3}\right) }=\widehat{S}_{\left(
A^{2}\right) \left( A^{3}-B^{0}\right) }  \notag \\
&=&\widehat{S}_{\left( A^{1}\right) \left( A^{3}-B^{0}\right) },
\end{eqnarray}

\begin{eqnarray}
\widehat{S}_{\left( A^{1}\right) \left( A^{3}-B^{0}\right) } &=&\frac{%
3\alpha _{em}\kappa g_{C}f_{A}}{32\pi M_{A}^{2}\sin ^{2}\theta _{W}}\Bigg\{%
\frac{\Lambda ^{4}}{2M_{A}^{2}} \\
&&+3\left[ \Lambda ^{2}-M_{A}^{2}\ln \left( \frac{\Lambda ^{2}+M_{A}^{2}}{%
M_{A}^{2}}\right) \right] \Bigg\},  \notag
\end{eqnarray}

\begin{eqnarray}
\widehat{S}_{\left( V^{2}\right) \left( W^{3}-A^{3}\right) } &=&\widehat{S}%
_{\left( V^{1}\right) \left( W^{3}-A^{3}\right) }=\widehat{S}_{\left(
V^{2}\right) \left( A^{3}-B^{0}\right) }  \notag \\
&=&\widehat{S}_{\left( V^{1}\right) \left( A^{3}-B^{0}\right) },
\end{eqnarray}

\begin{eqnarray}
\widehat{S}_{\left( V^{1}\right) \left( A^{3}-B^{0}\right) } &=&\frac{%
3\alpha _{em}\kappa g_{C}f_{A}}{32\pi M_{A}^{2}\sin ^{2}\theta _{W}}\Bigg\{%
\frac{\Lambda ^{4}}{2M_{V}^{2}} \\
&&+3\left[ \Lambda ^{2}-M_{V}^{2}\ln \left( \frac{\Lambda ^{2}+M_{V}^{2}}{%
M_{V}^{2}}\right) \right] \Bigg\},  \notag
\end{eqnarray}

\begin{eqnarray}
\widehat{S}_{\left( V^{2}\right) \left( W^{3}-V^{3}\right) } &=&\widehat{S}%
_{\left( V^{1}\right) \left( W^{3}-V^{3}\right) }=\widehat{S}_{\left(
V^{2}\right) \left( V^{3}-B^{0}\right) }  \notag \\
&=&\widehat{S}_{\left( V^{1}\right) \left( V^{3}-B^{0}\right) },
\end{eqnarray}

\begin{eqnarray}
\widehat{S}_{\left( V^{1}\right) \left( V^{3}-B^{0}\right) } &=&-\frac{%
3\alpha _{em}g_{C}f_{V}}{32\pi M_{V}^{2}\sin ^{2}\theta _{W}}\Bigg\{\frac{%
\Lambda ^{4}}{2M_{V}^{2}} \\
&&+3\left[ \Lambda ^{2}-M_{V}^{2}\ln \left( \frac{\Lambda ^{2}+M_{V}^{2}}{%
M_{V}^{2}}\right) \right] \Bigg\},  \notag
\end{eqnarray}

\begin{eqnarray}
\widehat{S}_{\left( A^{2}\right) \left( W^{3}-V^{3}\right) } &=&\widehat{S}%
_{\left( A^{1}\right) \left( W^{3}-V^{3}\right) }=\widehat{S}_{\left(
A^{2}\right) \left( V^{3}-B^{0}\right) }  \notag \\
&=&\widehat{S}_{\left( A^{1}\right) \left( V^{3}-B^{0}\right) },
\end{eqnarray}

\begin{eqnarray}
\widehat{S}_{\left( A^{1}\right) \left( V^{3}-B^{0}\right) } &=&-\frac{%
3\alpha _{em}g_{C}f_{V}}{32\pi M_{V}^{2}\sin ^{2}\theta _{W}}\Bigg\{\frac{%
\Lambda ^{4}}{2M_{A}^{2}} \\
&&+3\left[ \Lambda ^{2}-M_{A}^{2}\ln \left( \frac{\Lambda ^{2}+M_{A}^{2}}{%
M_{A}^{2}}\right) \right] \Bigg\},  \notag
\end{eqnarray}

\begin{eqnarray}
\widehat{S}_{\left( V^{1}V^{2}\right) }^{\left( W^{3}-V^{3}\right) }=%
\widehat{S}_{\left( V^{1}V^{2}\right) }^{\left( V^{3}-B^{0}\right) } &=&%
\frac{3\alpha _{em}g_{C}f_{V}}{64\pi M_{V}^{4}\sin ^{2}\theta _{W}}\Bigg[%
\Lambda ^{4}+2M_{V}^{2}\Lambda ^{2}  \notag \\
&&-6M_{V}^{4}\ln \left( \frac{\Lambda ^{2}+M_{V}^{2}}{M_{V}^{2}}\right) 
\notag \\
&&+\frac{4M_{V}^{4}\Lambda ^{2}}{\Lambda ^{2}+M_{V}^{2}}\Bigg],
\end{eqnarray}

\begin{eqnarray}
\widehat{S}_{\left( A^{1}A^{2}\right) }^{\left( W^{3}-V^{3}\right) }=%
\widehat{S}_{\left( A^{1}A^{2}\right) }^{\left( V^{3}-B^{0}\right) } &=&%
\frac{3\alpha _{em}g_{C}f_{V}}{64\pi M_{V}^{2}M_{A}^{2}\sin ^{2}\theta _{W}}%
\Bigg[\Lambda ^{4}  \notag \\
&&+2M_{A}^{2}\Lambda ^{2}-6M_{A}^{4}\ln \left( \frac{\Lambda ^{2}+M_{A}^{2}}{%
M_{A}^{2}}\right)  \notag \\
&&+\frac{4M_{A}^{4}\Lambda ^{2}}{\Lambda ^{2}+M_{A}^{2}}\Bigg],
\end{eqnarray}

\begin{equation}
\widehat{S}_{\left( A^{1}V^{2}\right) }^{\left( W^{3}-A^{3}\right) }=%
\widehat{S}_{\left( V^{1}A^{2}\right) }^{\left( W^{3}-A^{3}\right) }=%
\widehat{S}_{\left( A^{1}V^{2}\right) }^{\left( A^{3}-B^{0}\right) }=%
\widehat{S}_{\left( V^{1}A^{2}\right) }^{\left( A^{3}-B^{0}\right) },
\end{equation}

\begin{eqnarray}
\widehat{S}_{\left( V^{1}A^{2}\right) }^{\left( A^{3}-B^{0}\right) } &=&-%
\frac{3\alpha _{em}\kappa g_{C}f_{A}}{128\pi M_{A}^{2}\sin ^{2}\theta _{W}}%
\Bigg\{\frac{M_{V}^{2}+M_{A}^{2}}{M_{V}^{2}M_{A}^{2}}\Lambda ^{4}  \notag \\
&&+2\left[ 6-\frac{\left( M_{V}^{2}+M_{A}^{2}\right) ^{2}}{M_{V}^{2}M_{A}^{2}%
}\right] \Lambda ^{2} \\
&&-\frac{2M_{V}^{4}}{M_{V}^{2}-M_{A}^{2}}\left( 6-\frac{M_{V}^{2}+M_{A}^{2}}{%
M_{A}^{2}}\right) \ln \left( \frac{\Lambda ^{2}+M_{V}^{2}}{M_{V}^{2}}\right)
\notag \\
&&-\frac{2M_{A}^{4}}{M_{A}^{2}-M_{V}^{2}}\left( 6-\frac{M_{V}^{2}+M_{A}^{2}}{%
M_{V}^{2}}\right) \ln \left( \frac{\Lambda ^{2}+M_{A}^{2}}{M_{A}^{2}}\right) %
\Bigg\},  \notag
\end{eqnarray}

\begin{eqnarray}
\widehat{S}_{\left( h\right) \left( W^{3}-A^{3}\right) } &=&\widehat{S}%
_{\left( h\right) \left( A^{3}-B^{0}\right) } \\
&=&-\frac{\alpha _{em}g_{C}f_{A}\left( 1-5\kappa \right) }{64\pi
M_{A}^{2}\sin ^{2}\theta _{W}}\Bigg[\Lambda ^{2}\notag\\
&&-m_{h}^{2}\ln \left( \frac{\Lambda ^{2}+m_{h}^{2}}{m_{h}^{2}}\right)\Bigg],  \notag
\end{eqnarray}

\begin{eqnarray}
\widehat{S}_{\left( H\right) \left( W^{3}-A^{3}\right) } &=&\widehat{S}%
_{\left( H\right) \left( A^{3}-B^{0}\right) }  \label{SH3} \\
&=&-\frac{\alpha _{em}g_{C}f_{A}\left( 1-5\kappa \right) }{64\pi
M_{A}^{2}\sin ^{2}\theta _{W}}\Bigg[\Lambda ^{2}\notag\\
&&-m_{H}^{2}\ln \left( \frac{\Lambda ^{2}+m_{H}^{2}}{m_{H}^{2}}\right)\Bigg] ,\notag
\end{eqnarray}

\begin{eqnarray}
\widehat{S}_{\left( \eta \right) \left( W^{3}-A^{3}\right) } &=&\widehat{S}%
_{\left( \eta \right) \left( A^{3}-B^{0}\right) }  \label{Seta2} \\
&=&-\frac{\alpha _{em}g_{C}f_{A}\left( 1-\kappa \right) }{32\pi
M_{A}^{2}\sin ^{2}\theta _{W}}\Bigg[\Lambda ^{2}\notag\\
&&-m_{\eta }^{2}\ln \left(\frac{\Lambda ^{2}+m_{\eta }^{2}}{m_{\eta }^{2}}\right)\Bigg] ,  \notag
\end{eqnarray}

\begin{equation}
\widehat{S}_{\left( \pi ^{1}\right) }=\widehat{S}_{\left( \pi ^{2}\right) }=-%
\frac{\alpha _{em}\left( 2f_{V}^{2}-2c_{WB}-f_{A}^{2}\right) }{8\pi
v^{2}\sin ^{2}\theta _{W}}\Lambda ^{2}.
\end{equation}

\section{: Scalar masses.}
\label{Apendix3}
The masses of the scalars $h$ and $H$ and pseudoscalar $\eta$ receive contributions at tree- and at one-loop level corrections. These masses are given by
\begin{figure*}[tbh]
\resizebox{0.95\textwidth}{!}{
\includegraphics{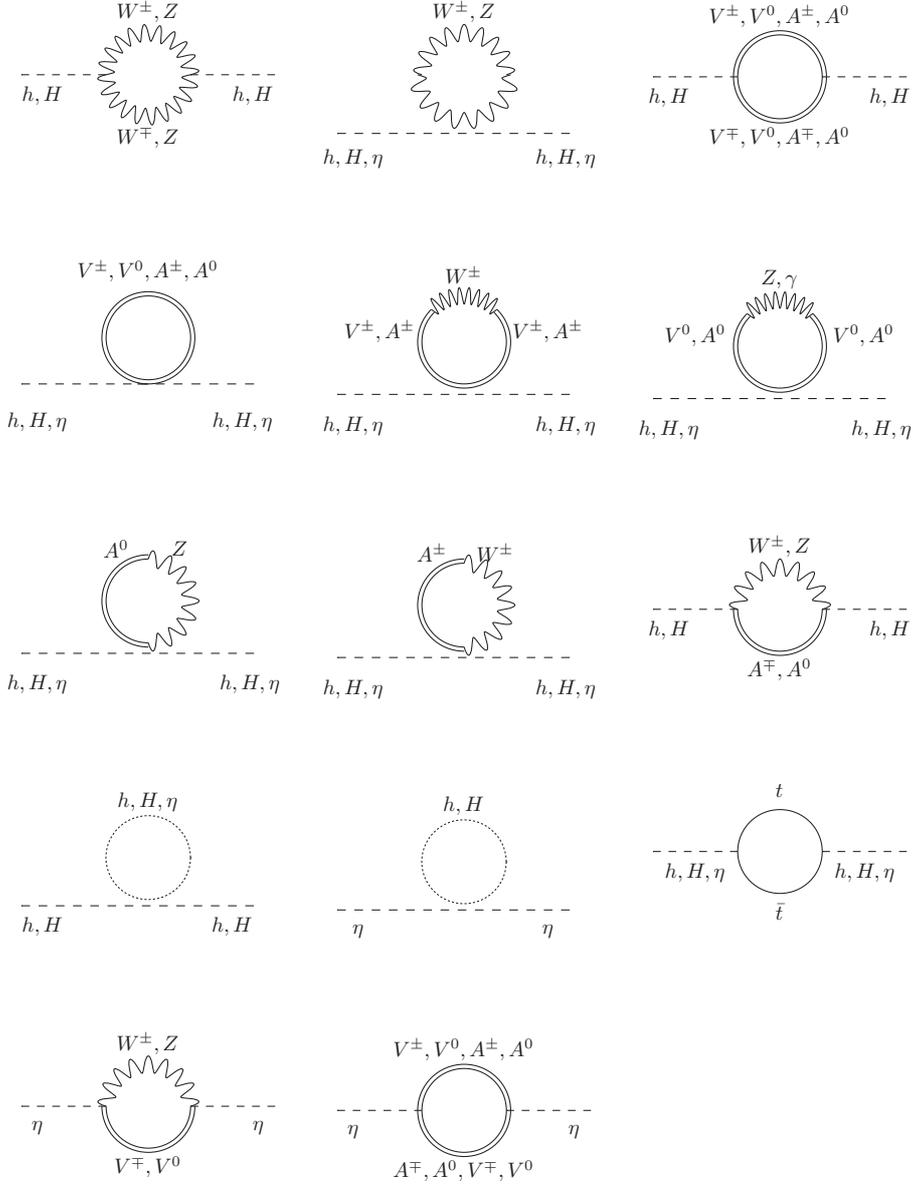}}\vspace{-4cm}
\caption{One-loop Feynman diagrams in the unitary gauge contributing to the masses of the parity even $h$ and $H$ and parity odd $\eta$ scalars.}
\label{Diagmscalars}
\end{figure*}
\begin{equation}
m_{h}^{2}=\left( m_{h}\right) _{three}^{2}+\Sigma _{h},
\end{equation}
\begin{equation}
m_{H}^{2}=\left( m_{H}\right) _{three}^{2}+\Sigma _{H},
\end{equation}
\begin{equation}
m_{\eta }^{2}=\left( m_{\eta }\right) _{three}^{2}+\Sigma _{\eta },
\end{equation}

where the tree-level contributions to the parity even and parity odd scalar masses, which are obtained from the scalar potential given in Eq. (\ref{scalarpotential}) are

\begin{equation}
\left( m_{h}^{2}\right) _{three}=\frac{1}{4}v_{LC}\left[ \left( 2\kappa
_{2}+\lambda _{3}^{2}\right) v_{LC}-2\sqrt{2}\kappa _{3}v_{CD}\right] ,
\end{equation}
\begin{equation}
\left( m_{H}^{2}\right) _{three}=\frac{1}{4}v_{LC}\left[ \left( 2\kappa
_{2}+\lambda _{3}^{2}\right) v_{LC}+2\sqrt{2}\kappa _{3}v_{CD}\right] ,
\end{equation}
\begin{equation}
\left( m_{\eta }^{2}\right) _{three}=\frac{1}{4}\left( -2\kappa _{2}+\lambda
_{3}^{2}\right) v_{LC}^{2}.
\end{equation}

with

\begin{equation}
v_{LC}=\frac{v}{2\sqrt{1-\kappa }},\quad v_{CD}=\frac{v}{2\sqrt{2\kappa }}%
,\quad \kappa=\frac{M^2_V}{M^2_A}.
\end{equation}
while the one-loop level contributions to the masses of the scalars $h$, $H$ and pseudoscalar $\eta$ can be decomposed as:
\begin{equation}
\Sigma _{h}=\Sigma _{h}^{\left( spin-0\right) }+\Sigma _{h}^{\left(
spin-1/2\right) }+\Sigma _{h}^{\left( spin-1\right) },
\end{equation}
\begin{equation}
\Sigma _{H}=\Sigma _{H}^{\left( spin-0\right) }+\Sigma _{H}^{\left(
spin-1/2\right) }+\Sigma _{H}^{\left( spin-1\right) },
\end{equation}
\begin{equation}
\Sigma _{\eta }=\Sigma _{\eta }^{\left( spin-0\right) }+\Sigma _{\eta
}^{\left( spin-1/2\right) }+\Sigma _{\eta }^{\left( spin-1\right) },
\end{equation}
These one-loop level contributions come from Feynman diagrams containing spin-$0$, spin-$1/2$ and spin-$1$ particles in the internal lines of the loops. For the contribution from the fermion loops we will only keep the dominant term, which is the one involving the top quark. From the Feynman diagrams shown in Figure \ref{Diagmscalars}, it follows that the spin-$0$, spin-$1/2$ and spin-$1$ particles give the following one-loop level contributions to the masses of the scalars $h$ and $H$ and pseudoscalar $\eta$: 
\begin{equation}
\Sigma _{h}^{\left( spin-0\right) }\simeq 12\lambda _{h^{4}}I_{1}\left(
m_{h}\right) +2\lambda _{h^{2}H^{2}}I_{1}\left( m_{H}\right) +2\lambda
_{h^{2}\eta ^{2}}I_{1}\left( m_{\eta }\right) ,
\end{equation}

\begin{equation}
\Sigma _{H}^{\left( spin-0\right) }\simeq 12\lambda _{H^{4}}I_{1}\left(
m_{H}\right) +2\lambda _{h^{2}H^{2}}I_{1}\left( m_{h}\right) +2\lambda
_{H^{2}\eta ^{2}}I_{1}\left( m_{\eta }\right) ,
\end{equation}

\begin{equation}
\Sigma _{\eta }^{\left( spin-0\right) }\simeq 2\lambda _{h^{2}\eta
^{2}}I_{1}\left( m_{h}\right) +2\lambda _{H^{2}\eta ^{2}}I_{1}\left(
m_{H}\right) ,
\end{equation}

\begin{equation}
\Sigma _{h}^{\left( spin-1/2\right) }\simeq -\frac{12a_{htt}^{2}}{v^{2}}%
\left[ I_{4}\left( m_{t}\right) +m_{t}^{2}I_{3}\left( m_{t}\right) \right]
m_{t}^{2},
\end{equation}

\begin{equation}
\Sigma _{H}^{\left( spin-1/2\right) }\simeq -\frac{12a_{Htt}^{2}}{v^{2}}%
\left[ I_{4}\left( m_{t}\right) +m_{t}^{2}I_{3}\left( m_{t}\right) \right]
m_{t}^{2},
\end{equation}

\begin{equation}
\Sigma _{\eta }^{\left( spin-1/2\right) }\simeq -\frac{12a_{\eta tt}^{2}}{%
v^{2}}\left[ -I_{4}\left( m_{t}\right) +m_{t}^{2}I_{3}\left( m_{t}\right) %
\right] m_{t}^{2},
\end{equation}

\begin{eqnarray}
\Sigma _{h}^{\left( spin-1\right) } &\simeq &2a_{hWW}^{2}F_{A}\left(
M_{W}\right) +2b_{hhWW}F_{B}\left( M_{W}\right)   \notag \\
&&+a_{hWW}^{2}F_{A}\left( M_{Z}\right) +b_{hhWW}F_{B}\left( M_{Z}\right)  
\notag \\
&&+3a_{hVV}^{2}F_{A}\left( M_{V}\right) +3b_{hhVV}F_{B}\left( M_{V}\right)  
\notag \\
&&+3a_{hAA}^{2}F_{A}\left( M_{A}\right) +3b_{hhAA}F_{B}\left( M_{A}\right)  
\notag \\
&&+J_{h}\left( M_{W},M_{Z},M_{A}\right) +\Sigma _{h\left( AW\right) }  \notag
\\
&&+\Sigma _{h\left( AZ\right) }+\Sigma _{h\left( VWV\right) }+\Sigma
_{h\left( VZV\right) }  \notag \\
&&+\Sigma _{h\left( V\gamma V\right) }+\Sigma _{h\left( AWA\right) }+\Sigma
_{h\left( AZA\right) },
\end{eqnarray}

\begin{eqnarray}
\Sigma _{H}^{\left( spin-1\right) } &\simeq &2a_{HWW}^{2}F_{A}\left(
M_{W}\right) +2b_{HHWW}F_{B}\left( M_{W}\right)   \notag \\
&&+a_{HWW}^{2}F_{A}\left( M_{Z}\right) +b_{HHWW}F_{B}\left( M_{Z}\right)  
\notag \\
&&+3a_{HVV}^{2}F_{A}\left( M_{V}\right) +3b_{HHVV}F_{B}\left( M_{V}\right)  
\notag \\
&&+3a_{HAA}^{2}F_{A}\left( M_{A}\right) +3b_{HHAA}F_{B}\left( M_{A}\right)  
\notag \\
&&+J_{H}\left( M_{W},M_{Z},M_{A}\right) +\Sigma _{H\left( AW\right) }  \notag
\\
&&+\Sigma _{H\left( AZ\right) }+\Sigma _{H\left( VWV\right) }+\Sigma
_{H\left( VZV\right) }  \notag \\
&&+\Sigma _{H\left( V\gamma V\right) }+\Sigma _{H\left( AWA\right) }+\Sigma
_{H\left( AZA\right) },
\end{eqnarray}

\begin{eqnarray}
\Sigma _{\eta }^{\left( spin-1\right) } &\simeq &2b_{\eta \eta
WW}F_{B}\left( M_{W}\right) +b_{\eta \eta WW}F_{B}\left( M_{Z}\right)  
\notag \\
&&+3b_{\eta \eta VV}F_{B}\left( M_{V}\right) +3b_{\eta \eta AA}F_{B}\left(
M_{A}\right)   \notag \\
&&+J_{\eta }\left( M_{W},M_{Z},M_{V},M_{A}\right) +\Sigma _{\eta \left(
AW\right) }  \notag \\
&&+\Sigma _{\eta \left( AZ\right) }+\Sigma _{\eta \left( VWV\right) }+\Sigma
_{\eta \left( VZV\right) }  \notag \\
&&+\Sigma _{\eta \left( V\gamma V\right) }+\Sigma _{\eta \left( AWA\right)
}+\Sigma _{\eta \left( AZA\right) },
\end{eqnarray}
where the different dimensionless couplings are given by
\begin{eqnarray}
\lambda _{h^{4}} &=&\lambda _{H^{4}} \\
&=&\frac{2v_{LC}^{2}\left( 2\kappa _{2}+\lambda _{3}^{2}\right)
+v_{CD}^{2}\left( 2\kappa _{2}+8\kappa _{3}+\lambda _{3}^{2}\right) }{%
1024v_{CD}^{2}},  \notag
\end{eqnarray}

\begin{equation}
\lambda _{h^{2}H^{2}}=\frac{6v_{LC}^{2}\left( 2\kappa _{2}+\lambda
_{3}^{2}\right) +v_{CD}^{2}\left( 6\kappa _{2}-8\kappa _{3}+3\lambda
_{3}^{2}\right) }{512v_{CD}^{2}},
\end{equation}

\begin{equation}
\lambda _{h^{2}\eta ^{2}}=\lambda _{H^{2}\eta ^{2}}=\frac{-2\kappa
_{2}+4\kappa _{3}+3\lambda _{3}^{2}}{256},
\end{equation}

\begin{equation}
a_{hWW}=\frac{2\kappa ^{\frac{3}{2}}-\left( 1-\kappa \right) ^{\frac{3}{2}}}{%
2\sqrt{2}},\quad a_{HWW}=\frac{2\kappa ^{\frac{3}{2}}+\left( 1-\kappa
\right) ^{\frac{3}{2}}}{2\sqrt{2}}  \label{ahff}
\end{equation}

\begin{equation}
a_{hVV}=-\frac{\left( 1-\kappa \right) ^{\frac{1}{2}}}{2\sqrt{2}},\quad
a_{hAA}=\kappa \left( 1-2\sqrt{\frac{1-\kappa }{\kappa }}\right) a_{hVV}.
\end{equation}

\begin{equation}
a_{HVV}=\frac{\left( 1-\kappa \right) ^{\frac{1}{2}}}{2\sqrt{2}},\quad
a_{HAA}=\kappa \left(1+2\sqrt{\frac{1-\kappa }{\kappa }}\right) a_{HVV}.
\end{equation}

\begin{equation}
b_{hhWW}=b_{HHWW}=\frac{\left( 1-\kappa \right) ^{2}+4\kappa ^{2}}{8},\quad
b_{\eta \eta WW}=\frac{\left( 1-\kappa \right) ^{2}}{4}
\label{b1}
\end{equation}

\begin{equation}
b_{hhVV}=b_{HHVV}=\frac{1-\kappa }{8},\quad b_{hhAA}=b_{HHAA}=\frac{5\kappa
\left( 1-\kappa \right) }{4},
\label{b2}
\end{equation}

\begin{equation}
b_{\eta \eta VV}=\frac{1-\kappa }{4},\quad b_{\eta \eta AA}=\frac{\kappa
\left( 1-\kappa \right) }{4},
\label{b3}
\end{equation}

and the following loop functions have been introduced:

\begin{equation}
F_{A}\left( M\right) =\frac{M^{4}}{v^{2}}I_{A}\left( M\right) ,\quad
F_{B}\left( M\right) =\frac{M^{2}}{v^{2}}I_{B}\left( M\right) ,
\end{equation}

\begin{eqnarray}
J_{h}\left( M_{W},M_{Z},M_{A}\right)  &=&\frac{\left( 1-\kappa \right)
\left( \sqrt{1-\kappa }+2\sqrt{\kappa }\right) ^{2}M_{W}^{2}M_{V}^{2}}{%
2v^{2}\cos ^{2}\theta _{W}}  \notag \\
&&\times I_{Agen}\left( M_{Z},M_{A}\right)   \notag \\
&&+\frac{\left( 1-\kappa \right) \left( \sqrt{1-\kappa }+2\sqrt{\kappa }%
\right) ^{2}M_{W}^{2}M_{V}^{2}}{v^{2}}  \notag \\
&&\times I_{Agen}\left( M_{W},M_{A}\right) ,
\end{eqnarray}

\begin{eqnarray}
J_{H}\left( M_{W},M_{Z},M_{A}\right)  &=&\frac{\left( 1-\kappa \right)
\left( \sqrt{1-\kappa }-2\sqrt{\kappa }\right) ^{2}M_{W}^{2}M_{V}^{2}}{%
2v^{2}\cos ^{2}\theta _{W}}  \notag \\
&&\times I_{Agen}\left( M_{Z},M_{A}\right)   \notag \\
&&+\frac{\left( 1-\kappa \right) \left( \sqrt{1-\kappa }-2\sqrt{\kappa }%
\right) ^{2}M_{W}^{2}M_{V}^{2}}{v^{2}}  \notag \\
&&\times I_{Agen}\left( M_{W},M_{A}\right) ,
\end{eqnarray}

\begin{eqnarray}
J_{\eta }\left( M_{W},M_{Z},M_{V},M_{A}\right) &=&\frac{3\left( 1-\kappa
\right) M_{V}^{4}}{v^{2}}I_{Agen}\left( M_{V},M_{A}\right)\notag \\
&&+\frac{\left( 1-\kappa \right)^{2}M_{W}^{2}M_{V}^{2}}{v^{2}\cos
^{2}\theta _{W}}\notag \\
&&\times I_{Agen}\left( M_{Z},M_{V}\right)\notag \\
&&+\frac{2\left( 1-\kappa \right)^{2}M_{W}^{2}M_{V}^{2}}{v^{2}}\notag \\
&&\times I_{Agen}\left( M_{W},M_{V}\right) ,
\end{eqnarray}

\begin{eqnarray}
\Sigma _{h\left( AW\right) } &=&\Sigma _{H\left( AW\right) } \\
&=&\frac{3\left( 1-5\kappa \right) \sqrt{1-\kappa }f_{A}M_{W}^{2}M_{V}}{%
2v^{3}}I_{4gen}\left( M_{W},M_{A}\right) ,  \notag
\end{eqnarray}

\begin{eqnarray}
\Sigma _{h\left( AZ\right) } &=&\Sigma _{H\left( AZ\right) } \\
&=&\frac{3\left( 1-5\kappa \right) \sqrt{1-\kappa }f_{A}M_{W}^{2}M_{V}}{%
4v^{3}\cos \theta _{W}}I_{4gen}\left( M_{Z},M_{A}\right) ,  \notag
\end{eqnarray}

\begin{eqnarray}
\Sigma _{h\left( VWV\right) } &=&\Sigma _{H\left( VWV\right) } \\
&=&\frac{6b_{hhVV}f_{V}^{2}M_{W}^{2}M_{V}^{2}}{v^{4}}I_{6gen}\left(
M_{V},M_{W}\right) ,  \notag
\end{eqnarray}

\begin{eqnarray}
\Sigma _{h\left( VZV\right) } &=&\Sigma _{H\left( VZV\right) }  \notag \\
&=&\frac{3b_{hhVV}f_{V}^{2}\left( \cos ^{2}\theta _{W}-\sin ^{2}\theta
_{W}\right) M_{W}^{2}M_{V}^{2}}{v^{4}\cos \theta _{W}}  \notag \\
&&\times I_{6gen}\left( M_{V},M_{Z}\right) ,
\end{eqnarray}

\begin{eqnarray}
\Sigma _{h\left( V\gamma V\right) } &=&\Sigma _{H\left( V\gamma V\right) } 
\notag \\
&=&\frac{6b_{hhVV}f_{V}^{2}\sin \theta _{W}M_{W}^{2}M_{V}^{2}}{v^{4}}%
I_{4}\left( M_{V}\right) ,
\end{eqnarray}

\begin{eqnarray}
\Sigma _{h\left( AWA\right) } &=&\Sigma _{H\left( AWA\right) } \\
&=&\frac{6b_{hhAA}f_{A}^{2}M_{W}^{2}M_{A}^{2}}{v^{4}}I_{6gen}\left(
M_{A},M_{W}\right) ,  \notag
\end{eqnarray}

\begin{eqnarray}
\Sigma _{h\left( AZA\right) } &=&\Sigma _{H\left( AZA\right) } \\
&=&\frac{3b_{hhAA}f_{A}^{2}M_{W}^{2}M_{A}^{2}}{v^{4}\cos \theta _{W}}%
I_{6gen}\left( M_{A},M_{Z}\right) ,  \notag
\end{eqnarray}

\begin{equation}
\Sigma _{\eta \left( AW\right) }=\frac{3\left( 1-\kappa \right) \sqrt{%
1-\kappa }f_{A}M_{W}^{2}M_{V}}{v^{3}}I_{4gen}\left( M_{W},M_{A}\right) ,
\end{equation}

\begin{equation}
\Sigma _{\eta \left( AZ\right) }=\frac{3\left( 1-\kappa \right) \sqrt{%
1-\kappa }f_{A}M_{W}^{2}M_{V}}{2v^{3}\cos \theta _{W}}I_{4gen}\left(
M_{W},M_{A}\right) ,
\end{equation}

\begin{equation}
\Sigma _{\eta \left( VWV\right) }=\frac{6b_{\eta \eta
VV}f_{V}^{2}M_{W}^{2}M_{V}^{2}}{v^{4}}I_{6gen}\left( M_{V},M_{W}\right) ,
\end{equation}

\begin{eqnarray}
\Sigma _{\eta \left( VZV\right) } &=&\frac{3b_{\eta \eta VV}f_{V}^{2}\left(
\cos ^{2}\theta _{W}-\sin ^{2}\theta _{W}\right) M_{W}^{2}M_{V}^{2}}{%
v^{4}\cos \theta _{W}}  \notag \\
&&\times I_{6gen}\left( M_{V},M_{Z}\right) ,
\end{eqnarray}

\begin{eqnarray}
\Sigma _{\eta \left( V\gamma V\right) } &=&\frac{6b_{\eta \eta VV}f_{V}^{2}\sin\theta_W M_{W}^{2}M_{V}^{2}}{%
v^{4}}I_{4}\left( M_{V}\right) ,
\end{eqnarray}

\begin{equation}
\Sigma _{\eta \left( AWA\right) }=\frac{6b_{\eta \eta
AA}f_{A}^{2}M_{W}^{2}M_{A}^{2}}{v^{4}}I_{6gen}\left( M_{A},M_{W}\right) ,
\end{equation}

\begin{equation}
\Sigma _{\eta \left( AZA\right) }=\frac{3b_{\eta \eta
AA}f_{A}^{2}M_{W}^{2}M_{A}^{2}}{v^{4}\cos \theta _{W}}I_{6gen}\left(
M_{A},M_{Z}\right) ,
\end{equation}

\begin{equation}
I_{B}\left( M\right) =4I_{1}\left( M\right) -\frac{1}{M^{2}}I_{2}\left(
M\right) ,
\end{equation}

\begin{equation}
I_{1}\left( M\right) =-\frac{1}{16\pi ^{2}}\left[ \Lambda ^{2}-M^{2}\ln
\left( \frac{\Lambda ^{2}+M^{2}}{M^{2}}\right) \right] ,
\end{equation}

\begin{equation}
I_{2}\left( M\right) =\frac{1}{32\pi ^{2}}\Lambda ^{4}-\frac{M^{2}}{16\pi
^{2}}\left[ \Lambda ^{2}-M^{2}\ln \left( \frac{\Lambda ^{2}+M^{2}}{M^{2}}%
\right) \right] ,
\end{equation}

\begin{equation}
I_{A}\left( M\right) =4I_{3}\left( M\right) -\frac{2}{M^{2}}I_{4}\left(
M\right) +\frac{1}{M^{4}}I_{5}\left( M\right) ,
\end{equation}

\begin{equation}
I_{3}\left( M\right) =\frac{1}{16\pi ^{2}}\left[ \ln \left( \frac{\Lambda
^{2}+M^{2}}{M^{2}}\right) -\frac{\Lambda ^{2}}{\Lambda ^{2}+M^{2}}\right] ,
\end{equation}

\begin{equation}
I_{4}\left( M\right) =-\frac{1}{16\pi ^{2}}\left[ \allowbreak \Lambda
^{2}-2M^{2}\ln \left( \frac{\Lambda ^{2}+M^{2}}{M^{2}}\right) +\frac{%
M^{2}\Lambda ^{2}}{\Lambda ^{2}+M^{2}}\right] ,
\end{equation}

\begin{eqnarray}
I_{5}\left( M\right) &=&\frac{1}{32\pi ^{2}}\Bigg[\Lambda ^{4}-4M^{2}\Lambda
^{2}  \notag \\
&&+6M^{4}\ln \left( \frac{\Lambda ^{2}+M^{2}}{M^{2}}\right) -\frac{%
2M^{4}\Lambda ^{2}}{\Lambda ^{2}+M^{2}}\Bigg],
\end{eqnarray}

\begin{eqnarray}
I_{Agen}\left( M_{1},M_{2}\right) &=&4I_{3gen}\left( M_{1},M_{2}\right) 
\notag \\
&&-\left( \frac{1}{M_{1}^{2}}+\frac{1}{M_{2}^{2}}\right) I_{4gen}\left(
M_{1},M_{2}\right)  \notag \\
&&+\frac{1}{M_{1}^{2}M_{2}^{2}}I_{5gen}\left( M_{1},M_{2}\right) ,
\end{eqnarray}
\begin{eqnarray}
I_{3gen}\left( M_{1},M_{2}\right) &=&\frac{1}{16\pi ^{2}}\Bigg[\frac{%
M_{1}^{2}}{M_{1}^{2}-M_{2}^{2}}\ln \left( \frac{\Lambda ^{2}+M_{1}^{2}}{%
M_{1}^{2}}\right)  \notag \\
&&+\frac{M_{2}^{2}}{M_{2}^{2}-M_{1}^{2}}\ln \left( \frac{\Lambda
^{2}+M_{2}^{2}}{M_{2}^{2}}\right) \Bigg],
\end{eqnarray}

\begin{eqnarray}
I_{4gen}\left( M_{1},M_{2}\right) &=&-\frac{1}{16\pi ^{2}}\Bigg[\Lambda ^{2}-%
\frac{M_{1}^{4}}{M_{1}^{2}-M_{2}^{2}}\ln \left( \frac{\Lambda ^{2}+M_{1}^{2}%
}{M_{1}^{2}}\right)  \notag \\
&&-\frac{M_{2}^{4}}{M_{2}^{2}-M_{1}^{2}}\ln \left( \frac{\Lambda
^{2}+M_{2}^{2}}{M_{2}^{2}}\right) \Bigg],
\end{eqnarray}

\begin{eqnarray}
I_{5gen}\left( M_{1},M_{2}\right) &=&\frac{1}{32\pi ^{2}}\Bigg[\Lambda
^{4}-2\left( M_{1}^{2}+M_{2}^{2}\right) \Lambda ^{2}  \notag \\
&&+\frac{2M_{1}^{6}}{M_{1}^{2}-M_{2}^{2}}\ln \left( \frac{\Lambda
^{2}+M_{1}^{2}}{M_{1}^{2}}\right) \\
&&+\frac{2M_{2}^{6}}{M_{2}^{2}-M_{1}^{2}}\ln \left( \frac{\Lambda
^{2}+M_{2}^{2}}{M_{2}^{2}}\right) \Bigg],  \notag
\end{eqnarray}

\begin{eqnarray}
I_{6gen}\left( M_{1},M_{2}\right) &=&-\frac{1}{16\pi ^{2}}\Bigg[\Lambda ^{2}
\notag \\
&&+\frac{M_{1}^{4}\left( 3M_{2}^{2}-2M_{1}^{2}\right) }{\left(
M_{1}^{2}-M_{2}^{2}\right) ^{2}}\ln \left( \frac{\Lambda ^{2}+M_{1}^{2}}{%
M_{1}^{2}}\right)  \notag \\
&&-\frac{M_{2}^{6}}{\left( M_{2}^{2}-M_{1}^{2}\right) ^{2}}\ln \left( \frac{%
\Lambda ^{2}+M_{2}^{2}}{M_{2}^{2}}\right)  \notag \\
&&+\frac{M_{1}^{4}\Lambda ^{2}}{\left( M_{1}^{2}-M_{2}^{2}\right) \left(
\Lambda ^{2}+M_{1}^{2}\right) }\Bigg].
\end{eqnarray}

\end{document}